\documentclass[apj]{emulateapj}

\usepackage{graphicx}
\usepackage{natbib}
\usepackage{adjustbox}

\shorttitle{Chemical Abundances of PNe in the Substructures of M31 
-- II.}
\shortauthors{X.\ Fang et al.}

\begin{document}

\title{Chemical Abundances of Planetary Nebulae in the Substructures 
of M31 -- II.\\
The Extended Sample and A Comparison Study with the Outer-disk 
Group\footnotemark[$\ast$]}

\footnotetext[$\ast$]{Based on observations made with the Gran 
Telescopio Canarias, installed at the Spanish Observatorio del 
Roque de los Muchachos of Instituto de Astrof\'{i}sica de Canarias, 
in the island of La Palma.  The observations presented in this 
paper are associated with GTC programs \#GTC66-16A and \#GTC25-16B.}

\author{Xuan Fang$^{1,2}$\footnotemark[$\dagger$],
Rub\'{e}n Garc\'{i}a-Benito$^{3}$,
Mart\'{i}n A.\ Guerrero$^{3}$,
Yong Zhang$^{1,4,5}$,
Xiaowei Liu$^{6,7,8}$,\\
Christophe Morisset$^{9}$, 
Amanda I.\ Karakas$^{10}$, 
Marcelo M.\ Miller Bertolami$^{11}$, 
Haibo Yuan$^{12}$,\\
and Antonio Cabrera-Lavers$^{13,14}$}
\affil{
$^{1}$Laboratory for Space Research, Faculty of Science, University of Hong Kong, Pokfulam Road, Hong Kong, China\\
$^{2}$Department of Earth Sciences, Faculty of Science, University of Hong Kong, Pokfulam Road, Hong Kong, China\\
$^{3}$Instituto de Astrof\'{i}sica de Andaluc\'{i}a (IAA-CSIC), Glorieta de la Astronom\'{i}a s/n, E-18008 Granada, Spain\\
$^{4}$Department of Physics, Faculty of Science, University of Hong Kong, Pokfulam Road, Hong Kong, China\\
$^{5}$School of Physics and Astronomy, Sun Yat-Sen University, Zhuhai 519082, China\\
$^{6}$South-Western Institute for Astronomy Research, Yunnan University, Kunming 650500, China\\
$^{7}$Department of Astronomy, School of Physics, Peking University, Beijing 100871, China\\
$^{8}$Kavli Institute for Astronomy and Astrophysics, Peking University, Beijing 100871, China\\
$^{9}$Instituto de Astronom\'{i}a, Universidad Nacional Aut\'{o}noma de M\'{e}xico, Apdo. Postal 70264, Mexico CDMX 04510, Mexico\\
$^{10}$Monash Centre for Astrophysics, School of Physics and Astronomy, Monash University, VIC 3800, Australia\\
$^{11}$Instituto de Astrof\'{i}sica de La Plata, UNLP-CONICET, Paseo del Bosque s/n, 1900 La Plata, Argentina\\
$^{12}$Department of Astronomy, Beijing Normal University, Beijing 100875, China\\
$^{13}$GRANTECAN, Cuesta de San Jos\'{e} s/n, E-38712, Bre\~{n}a Baja, La Palma, Spain\\
$^{14}$Instituto de Astrof\'{i}sica de Canarias, V\'{i}a L\'{a}ctea s/n, La Laguna, E-38205 Tenerife, Spain} 

\footnotetext[$\dagger$]{Visiting Astronomer, Key Laboratory of 
Optical Astronomy, National Astronomical Observatories, Chinese 
Academy of Sciences (NAOC), 20A Datun Road, Chaoyang District, 
Beijing 100012, China}

\email{fangx@hku.hk}

\begin{abstract}

We report deep spectroscopy of ten planetary nebulae (PNe) in the 
Andromeda Galaxy (M31) using the 10.4\,m GTC.  Our targets reside 
in different regions of M31, including halo streams and dwarf 
satellite M32, and kinematically deviate from the extended disk. 
The temperature-sensitive [O~{\sc iii}] $\lambda$4363 line is 
observed in all PNe.  For four PNe, the GTC spectra extend 
beyond 1\,$\mu$m, enabling explicit detection of the [S~{\sc iii}] 
$\lambda$6312 and $\lambda\lambda$9069,9531 lines and thus 
determination of the [S~{\sc iii}] temperature.  Abundance ratios 
are derived and generally consistent with AGB model predictions. 
Our PNe probably all evolved from low-mass ($<$2\,$M_{\sun}$) stars, 
as analyzed with the most up-to-date post-AGB evolutionary models, 
and their main-sequence ages are mostly $\sim$2--5~Gyr.  Compared 
to the underlying, smooth, metal-poor halo of M31, our targets are 
uniformly metal-rich ([O/H]$\gtrsim-$0.4), and seem to resemble the 
younger population in the stream.  We thus speculate that our halo 
PNe formed in the Giant Stream's progenitor through extended star 
formation.  Alternatively, they might have formed from the same 
metal-rich gas as did the outer-disk PNe, but was displaced into 
their present locations as a result of galactic interactions. 
These interpretations are, although speculative, qualitatively in 
line with the current picture, as inferred from previous wide-field 
photometric surveys, that M31's halo is the result of complex 
interactions and merger processes.  The behavior of N/O of the 
combined sample of the outer-disk and our halo/substructure PNe 
signifies that hot bottom burning might actually occur at 
$<$3\,$M_{\sun}$, but careful assessment is needed.

\end{abstract}


\section{Introduction} 
\label{sec1}

In the cold dark matter ($\Lambda$CDM)-dominated universe, large 
galaxies formed hierarchically \citep[e.g.,][]{white78,wr78} 
through accretion/merger of smaller subsystems.  Such interactions 
tidally disrupt smaller galaxies and result in extended stellar 
halo surrounding the central galaxy \citep[e.g.,][]{ibata07,
ibata14}.  The relics of galaxy interaction and assemblage are 
registered into the extended halo in forms of stellar streams which, 
if detected, can be used to study the properties of galaxies and 
backtrack past interactions \citep[e.g.,][]{ibata01a,ibata01b,
ibata01c,ferguson02,majewski03,mccon09}.

\begin{table*}
\begin{center}
\caption{Observing Log and Properties of PNe}
\label{targets}
\begin{tabular}{lcccrccrccc}
\hline\hline
PN ID\,$^{\rm a}$ & R.A. & Decl. & $m_{\lambda5007}\,^{\rm b}$ 
& $v_{\rm helio}\,^{\rm c}$ & $\xi$ & $\eta$ & $R_{\rm gal}\,^{\rm d}$ 
& Location\,$^{\rm e}$ & \multicolumn{2}{c}{GTC Obs.}\\
 & (J2000.0) & (J2000.0) &  & (km\,s$^{-1}$) & ($\degr$) & ($\degr$) 
& (kpc) &  & Grism & Expos.\\
\hline
PN8  (M2430)   & 00:47:25.9 & $+$42:58:59.7 & 21.32 & $-$135.1 &    0.858 &    1.720 & 26.3 & Northern Spur & R1000B & 2$\times$2400\,s\\
PN9  (M2449)   & 00:46:13.8 & $+$42:40:28.5 & 20.88 & $-$70.6  &    0.642 &    1.408 & 21.2 & Northern Spur & R1000B & 4$\times$1200\,s\\
               &            &               &       &          &          &          &      &               & R1000R & 2$\times$1200\,s\\
PN10 (LAMOST)  & 00:44:03.1 & $+$42:27:46.6 & 20.74 & $-$234.0 &    0.242 &    1.194 & 16.7 & Northern Spur & R1000B & 4$\times$1200\,s\\
PN11 (M2432)   & 00:47:30.3 & $+$43:03:40.9 & 20.69 & $-$411.0 &    0.871 &    1.798 & 27.4 & Giant Stream  & R1000B & 4$\times$1200\,s\\
               &            &               &       &          &          &          &      &               & R1000R & 2$\times$1200\,s\\
PN12 (M2466)   & 00:49:08.0 & $+$42:28:44.3 & 21.96 & $-$392.3 &    1.179 &    1.220 & 23.3 &  Giant Stream & R1000B & 4$\times$2400\,s\\
PN13 (LAMOST)  & 00:49:55.2 & $+$38:32:49.0 & 21.91 & $-$362.0 &    1.404 & $-$2.707 & 41.8 &       SE Halo & R1000B & 4$\times$2400\,s\\
               &            &               &       &          &          &          &      &               & R1000R & 2$\times$1890\,s\\
PN14 (M2507)$^{\rm f}$
               & 00:48:27.2 & $+$39:55:34.3 & 21.23 & $-$146.9 &    1.095 & $-$1.334 & 23.7 &  Giant Stream & R1000B & 4$\times$2400\,s\\
PN15 (M2512)   & 00:45:58.5 & $+$39:13:25.4 & 21.10 & $-$318.2 &    0.627 & $-$2.042 & 29.3 &       SE Halo & R1000B & 8$\times$1200\,s\\
PN16 (M2895)   & 00:42:42.2 & $+$40:51:39.8 & 20.78 & $-$193.3 & $-$0.007 & $-$0.408 & 5.59 &           M32 & R1000B & 6$\times$1200\,s\\
PN17 (LAMOST)  & 00:53:38.6 & $+$41:09:32.1 & 21.15 & $-$437.0 &    2.052 & $-$0.078 & 28.1 & Eastern Halo\,$^{\rm g}$ & R1000B & 5$\times$2100\,s\\
               &            &               &       &          &          &          &      &               & R1000R & 2$\times$1800\,s\\
PN18 (M2234)   & 00:42:42.3 & $+$40:51:49.5 & 20.13 & $-$147.3 & $-$0.006 & $-$0.405 & 5.56 &           M32 & R1000B & 6$\times$1000\,s\\
\tableline
\multicolumn{11}{l}{NOTE. -- PN18 was discarded from analysis because 
no nebular emission lines were detected in its spectrum.} 
\end{tabular}
\begin{description}
\item[$^{\rm a}$] Number in the bracket is the ID from 
\citet{merrett06} except PN10, PN13 and PN17, which were discovered 
and identified in the LAMOST survey \citep{yuan10}.
\item[$^{\rm b}$] From \citet{merrett06}, except PN10, PN13 and PN17,
whose $m_{\lambda5007}$ are adopted from \citet{yuan10}.
\item[$^{\rm c}$] From \citet{merrett06}, except PN10, PN13 and PN17,
whose $v_{\rm helio}$ are adopted from \citet{yuan10}.
\item[$^{\rm d}$] Sky-projected galactocentric distance estimated 
at a distance of 785~kpc to M31 \citep{mccon05}.
\item[$^{\rm e}$] Here ``Halo'' means that the PN belongs to the 
outer halo, or is associated with some substructure. 
\item[$^{\rm f}$] PN nature confirmed by the LAMOST survey 
\citep{yuan10}.
\item[$^{\rm g}$] Might be associated with the NE Shelf, as 
explained in Section~\ref{sec4:part4}. 
\end{description}
\end{center}
\end{table*}

The Andromeda Galaxy (M31) is a nearby \citep[785~kpc,][]{mccon05} 
large spiral system and an ideal candidate for studying galaxy 
formation and evolution.  Wide-field surveys, such as 
PAndAS\footnote{The Pan-Andromeda Archeological Survey. URL: 
https://www.astrosci.ca/users/alan/PANDAS/Home.html}, have 
revealed in M31's outer halo a wealth of large-scale stellar 
substructures extending to nearly 150~kpc from the galactic centre 
\citep[e.g.,][]{ibata01a,ibata07,ferguson02,mccon03,mccon04,
mccon09,irwin05,tanaka10}, with the Northern Spur and the southern 
Giant Stellar Stream (hereafter the Giant Stream, \citealt{ibata01a,
caldwell10}) among the first discovered.  The Giant Stream threads 
to the southeast halo, as far as $>$4$\degr$ from the centre of 
M31 \citep{ibata01a,mccon03}.  The Northern Spur is a feature with 
enhanced density in metal-rich red giant branch (RGB) stars, 
located at $\sim$2$\degr$ towards the north \citep{ferguson02}. 

Planetary nebulae (PNe) are descendants of low- and 
intermediate-mass ($\sim$1--8\,$M_{\sun}$) stars, which account for
the majority of stellar populations in our universe.  Given their 
bright, narrow emission lines, PNe are excellent tracers of the 
chemistry, dynamics and stellar populations of their host galaxies.
In the optical spectrum of a PN, the bright [O~{\sc iii}] 
$\lambda$5007 nebular line alone can carry $\sim$10\% of the 
central star's energy \citep[e.g.,][]{schoen07}.  PNe thus are 
well detected in distant galaxies, even as far as $>$100~Mpc 
\citep[e.g.,][]{gerhard05,gerhard07,longobardi15a,longobardi15b}. 
Spectroscopy of PNe in M31, mainly in the bulge and disk 
\citep[e.g.,][]{jc99,richer99,kwitter12} has found a slightly 
negative gradient in the oxygen abundance within 50~kpc in the 
disk \citep{kwitter12}.  However, recent observations with large 
(8--10\,m) telescopes found that the outer-disk PNe, as far as 
100~kpc from the centre of M31, have nearly solar abundances 
\citep{balick13,corradi15}.  Even some of the PNe associated with 
the substructures have O/H close to the Sun \citep{fang13,fang15}.
These metal-rich PNe in the outskirts of M31 seem to have 
different origins from the ancient halo, which formed through 
galaxy mergers long time ago \citep[e.g.,][]{ibata07,ibata14}. 

One long-standing, unresolved question is what the origin 
of M31's stellar substructure is.  It has been proposed that the 
Northern Spur and the Giant Stream might be connected by a stellar 
stream \citep{ferguson02,merrett03}, of which the dwarf satellite 
M32 could be the origin \citep{ibata01a,merrett03}, but this 
hypothesis needs assessment.  In pursuit of answering this question, 
we have carried out deep spectroscopic observations of ten bright 
PNe associated with the two substructures and mostly located in the 
outer halo \citep[][hereafter Papers~I and II, 
respectively]{fang13,fang15}, and found that they are in overall 
metal-rich ([O/H]$\sim-$0.3--0) and their oxygen abundances are 
consistent within the errors (although some internal scatter 
exists). These abundance analyses led to a tempting, yet tentative, 
conclusion that the Giant Stream and the Northern Spur might have 
the same origin.  Given the vast extension and complexity of M31's 
halo \citep{ibata07,ibata14}, our sample of PNe so far observed, 
although representative, is still too limited for us to draw any 
definite conclusion. 

That both the PNe on the halo streams and those kinematically 
belonging to the extended disk of M31 have been found to be 
metal-rich ($\sim$solar) is unexpected for a classical, metal-poor 
halo, and leads to a question of whether they have the same origin 
(or even population).  A comparison study between our halo sample 
and the disk objects may shed light on this conundrum.  Previous 
attempts have proved that PNe are a very efficient probe of 
different regions of M31.  It is thus possible not only to assess 
the connection and origin of different substructures, which has 
been the motivation of our observations so far, but also to make a 
census study of the extended halo of M31, using PNe as a tool. 

In order to better understand the merging history of M31's halo, we 
recently carried out deep spectroscopy of eleven PNe: three in the 
Northern Spur, three associated with the Giant Stream, two in M32, 
and another three located in the eastern and southeast halo regions. 
The immediate objectives of the new observations are 1) to obtain 
accurate abundances (mainly oxygen) for an extended sample, 2) to 
make a comparison study with the outer-disk PNe in terms of abundance 
and stellar population, and 3) to assess whether M32 is related to 
the Northern Spur and the Giant Stream.  Being the third paper 
targeting the PNe in the substructures of M31, this paper is the 
second one of a series to report deep spectroscopy with a 10\,m class 
telescope.  Section~\ref{sec2} introduces target selection and 
describes the observations and data reduction.  Section~\ref{sec3} 
present emission line measurements, plasma diagnostics and abundance 
determinations. We present in-depth discussion in Section~\ref{sec4} 
based on the results, and give summary and conclusions in 
Section~\ref{sec5}.

\section{Observations and Data Reduction}
\label{sec2}

\subsection{Target Selection:\\ 
The Spatial and Kinematical Distribution}
\label{sec2:part1}

Before introducing target selection, we brief some definitions 
in terms of boundaries in M31 structurre.  We adopted the M31 bulge 
radius ($\sim$3.4~kpc) from the surface brightness fitting by 
\citet{irwin05}.  The inner disk of M31 is defined at $R_{25}$ 
\citep[=95\arcmin,][]{dev91}, which corresponds to 21.7~kpc at the 
distance of M31; this radius well encompasses the optical disk of 
M31.  Beyond $R_{25}$ lies the extended disk that stretches to 
40~kpc, with detections as far as $\sim$70~kpc \citep{ibata05}. 
In the current paper, all M31 PNe beyond $R_{25}$ but with 
kinematics consistent with the extended disk are dubbed the 
outer-disk PNe.  Previous spectroscopic observations of 
\citet{kwitter12}, \citet{balick13} and \citet{corradi15} all 
focused on the outer-disk PNe in M31.

\begin{figure}[ht!]
\begin{center}
\includegraphics[width=1.0\columnwidth,angle=0]{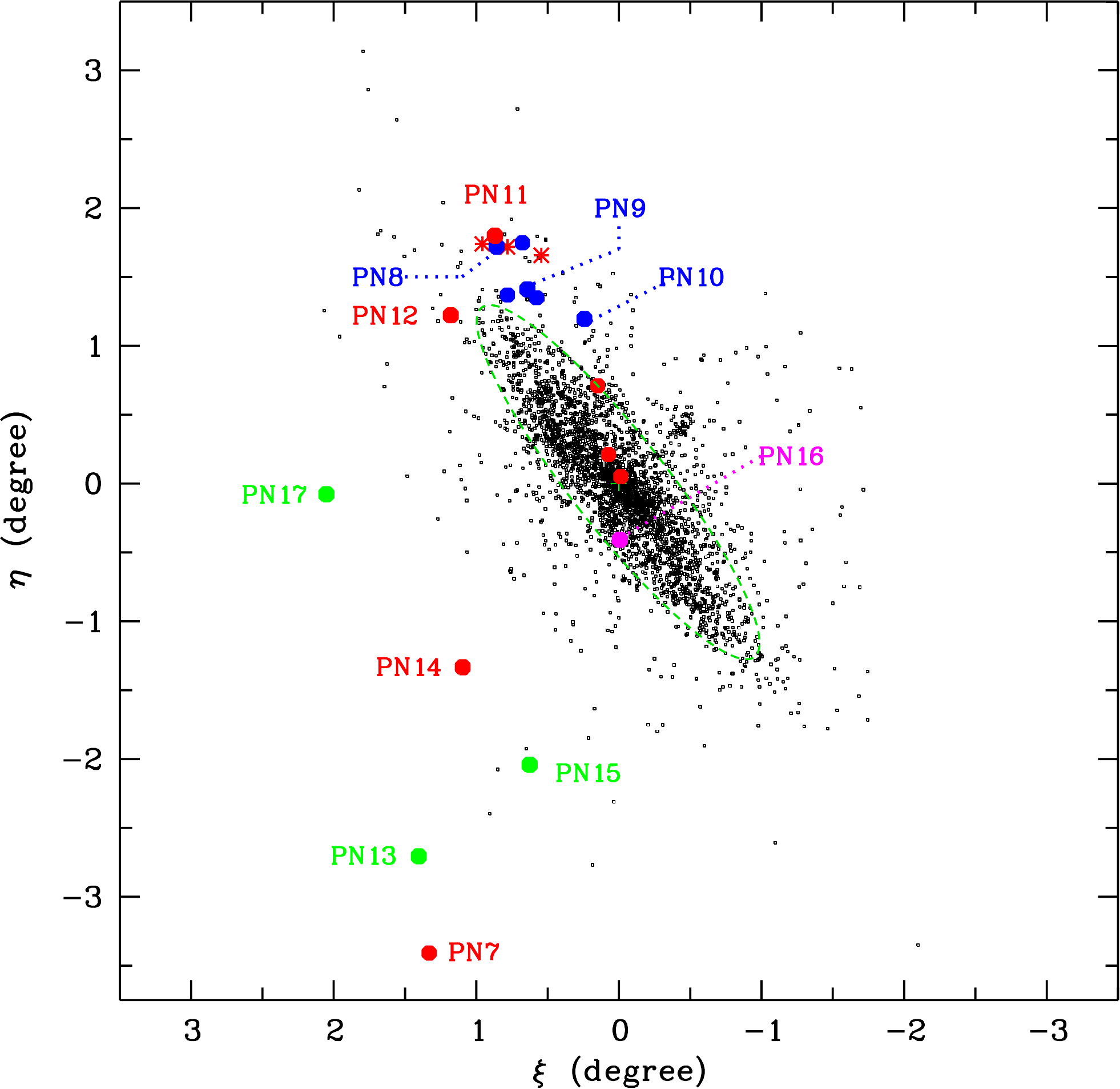}
\caption{Spatial distribution of PNe in M31.  Objects are from 
\citet{merrett06}, \citet{hkpn04}, \citet{yuan10} and 
\citet{kniazev14}.  Our GTC targets (including those in Paper~II) 
are highlighted with color-filled circles, which are color-coded 
according to their locations (see Table~\ref{targets}). 
The three Northern Spur PNe from Paper~I are indicated by red 
asterisks.  Coordinates $\xi$ and $\eta$ (given in 
Table~\ref{targets}) are the M31-based reference frame defined by 
\citet{hbk91}.  The green dashed ellipse represents the optical 
disk of M31 with radius $R_{25}$=95\arcmin\ \citep{dev91}, 
assuming an inclination angle of 77\fdg7 \citep{dev58} and a 
position angle of 37\fdg7 \citep{merrett06} for the M31 disk.} 
\label{fig1}
\end{center}
\end{figure}

In Papers~I and II, we targeted the PNe in the Northern Spur 
and the Giant Stream substructures.  Since our targets 
kinematically deviate from the extended disk of M31 and are mostly 
located in the halo, hereafter we call them the halo PNe, to avoid 
possible confusion with the outer-disk PNe. 
For the new GTC observations, we selected a sample that covers 
not only the substructures, but also more extended areas in the 
M31 system such as the eastern and the southeast halo regions and 
dwarf satellite M32.  The locations/hosts of our targets are given 
in Table~\ref{targets}, where other properties such as target 
positions (right ascension--R.A., declination--Decl.), visual 
magnitudes in [O~{\sc iii}] $\lambda$5007 ($m_{\lambda5007}$), 
heliocentric velocities $v_{\rm helio}$ (in km\,s$^{-1}$), angular 
distances to the centre of M31, and the sky-projected 
galactocentric distances (in kpc) are also presented.  Spatial 
locations of our targets, including those studied in Papers~I and 
II, are shown in Figure~\ref{fig1}.  Our halo nebulae are 
mostly outside $R_{25}$. 

We selected eight PNe from the catalog of \citet{merrett06} and 
three from \citet{yuan10}; the latter was based on a spectroscopic 
survey at the Large Sky Area Multi-Object Fiber Spectroscopic 
Telescope\footnote{Also named the Guoshoujing Telescope (GSJT). 
URL http://www.lamost.org} \citep[LAMOST,][]{su98,cui04,cui10,
cui12,zhao12}.  These new targets were named PN8--PN18 (see 
Table~\ref{targets} and Figure~\ref{fig1}), following the target 
naming (PN1--PN7) in Paper~II.  According to their locations in 
M31, our GTC samples PN1--PN17 are highlighted with different 
colors in Figure~\ref{fig1}, where PN16 and PN18 are too close to
each other and visually indistinguishable.  The [O~{\sc iii}] 
brightnesses of the new sample are $m_{\lambda5007}$ 
$\sim$20.48--21.96, extending down to nearly 1.8~mag from the 
bright-end cut-off of the planetary nebula luminosity function 
(PNLF) of M31 \citep{merrett06,ciardullo89,ciardullo10}.

\begin{figure*}
\begin{center}
\includegraphics[width=17.5cm,angle=0]{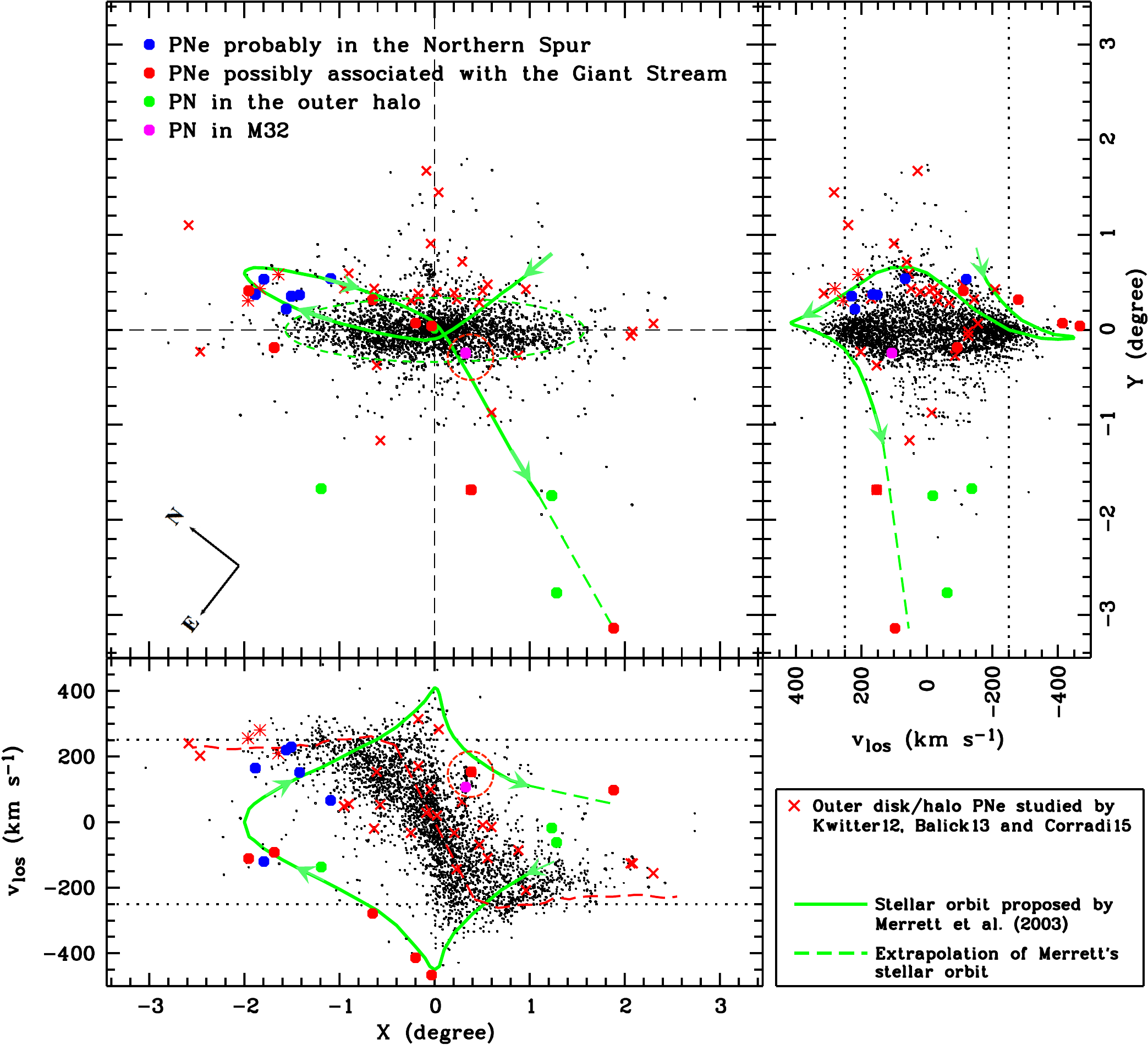}
\caption{Spatial and kinematical distribution of PNe in M31 
(description of the $X$ and $Y$ coordinates is given the text). 
Source of the PN samples are the same as in Figure~\ref{fig1}. 
The three Northern Spur PNe studied in Paper~I indicated by red 
asterisks.  The seven PNe studied in Paper~II and the ten PNe in 
this work are highlighted (see the legend).  The red crosses 
``{\bf $\times$}'' are the outer-disk PNe studied by 
\citet[][Kwitter12]{kwitter12}, \citet[][Balick13]{balick13}, and 
\citet[][Corradi15]{corradi15}. 
In the upper-left panel: over-plotted is the stellar orbit 
proposed by \citet[][thick green curve; reproduced with kind 
permission of the authors]{merrett03}. 
In the two side panels (bottom and right): a projection of the 
orbit in line-of-sight velocity with respect to M31, $v_{\rm los}$,
versus distance along the major and minor axes of M31 is 
superimposed on the PNe data.  Velocities of PNe have been 
corrected for the systemic velocity $-$306 km\,s$^{-1}$ of M31 
\citep{corbelli10}.  The red dashed curve in the bottom panel is 
the H~{\sc i} rotation curve from \citet{carignan06}.} 
\label{fig2}
\end{center}
\end{figure*}

\citet{yuan10} did not assign their newly discovered PNe to any  
locations (i.e., substructure or the extended disk).  We 
identified the locations of the three LAMOST targets (PN10, PN13 
and PN17) according to their kinematics shown in Figure~\ref{fig2}, 
which also presents the distribution of the line-of-sight velocity 
with respect to the centre of M31, $v_{\rm los}$, versus distance 
along the major and minor axes of M31.  The kinematics of PN10 
obviously deviates from the extended disk of M31 and is somewhat 
close to the Northern Spur sample identified by \citet[][Figure~32 
therein]{merrett06}.  We thus identified PN10 as a possible 
Northern Spur object.  PN13 visually resides on the southeast (SE) 
extension of the Giant Stream; PN15 also seems to be on the stream. 
However, the velocities of these two PNe, although both deviating 
from the kinematics of the extended disk, are inconsistent with the 
stellar orbit of \citet{merrett03}. 
PN17 is located in the eastern halo, 2\fdg05 from the centre 
of M31.  Its velocity differs significantly from the disk, and its
location seems to be very close to the NE Shelf \citep{ferguson05}.
We temporarily assign PN13, PN15 and PN17 to be the halo nebulae; 
detailed discussion is in Section~\ref{sec4}.  The PN nature of 
PN14 (ID~2507 in \citealt{merrett06}) was confirmed in the LAMOST 
survey; it might be associated with the Giant Stream.  For the 
other targets selected from \citet{merrett06}, we adopted their 
locations identified by the authors (Table~\ref{targets}).

\subsection{Spectroscopic Observations}
\label{sec2:part2}

Deep spectroscopy of M31 PNe were carried out with the Optical 
System for Imaging and low-intermediate-Resolution Integrated 
Spectroscopy (OSIRIS) spectrograph on the 10.4\,m Gran Telescopio 
Canarias (GTC) at Observatorio de El Roque de los Muchachos (ORM, 
La Palma).  These observations were obtained from 2016 September 2 
to 2016 September 11 for GTC program No. GTC25-16B (PI: X. Fang) 
in service mode.  The OSIRIS grism R1000B (1000 lines\,mm$^{-1}$), 
which covers $\sim$3630--7850\,{\AA}, and a long slit with 1\farcs0 
width were used.  The OSIRIS detector is a combination of two 
2048$\times$4096 CCDs.  The pixel size is 15\,$\mu$m, corresponding 
to 0\farcs127 in angular size.  We adopted the standard observing 
mode where the output images were binned by 2$\times$2.  The above 
instrument setup produces a spectral resolution of $\sim$5.5\,{\AA} 
(full width at half-maximum, FWHM) in the blue part of the spectrum 
and 6.4\,{\AA} in the red, at a dispersion of 
$\sim$2.072\,{\AA}\,pixel$^{-1}$. The ideal observing 
conditions at the ORM provided photometric and clear nights, and 
excellent seeing (0\farcs6--0\farcs8) for most of the observations.
Moon was also close to dark during the observations.  Throughout 
the observations, the long slit was placed along the parallactic 
angles to minimize light loss due to atmospheric diffraction.  The 
typical physical sizes of PNe are $\lesssim$0.5~pc 
\citep[e.g.,][]{frew16}, corresponding to $\lesssim$0\farcs13 in 
angular size at the distance of M31.  This is smaller than the 
binned CCD pixel size (0\farcs254) of OSIRIS, and thus our targets 
are all point sources and supposed to be well accommodated within 
the GTC 1\arcsec-wide long slit. 

\begin{figure}
\begin{center}
\includegraphics[width=1.0\columnwidth,angle=0]{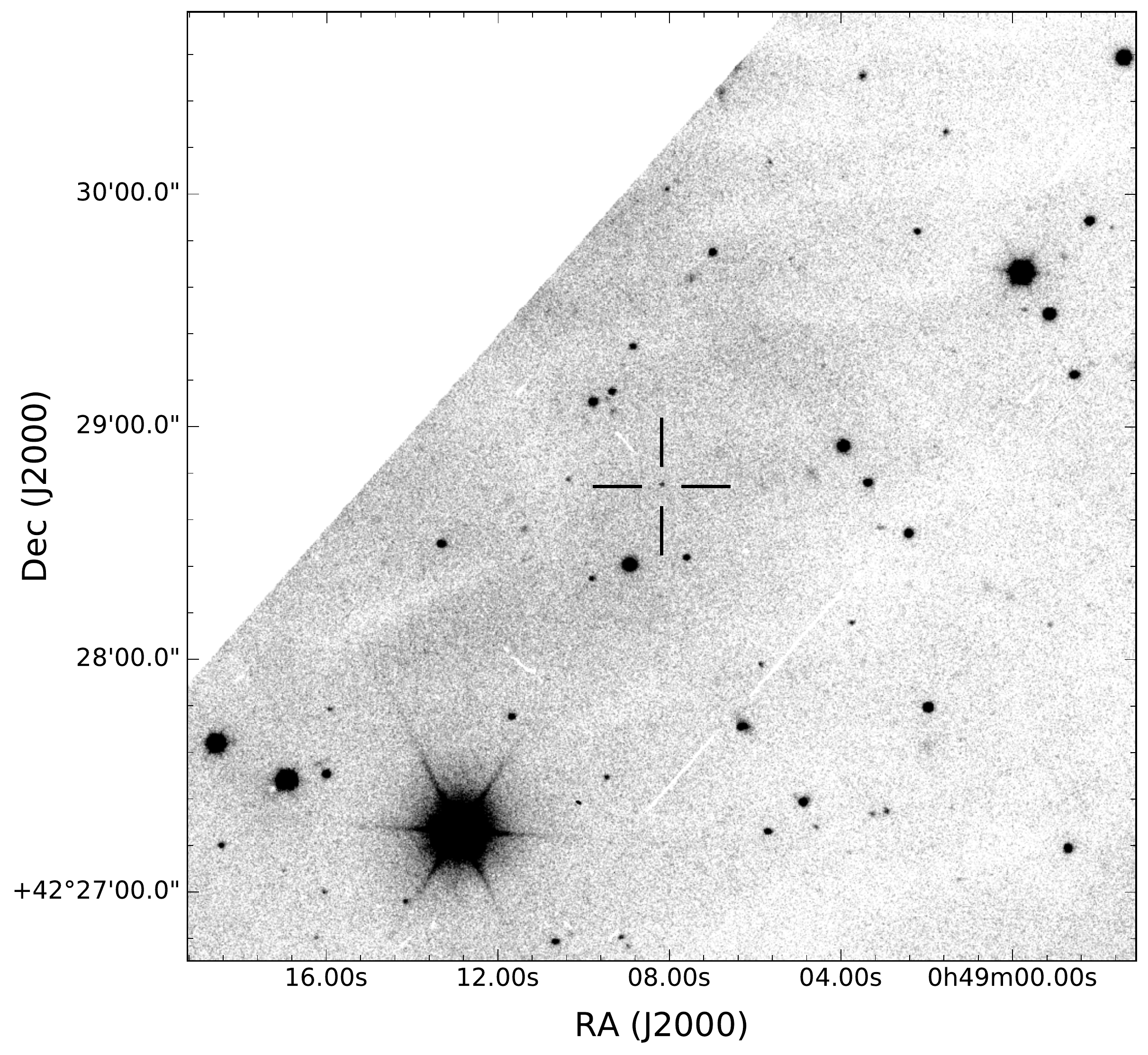}
\caption{Negative grey-scale GTC OSIRIS $g$-band acquisition image 
of PN12 (marked with the black crosshair) taken with an exposure 
of 5\,s.}
\label{fig3}
\end{center}
\end{figure}

In order to remove cosmic rays and to avoid saturation of strong 
emission lines, multiple exposures were made for each target PN. 
These exposures are summarized in Table~\ref{targets}.  In total, 
30 hours observations were completed at the GTC for eleven targets. 
Thanks to the large light-collecting area of the GTC, we could 
clearly see almost all the PNe in the direct acquisition CCD image 
with an exposure of a few seconds (e.g., Figure~\ref{fig3}), and 
then placed the GTC long slit on the targets.  Blind offset was 
utilized only for the two PNe in M32 due to their close proximity 
($\sim$7\farcs4 and 15\farcs4) to the centre of M32.  Exposures of 
spectrophotometric standard stars Ross\,640 and G191-B2B 
\citep{oke74,oke90} were made in each night to calibrate fluxes 
for the target spectra, using a slit width of 2\farcs52.  The HgAr 
and neon arc line images were obtained (with both 1\farcs0 and 
2\farcs52 slit widths) for wavelength calibration and geometric 
rectification.  Other basic calibration files, such as bias and 
spectral flats, were also obtained for both the target PN and the 
standard spectrophotometric star on each night. 

We also obtained long-slit spectroscopy of four PNe (PN9, PN11, 
PN13 and PN17) using the GTC OSIRIS red grism R1000R that covers 
$\sim$5080--10370\,{\AA}.  These observations were obtained on 2016 
August 23--24 for program No.\ GTC66-16A (PI: X. Fang).  Slit 
width was 1\farcs0, and spectral resolution FWHM$\sim$6.7\,{\AA}
in the blue region and 8.5\,{\AA} in the red, with a dispersion of 
2.59\,{\AA}\,pixel$^{-1}$.  The R1000R exposures are summarized in 
Table~\ref{targets}.  The HgAr, neon and xenon arc lines were used 
for wavelength calibration.  Spectrophotometric standard stars for 
flux calibration were the same as in the R1000B spectroscopy.  Data 
were obtained under photometric conditions, with seeing 
$\sim$0\farcs8--1\farcs0.

\subsection{Data Reduction}
\label{sec2:part3}

The GTC OSIRIS long-slit spectra were reduced using {\sc 
iraf}\footnote{{\sc iraf}, the Image Reduction and Analysis Facility,
is distributed by the National Optical Astronomy Observatory, which 
is operated by the Association of Universities for Research in 
Astronomy under cooperative agreement with the National Science 
Foundation.} v2.16.  Data reduction generally followed the standard
procedure, similar to what has been described in \citet{fang15}. 
The raw PN spectral images were first bias-subtracted and corrected 
for flat-field.  We then performed wavelength calibration using 
HgAr arc lines for the PN spectra obtained with the R1000B grism, 
and HgAr+Xe for the R1000R spectra.  Although geometry distortion 
along the long slit does not affect the nebular emission lines of 
our targets, which are point sources on CCD, such distortion of the 
sky lines must be corrected for so that background subtraction can 
be properly done.  During the wavelength calibration, we rectified 
geometry distortion by fitting the arc lines using third-order 
polynomial functions in the two-dimensional (2D) spectrogram.  This 
geometry rectification ``straightened'' the sky lines along the slit.

We subtracted the background from each single exposure of the 
target frame by fitting the background emission along the slit 
direction using high-order cubic spline functions (see more 
details in \citealt{fang15}).  We then combined the 
background-subtracted 2D frames of the same PN to remove the cosmic 
rays.  We then used the FILTER/COSMIC task in the software {\sc 
midas}\footnote{{\sc midas}, Munich Image Data Analysis System, is 
developed and distributed by the European Southern Observatory} 
v13SEPpl1.2 to further eliminate any possible cosmic residuals in 
the CCD images.  The above procedures produced a well ``cleaned'' 
spectral image for each PN, which was then flux-calibrated (and 
also corrected for the atmospheric extinction) using the spectrum 
of spectrophotometric standards.

\begin{figure*}
\begin{center}
\includegraphics[width=17.0cm,angle=0]{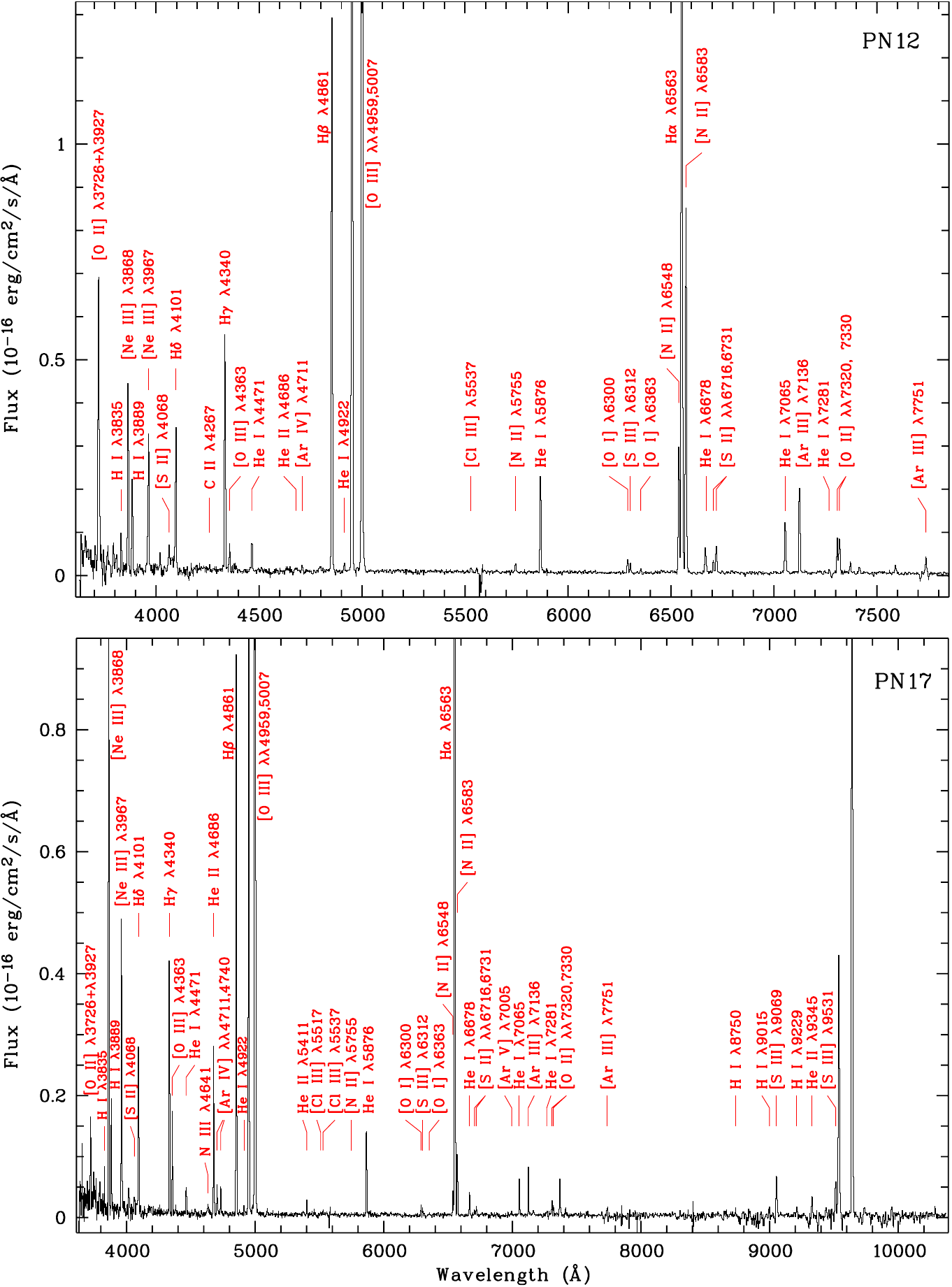}
\caption{GTC OSIRIS spectrum of PN12 (top) and PN17 (bottom). The
spectrum of PN12 was obtained using the R1000B grism, while the
spectrum of PN17 is a combination of the R1000B ($\leq$7500\,{\AA})
and R1000R ($>$7500\,{\AA}) spectra.  Vertical ranges of both 
panels are set to accommodate the intensity of H$\beta$.  All 
important emission lines are labeled.  Extinction has not been 
corrected for.  For PN12, the weak features between [O~{\sc ii}] 
$\lambda\lambda$7320,7330 and [Ar~{\sc iii}] $\lambda$7751 are the 
second-order effect.  In the spectrum of PN17, the strong emission 
features redward of [S~{\sc iii}] $\lambda$9531 are also due to the 
second-order contamination.} 
\label{fig4}
\end{center}
\end{figure*}

\begin{table*}
\begin{center}
\caption{Fluxes and Intensities}
\label{lines}
\begin{tabular}{lllcccccccccc}
\hline\hline
Ion  & $\lambda$  & Transition &
\multicolumn{2}{c}{\underline{~~~~~~~PN8~~~~~~~}} &
\multicolumn{2}{c}{\underline{~~~~~~~PN9~~~~~~~}} &
\multicolumn{2}{c}{\underline{~~~~~~PN10~~~~~~}} &
\multicolumn{2}{c}{\underline{~~~~~~PN11~~~~~~}} &
\multicolumn{2}{c}{\underline{~~~~~~PN12~~~~~~}} \\
     &  (\AA)     &            &
$F$($\lambda$) & $I$($\lambda$) &
$F$($\lambda$) & $I$($\lambda$) &
$F$($\lambda$) & $I$($\lambda$) &
$F$($\lambda$) & $I$($\lambda$) &
$F$($\lambda$) & $I$($\lambda$) \\
\hline
$[$O~{\sc ii}$]$   & 3727$^{\rm a}$ & 2p$^{3}$\,$^{4}$S$^{\rm o}$--2p$^{3}$\,$^{2}$D$^{\rm o}$         &   19.6   &  22.4$\pm$2.5  &    50.3  &   57.3$\pm$5.2  &    29.3  &   35.1$\pm$3.9  &    32.9  &   39.4$\pm$4.3  &    59.7  &   68.2$\pm$5.5 \\
H~{\sc i}          & 3798       & 2p\,$^{2}$P$^{\rm o}$--10d\,$^{2}$D                                  & $\cdots$ &      $\cdots$  &     3.72 &   4.22$\pm$0.94 &     3.34 &   3.96$\pm$0.90 &     3.15 &   3.74$\pm$0.83 &     4.55 &   5.17$\pm$1.05\\
H~{\sc i}          & 3835       & 2p\,$^{2}$P$^{\rm o}$--9d\,$^{2}$D                                   &    3.12  &  3.53$\pm$1.00 &     5.74 &   6.48$\pm$1.83 &     4.78 &   5.65$\pm$1.60 &     6.24 &   7.38$\pm$2.01 &     4.93 &   5.58$\pm$1.17\\
$[$Ne~{\sc iii}$]$ & 3868       & 2p$^{4}$\,$^{3}$P$_{2}$--2p$^{4}$\,$^{1}$D$_{2}$                     &  107     &   120$\pm$8    &    82.4  &   92.8$\pm$6.2  &    76.5  &   90.0$\pm$6.0  &   100    &    117$\pm$8    &    29.2  &   33.0$\pm$2.2 \\
H~{\sc i}          & 3889$^{\rm b}$ & 2p\,$^{2}$P$^{\rm o}$--8d\,$^{2}$D                               &    5.75  &  6.48$\pm$0.97 &    15.5  &   17.4$\pm$2.6  &    12.4  &   14.6$\pm$2.20 &    15.1  &   17.8$\pm$2.6  &    14.4  &   16.2$\pm$2.4 \\
$[$Ne~{\sc iii}$]$ & 3967$^{\rm c}$ & 2p$^{4}$\,$^{3}$P$_{1}$--2p$^{4}$\,$^{1}$D$_{2}$                 &   46.2   &  51.6$\pm$4.0  &    38.5  &   43.0$\pm$3.3  &    36.0  &   41.8$\pm$3.2  &    50.0  &   58.2$\pm$4.5  &    27.7  &   31.0$\pm$2.3 \\
He~{\sc i}         & 4026       & 2p\,$^{3}$P$^{\rm o}$--5d\,$^{3}$D                                   &    1.48  &  1.64$\pm$0.85 &     2.53 &   2.79$\pm$1.45 &     1.50 &   1.73$\pm$0.89 &     0.46 &   0.53$\pm$:    &     2.21 &   2.45$\pm$1.27\\
$[$S~{\sc ii}$]$   & 4068$^{\rm d}$ & 3p$^{3}$\,$^{4}$S$^{\rm o}_{3/2}$--3p$^{3}$\,$^{2}$P$^{\rm o}_{3/2}$ & 9.05 &  10.0$\pm$1.5  &     2.73 &   3.01$\pm$0.45 &     1.91 &   2.18$\pm$0.33 &     4.88 &   5.58$\pm$0.83 &     5.32 &   5.88$\pm$0.87\\
H~{\sc i}          & 4101       & 2p\,$^{2}$P$^{\rm o}$--6d\,$^{2}$D                                   &   27.0   &  29.8$\pm$3.1  &    23.0  &   25.2$\pm$2.6  &    22.8  &   25.9$\pm$2.7  &    29.4  &   33.5$\pm$3.4  &    27.5  &   30.3$\pm$3.1 \\
C~{\sc ii}         & 4267       & 3d\,$^{2}$D--4f\,$^{2}$F$^{\rm o}$                                   & $\cdots$ &      $\cdots$  & $\cdots$ &       $\cdots$  & $\cdots$ &       $\cdots$  &     1.38 &   1.52$\pm$0.27 &     0.79 &   0.85$\pm$0.31\\
H~{\sc i}          & 4340$^{\rm e}$ & 2p\,$^{2}$P$^{\rm o}$--5d\,$^{2}$D                               &   42.5   &  45.3$\pm$3.3  &    43.5  &   46.4$\pm$3.5  &    42.0  &   45.7$\pm$3.3  &    40.1  &   43.7$\pm$3.1  &    42.2  &   45.0$\pm$3.2 \\
$[$O~{\sc iii}$]$  & 4363       & 2p$^{2}$\,$^{1}$D$_{2}$--2p$^{2}$\,$^{1}$S$_{0}$                     &    9.43  &  10.0$\pm$1.2  &     8.20 &   8.71$\pm$1.04 &     8.19 &   8.90$\pm$1.07 &    12.4  &   13.5$\pm$1.6  &     4.20 &   4.47$\pm$0.53\\
He~{\sc i}         & 4388       & 2p\,$^{1}$P$^{\rm o}_{1}$--5d\,$^{1}$D$_{2}$                         & $\cdots$ &      $\cdots$  & $\cdots$ &       $\cdots$  & $\cdots$ &       $\cdots$  & $\cdots$ &       $\cdots$  &     1.13 &   1.20$\pm$0.54\\
He~{\sc i}         & 4471       & 2p\,$^{3}$P$^{\rm o}$--4d\,$^{3}$D                                   &    5.61  &  5.88$\pm$0.82 &     4.38 &   4.59$\pm$0.74 &     4.47 &   4.76$\pm$0.66 &     4.88 &   5.21$\pm$0.84 &     5.47 &   5.74$\pm$0.82\\
N~{\sc iii}        & 4641$^{\rm f}$ & 3p\,$^{2}$P$^{\rm o}_{3/2}$--3d\,$^{2}$D$_{5/2}$                 & $\cdots$ &      $\cdots$  & $\cdots$ &       $\cdots$  & $\cdots$ &       $\cdots$  &     3.44 &   3.57$\pm$0.72 & $\cdots$ &       $\cdots$ \\
C~{\sc iii}        & 4649$^{\rm g}$ & 3s\,$^{3}$S--3p\,$^{3}$P$^{\rm o}$                               & $\cdots$ &      $\cdots$  &     1.46 &   1.50$\pm$0.67 & $\cdots$ &       $\cdots$  & $\cdots$ &       $\cdots$  & $\cdots$ &       $\cdots$ \\
He~{\sc ii}        & 4686       & 3d\,$^{2}$D--4f\,$^{2}$F$^{\rm o}$                                   &    1.21  &  1.23$\pm$0.40 &     1.52 &   1.55$\pm$0.50 & $\cdots$ &       $\cdots$  &    13.0  &   13.2$\pm$1.1  &     0.87 &   0.89$\pm$0.18\\
$[$Ar~{\sc iv}$]$  & 4711$^{\rm h}$ & 3p$^{3}$\,$^{4}$S$^{\rm o}_{3/2}$--3p$^{3}$\,$^{2}$D$^{\rm o}_{5/2}$ & 1.72 &  1.75$\pm$0.53 &     1.14 &   1.16$\pm$0.35 &     1.34 &   1.37$\pm$0.41 &     4.28 &   4.38$\pm$0.66 &     0.81 &   0.83$\pm$0.20\\
$[$Ar~{\sc iv}$]$  & 4740       & 3p$^{3}$\,$^{4}$S$^{\rm o}_{3/2}$--3p$^{3}$\,$^{2}$D$^{\rm o}_{3/2}$ &    2.87  &  2.91$\pm$0.55 &     1.38 &   1.40$\pm$0.26 &     1.18 &   1.20$\pm$0.35 &     4.32 &   4.40$\pm$0.82 & $\cdots$ &       $\cdots$ \\
H~{\sc i}          & 4861$^{\rm e}$ & 2p\,$^{2}$P$^{\rm o}$--4d\,$^{2}$D                               &  100     &   100          &   100    &    100          &   100    &    100          &   100    &    100          &   100    &    100         \\
He~{\sc i}         & 4922       & 2p\,$^{1}$P$^{\rm o}_{1}$--4d\,$^{1}$D$_{2}$                         &    0.74  &  0.73$\pm$:    &     0.72 &   0.71$\pm$:    &     1.16 &   1.15$\pm$0.63 &     1.03 &   1.02$\pm$0.61 &     1.49 &   1.48$\pm$0.67\\
$[$O~{\sc iii}$]$  & 4959       & 2p$^{2}$\,$^{1}$P$_{1}$--2p$^{2}$\,$^{1}$D$_{2}$                     &  438     &   433$\pm$17   &   425    &    420$\pm$16   &   354    &    349$\pm$13   &   474.1  &    467$\pm$18   &   189    &    187$\pm$7   \\
$[$O~{\sc iii}$]$  & 5007       & 2p$^{2}$\,$^{1}$P$_{2}$--2p$^{2}$\,$^{1}$D$_{2}$                     & 1307     &  1286$\pm$25   &  1286    &   1266$\pm$24   &  1068    &   1046$\pm$20   &  1450    &   1420$\pm$26   &   569    &    560$\pm$11  \\
$[$N~{\sc i}$]$    & 5198$^{\rm i}$ & 2p$^{3}$\,$^{4}$S$^{\rm o}_{3/2}$--2p$^{3}$\,$^{2}$D$^{\rm o}_{3/2}$ & 0.77 &  0.74$\pm$:    & $\cdots$ &       $\cdots$  &     0.56 &   0.53$\pm$:    & $\cdots$ &       $\cdots$  & $\cdots$ &       $\cdots$ \\
He~{\sc ii}        & 5411       & 4f\,$^{2}$F$^{\rm o}$--7g\,$^{2}$G                                   & $\cdots$ &      $\cdots$  & $\cdots$ &       $\cdots$  & $\cdots$ &       $\cdots$  &     1.47 &   1.37$\pm$0.36 & $\cdots$ &       $\cdots$ \\
$[$Cl~{\sc iii}$]$ & 5537       & 3p$^{3}$\,$^{4}$S$^{\rm o}_{3/2}$--3p$^{3}$\,$^{2}$D$^{\rm o}_{3/2}$ &    1.71  &  1.60$\pm$0.47 & $\cdots$ &       $\cdots$  & $\cdots$ &       $\cdots$  &     0.55 &   0.51$\pm$0.28 &     0.58 &   0.54$\pm$0.31\\
$[$N~{\sc ii}$]$   & 5755       & 2p$^{2}$\,$^{1}$D$_{2}$--2p$^{2}$\,$^{1}$S$_{0}$                     &    2.42  &  2.24$\pm$0.49 &     1.46 &   1.36$\pm$0.30 &     1.10 &   1.00$\pm$0.22 &     1.04 &   0.94$\pm$0.20 &     1.47 &   1.36$\pm$0.26\\
C~{\sc iv}         & 5805       & 3s\,$^{2}$S--3p\,$^{2}$P$^{\rm o}$                                   & $\cdots$ &      $\cdots$  &     9.29 &   8.59$\pm$0.95 & $\cdots$ &       $\cdots$  & $\cdots$ &       $\cdots$  & $\cdots$ &       $\cdots$ \\
He~{\sc i}         & 5876       & 2p\,$^{3}$P$^{\rm o}$--3d\,$^{3}$D                                   &   17.5   &  16.1$\pm$1.8  &    18.5  &   17.1$\pm$1.8  &    15.6  &   14.0$\pm$1.2  &    16.4  &   14.6$\pm$1.3  &    17.2  &   15.8$\pm$1.7 \\
$[$O~{\sc i}$]$    & 6300       & 2p$^{4}$\,$^{3}$P$_{2}$--2p$^{4}$\,$^{1}$D$_{2}$                     &    4.22  &  3.78$\pm$1.24 &     7.19 &   6.46$\pm$2.12 &     5.12 &   4.42$\pm$1.45 &     2.38 &   2.05$\pm$0.67 &     2.38 &   2.13$\pm$0.50\\
$[$S~{\sc iii}$]$  & 6312$^{\rm j}$ & 3p$^{2}$\,$^{1}$D$_{2}$--3p$^{2}$\,$^{1}$S$_{0}$                 &    2.87  &  2.57$\pm$0.64 &     1.25 &   1.13$\pm$0.41 &     1.31 &   1.13$\pm$0.41 &     1.75 &   1.50$\pm$0.54 &     1.45 &   1.30$\pm$0.47\\
$[$O~{\sc i}$]$    & 6363       & 2p$^{4}$\,$^{3}$P$_{1}$--2p$^{4}$\,$^{1}$D$_{2}$                     &    1.60  &  1.43$\pm$1.20 &     1.52 &   1.36$\pm$1.14 &     1.64 &   1.41$\pm$1.18 &     0.62 &   0.53$\pm$0.44 &     0.51 &   0.46$\pm$0.40\\
$[$N~{\sc ii}$]$   & 6548       & 2p$^{2}$\,$^{3}$P$_{1}$--2p$^{2}$\,$^{1}$D$_{2}$                     &   15.8   &  14.0$\pm$1.5  &    19.8  &   17.5$\pm$2.0  &    10.2  &   8.65$\pm$0.92 &    13.0  &   11.0$\pm$1.2  &    23.6  &   20.9$\pm$2.4 \\
H~{\sc i}          & 6563$^{\rm e}$ & 2p\,$^{2}$P$^{\rm o}$--3d\,$^{2}$D                               &  304     &   284$\pm$13   &   322    &    285$\pm$15   &   287    &    283$\pm$12   &   293    &    283$\pm$13   &   301    &    284$\pm$14  \\
$[$N~{\sc ii}$]$   & 6583       & 2p$^{2}$\,$^{3}$P$_{2}$--2p$^{2}$\,$^{1}$D$_{2}$                     &   53.7   &  47.4$\pm$4.2  &    62.0  &   55.0$\pm$4.9  &    29.8  &   25.2$\pm$2.2  &    39.8  &   33.6$\pm$2.9  &    73.0  &   64.4$\pm$5.7 \\
He~{\sc i}         & 6678$^{\rm k}$ & 2p\,$^{1}$P$^{\rm o}_{1}$--3d\,$^{1}$D$_{2}$                     &    5.26  &  4.62$\pm$0.65 &     4.73 &   4.17$\pm$0.58 &     4.00 &   3.36$\pm$0.47 &     4.35 &   3.65$\pm$0.50 &     4.64 &   4.07$\pm$0.56\\
$[$S~{\sc ii}$]$   & 6716       & 3p$^{3}$\,$^{4}$S$^{\rm o}_{3/2}$--3p$^{3}$\,$^{2}$D$^{\rm o}_{5/2}$ &    1.87  &  1.64$\pm$0.34 &     2.42 &   2.13$\pm$0.44 &     1.03 &   0.86$\pm$0.20 &     2.23 &   1.86$\pm$0.38 &     2.38 &   2.10$\pm$0.43\\
$[$S~{\sc ii}$]$   & 6731       & 3p$^{3}$\,$^{4}$S$^{\rm o}_{3/2}$--3p$^{3}$\,$^{2}$D$^{\rm o}_{3/2}$ &    2.31  &  2.02$\pm$0.38 &     4.23 &   3.71$\pm$0.50 &     1.68 &   1.41$\pm$0.26 &     4.24 &   3.54$\pm$0.42 &     4.98 &   4.36$\pm$0.44\\
He~{\sc i}         & 7065       & 2p\,$^{3}$P$^{\rm o}$--3s\,$^{3}$S                                   &    9.96  &  8.55$\pm$1.03 &    11.5  &   9.93$\pm$1.20 &    10.7  &   8.76$\pm$1.05 &     8.25 &   6.72$\pm$0.81 &    10.7  &   9.23$\pm$1.11\\
$[$Ar~{\sc iii}$]$ & 7136       & 3p$^{4}$\,$^{3}$P$_{2}$--3p$^{4}$\,$^{1}$D$_{2}$                     &   19.3   &  16.5$\pm$1.6  &    14.8  &   12.7$\pm$1.23 &    12.3  &   10.0$\pm$1.10 &    18.0  &   14.5$\pm$1.4  &    17.1  &   14.7$\pm$1.4 \\
He~{\sc i}         & 7281       & 2p\,$^{1}$P$^{\rm o}_{1}$--3s\,$^{1}$S$_{0}$                         & $\cdots$ &      $\cdots$  &     0.43 &   0.36$\pm$:    & $\cdots$ &       $\cdots$  &     0.44 &   0.36$\pm$:    &     0.34 &   0.29$\pm$:   \\
$[$O~{\sc ii}$]$   & 7320       & 2p$^{3}$\,$^{2}$D$^{\rm o}_{5/2}$--2p$^{3}$\,$^{2}$P$^{\rm o}_{3/2}$ &   10.1   &  8.59$\pm$1.11 &     6.88 &   5.84$\pm$0.75 &     7.91 &   6.33$\pm$0.82 &     2.36 &   1.88$\pm$0.24 &     6.55 &   5.55$\pm$0.71\\
$[$O~{\sc ii}$]$   & 7330       & 2p$^{3}$\,$^{2}$D$^{\rm o}_{3/2}$--2p$^{3}$\,$^{2}$P$^{\rm o}_{3/2}$ &    8.54  &  7.23$\pm$1.10 &     5.78 &   4.91$\pm$0.74 &     6.19 &   4.95$\pm$0.75 &     1.83 &   1.46$\pm$0.22 &     7.20 &   6.10$\pm$0.91\\
$[$Ar~{\sc iii}$]$ & 7751       & 3p$^{4}$\,$^{3}$P$_{1}$--3p$^{4}$\,$^{1}$D$_{2}$                     &    3.05  &  2.52$\pm$0.63 &     2.24 &   1.86$\pm$0.46 &     2.24 &   1.74$\pm$0.78 &     3.89 &   3.00$\pm$0.74 &     2.96 &   2.45$\pm$0.60\\
H~{\sc i}          & 8750       & 3d\,$^{2}$D--12f\,$^{2}$F$^{\rm o}$                                  & $\cdots$ &      $\cdots$  &     1.72 &   1.36$\pm$0.41 & $\cdots$ &       $\cdots$  &     1.57 &   1.14$\pm$0.34 & $\cdots$ &       $\cdots$ \\
H~{\sc i}          & 9015       & 3d\,$^{2}$D--10f\,$^{2}$F$^{\rm o}$                                  & $\cdots$ &      $\cdots$  &     1.85 &   1.45$\pm$0.43 & $\cdots$ &       $\cdots$  &     1.31 &   0.94$\pm$0.27 & $\cdots$ &       $\cdots$ \\
$[$S~{\sc iii}$]$  & 9069       & 3p$^{2}$\,$^{3}$P$_{1}$--3p$^{2}$\,$^{1}$D$_{2}$                     & $\cdots$ &      $\cdots$  &    21.5  &   17.0$\pm$1.4  & $\cdots$ &       $\cdots$  &    22.7  &   16.2$\pm$1.3  & $\cdots$ &       $\cdots$ \\
H~{\sc i}          & 9229       & 3d\,$^{2}$D--9f\,$^{2}$F$^{\rm o}$                                   & $\cdots$ &      $\cdots$  &     2.26 &   1.76$\pm$0.62 & $\cdots$ &       $\cdots$  &     7.52 &   5.34$\pm$1.88 & $\cdots$ &       $\cdots$ \\
He~{\sc ii}        & 9345       & 5g\,$^{2}$G--8h\,$^{2}$H$^{\rm o}$                                   & $\cdots$ &      $\cdots$  &     5.14 &   4.00$\pm$1.3  & $\cdots$ &       $\cdots$  &     3.55 &   2.51$\pm$0.81 & $\cdots$ &       $\cdots$ \\
$[$S~{\sc iii}$]$  & 9531$^{\rm l}$ & 3p$^{2}$\,$^{3}$P$_{1}$--3p$^{2}$\,$^{1}$D$_{2}$                 & $\cdots$ &      $\cdots$  &    17.4  &   13.5$\pm$1.6  & $\cdots$ &       $\cdots$  &    14.4  &   10.1$\pm$1.2  & $\cdots$ &       $\cdots$ \\
\\
$c$(H$\beta$) & &                         & \multicolumn{2}{c}{0.181} & \multicolumn{2}{c}{0.177} & \multicolumn{2}{c}{0.243} & \multicolumn{2}{c}{0.245} & \multicolumn{2}{c}{0.180}\\
\multicolumn{2}{l}{$\log$\,$F$(H$\beta$)$^{\rm m}$}  &  & \multicolumn{2}{c}{$-$15.29} & \multicolumn{2}{c}{$-$15.09} & \multicolumn{2}{c}{$-$14.92} & \multicolumn{2}{c}{$-$15.04} & \multicolumn{2}{c}{$-$15.09}\\
\hline\hline
Ion  & $\lambda$  & Transition &
\multicolumn{2}{c}{\underline{~~~~~~PN13~~~~~~}} &
\multicolumn{2}{c}{\underline{~~~~~~PN14~~~~~~}} &
\multicolumn{2}{c}{\underline{~~~~~~PN15~~~~~~}} &
\multicolumn{2}{c}{\underline{~~~~~~PN16~~~~~~}} &
\multicolumn{2}{c}{\underline{~~~~~~PN17~~~~~~}} \\
     &  (\AA)     &            &
$F$($\lambda$) & $I$($\lambda$) &
$F$($\lambda$) & $I$($\lambda$) &
$F$($\lambda$) & $I$($\lambda$) &
$F$($\lambda$) & $I$($\lambda$) &
$F$($\lambda$) & $I$($\lambda$) \\
\hline
$[$O~{\sc ii}$]$   & 3727$^{\rm a}$ & 2p$^{3}$\,$^{4}$S$^{\rm o}$--2p$^{3}$\,$^{2}$D$^{\rm o}$          &   44.1  &   50.1$\pm$5.8  &   24.2  &   27.5$\pm$3.0  &    63.2  &   74.4$\pm$8.2  &    71.5  &   76.3$\pm$9.2  &    23.5  &   27.1$\pm$2.4 \\
H~{\sc i}          & 3798       & 2p\,$^{2}$P$^{\rm o}$--10d\,$^{2}$D                                  & $\cdots$ &       $\cdots$  &    5.68 &   6.44$\pm$1.43 & $\cdots$ &       $\cdots$ &     13.0  &   13.8$\pm$2.8  &     4.21 &   4.83$\pm$1.01\\
H~{\sc i}          & 3835       & 2p\,$^{2}$P$^{\rm o}$--9d\,$^{2}$D                                    &    3.72 &   4.19$\pm$1.10 &    3.82 &   4.32$\pm$1.00 &     3.07 &   3.57$\pm$0.91 & $\cdots$ &       $\cdots$  &     5.63 &   6.44$\pm$1.14\\
$[$Ne~{\sc iii}$]$ & 3868       & 2p$^{4}$\,$^{3}$P$_{2}$--2p$^{4}$\,$^{1}$D$_{2}$                      &   43.0  &   48.1$\pm$3.1  &   40.2  &   45.2$\pm$2.8  &    80.5  &   93.2$\pm$4.7  &   122    &    130$\pm$10.4 &    93.0  &    106$\pm$8   \\
H~{\sc i}          & 3889$^{\rm b}$ & 2p\,$^{2}$P$^{\rm o}$--8d\,$^{2}$D                                &   12.6  &   14.1$\pm$2.0  &   13.3  &   15.0$\pm$2.1  &    12.3  &   14.2$\pm$1.9  &     3.87 &   4.10$\pm$2.1  &    16.4  &   18.6$\pm$2.6 \\
$[$Ne~{\sc iii}$]$ & 3967$^{\rm c}$ & 2p$^{4}$\,$^{3}$P$_{1}$--2p$^{4}$\,$^{1}$D$_{2}$                  &   28.2  &   31.3$\pm$2.4  &   27.3  &   30.4$\pm$2.5  &    41.3  &   47.2$\pm$2.8  &    60.2  &   63.5$\pm$6.4  &    51.2  &   57.7$\pm$4.7 \\
He~{\sc i}         & 4026       & 2p\,$^{3}$P$^{\rm o}$--5d\,$^{3}$D                                    &    1.73 &   1.91$\pm$0.78 &    1.89 &   2.09$\pm$0.85 &     1.03 &   1.17$\pm$0.67 & $\cdots$ &       $\cdots$  &     3.18 &   3.56$\pm$0.64\\
$[$S~{\sc ii}$]$   & 4068$^{\rm d}$ & 3p$^{3}$\,$^{4}$S$^{\rm o}_{3/2}$--3p$^{3}$\,$^{2}$P$^{\rm o}_{3/2}$ & 2.56 &   2.81$\pm$0.62 &    5.57 &   6.14$\pm$0.91 &     4.10 &   4.61$\pm$0.86 &     8.41 &   8.82$\pm$1.06 &     2.40 &   2.67$\pm$0.53\\
\end{tabular}
\end{center}
\end{table*}

\addtocounter{table}{-1}
\begin{table*}
\begin{center}
\caption{(Continued)}
\label{lines}
\begin{tabular}{lllcccccccccc}
\hline\hline
H~{\sc i}          & 4101       & 2p\,$^{2}$P$^{\rm o}$--6d\,$^{2}$D                                    &   22.8  &   25.0$\pm$2.4  &   24.9  &   27.4$\pm$2.8  &    24.4  &   27.4$\pm$2.0  &    25.2  &   26.4$\pm$3.7  &    28.0  &   31.1$\pm$2.2 \\
C~{\sc iii}        & 4187       & 4f\,$^{1}$F$^{\rm o}_{3}$--5g\,$^{1}$G$_{4}$                             & $\cdots$ &       $\cdots$  & $\cdots$ &       $\cdots$  & $\cdots$ &       $\cdots$  & $\cdots$ &       $\cdots$  &     1.03 &   1.13$\pm$0.24\\
He~{\sc ii}        & 4199       & 4f\,$^{2}$F$^{\rm o}$--11g\,$^{2}$G                                      & $\cdots$ &       $\cdots$  & $\cdots$ &       $\cdots$  & $\cdots$ &       $\cdots$  & $\cdots$ &       $\cdots$  &     1.05 &   1.15$\pm$0.27\\
H~{\sc i}          & 4340$^{\rm e}$ & 2p\,$^{2}$P$^{\rm o}$--5d\,$^{2}$D                                   &    42.0  &   44.5$\pm$3.2  &    40.8  &   43.5$\pm$3.0  &    39.3  &   42.5$\pm$3.1  &    60.2  &   62.1$\pm$6.5  &    41.2  &   44.2$\pm$3.2 \\
$[$O~{\sc iii}$]$  & 4363       & 2p$^{2}$\,$^{1}$D$_{2}$--2p$^{2}$\,$^{1}$S$_{0}$                         &     4.02 &   4.26$\pm$0.51 &     5.62 &   5.97$\pm$0.67 &     5.31 &   5.73$\pm$0.68 &    11.0  &   11.4$\pm$1.3  &    15.6  &   16.7$\pm$1.7 \\
He~{\sc i}         & 4388       & 2p\,$^{1}$P$^{\rm o}_{1}$--5d\,$^{1}$D$_{2}$                             & $\cdots$ &       $\cdots$  & $\cdots$ &       $\cdots $ & $\cdots$ &       $\cdots$  & $\cdots$ &       $\cdots$  &     1.27 &   1.36$\pm$0.31\\
He~{\sc i}         & 4471       & 2p\,$^{3}$P$^{\rm o}$--4d\,$^{3}$D                                       &     5.23 &   5.47$\pm$0.88 &     4.31 &   4.52$\pm$0.70 &     4.45 &   4.72$\pm$0.76 &     4.91 &   5.02$\pm$0.71 &     4.03 &   4.24$\pm$0.68\\
N~{\sc iii}        & 4641$^{\rm f}$ & 3p\,$^{2}$P$^{\rm o}_{3/2}$--3d\,$^{2}$D$_{5/2}$                     & $\cdots$ &       $\cdots$  & $\cdots$ &       $\cdots$  & $\cdots$ &       $\cdots$  & $\cdots$ &       $\cdots$  &     2.63 &   2.71$\pm$0.52\\
C~{\sc iv}         & 4658       & 5g\,$^{2}$G--6h\,$^{2}$H$^{\rm o}$                                       & $\cdots$ &       $\cdots$  & $\cdots$ &       $\cdots$  &     1.52 &   1.57$\pm$0.22 & $\cdots$ &       $\cdots$  & $\cdots$ &       $\cdots$ \\
He~{\sc ii}        & 4686       & 3d\,$^{2}$D--4f\,$^{2}$F$^{\rm o}$                                       & $\cdots$ &       $\cdots$  & $\cdots$ &       $\cdots$  &     2.03 &   2.08$\pm$0.24 &    21.5  &   21.7$\pm$2.3  &    26.0  &   26.6$\pm$2.4 \\
$[$Ar~{\sc iv}$]$  & 4711$^{\rm h}$ & 3p$^{3}$\,$^{4}$S$^{\rm o}_{3/2}$--3p$^{3}$\,$^{2}$D$^{\rm o}_{5/2}$ &     0.94 &   0.96$\pm$0.17 & $\cdots$ &       $\cdots$  &     1.06 &   1.10$\pm$0.17 & $\cdots$ &       $\cdots$  &     4.95 &   5.05$\pm$0.90\\
$[$Ar~{\sc iv}$]$  & 4740       & 3p$^{3}$\,$^{4}$S$^{\rm o}_{3/2}$--3p$^{3}$\,$^{2}$D$^{\rm o}_{3/2}$     & $\cdots$ &       $\cdots$  & $\cdots$ &       $\cdots$  &     1.40 &   1.43$\pm$0.20 & $\cdots$ &       $\cdots$  &     4.76 &   4.83$\pm$0.91\\
H~{\sc i}          & 4861$^{\rm e}$ & 2p\,$^{2}$P$^{\rm o}$--4d\,$^{2}$D                                   &   100    &    100          &   100    &    100          &   100    &    100          &   100    &    100          &   100    &    100         \\
He~{\sc i}         & 4922       & 2p\,$^{1}$P$^{\rm o}_{1}$--4d\,$^{1}$D$_{2}$                             &     1.48 &   1.47$\pm$0.87 &     0.73 &   0.73$\pm$0.43 &     1.96 &   1.94$\pm$0.64 & $\cdots$ &       $\cdots$  &     1.01 &   1.00$\pm$0.40\\
$[$O~{\sc iii}$]$  & 4959       & 2p$^{2}$\,$^{1}$P$_{1}$--2p$^{2}$\,$^{1}$D$_{2}$                         &   295    &    292$\pm$11   &   232    &    230$\pm$9    &   427    &    421$\pm$16   &   564    &    561$\pm$28   &   497    &    491$\pm$18  \\
$[$O~{\sc iii}$]$  & 5007       & 2p$^{2}$\,$^{1}$P$_{2}$--2p$^{2}$\,$^{1}$D$_{2}$                         &   891    &    878$\pm$16   &   710    &    699$\pm$13   &  1275    &   1250$\pm$23   &  1693    &   1680$\pm$31   &  1519    &   1493$\pm$27  \\
$[$N~{\sc i}$]$    & 5198$^{\rm i}$ & 2p$^{3}$\,$^{4}$S$^{\rm o}_{3/2}$--2p$^{3}$\,$^{2}$D$^{\rm o}_{3/2}$ & $\cdots$ &       $\cdots$  &     0.65 &   0.63$\pm$:    & $\cdots$ &       $\cdots$  & $\cdots$ &       $\cdots$  & $\cdots$ &       $\cdots$ \\
He~{\sc ii}        & 5411       & 4f\,$^{2}$F$^{\rm o}$--7g\,$^{2}$G                                       & $\cdots$ &       $\cdots$  & $\cdots$ &       $\cdots$  & $\cdots$ &       $\cdots$  & $\cdots$ &       $\cdots$  &     1.95 &   1.85$\pm$0.33\\
$[$Cl~{\sc iii}$]$ & 5517       & 3p$^{3}$\,$^{4}$S$^{\rm o}_{3/2}$--3p$^{3}$\,$^{2}$D$^{\rm o}_{5/2}$     & $\cdots$ &       $\cdots$  & $\cdots$ &       $\cdots$  & $\cdots$ &       $\cdots$  & $\cdots$ &       $\cdots$  &     0.46 &   0.43$\pm$:   \\
$[$Cl~{\sc iii}$]$ & 5537       & 3p$^{3}$\,$^{4}$S$^{\rm o}_{3/2}$--3p$^{3}$\,$^{2}$D$^{\rm o}_{3/2}$     & $\cdots$ &       $\cdots$  & $\cdots$ &       $\cdots$  & $\cdots$ &       $\cdots$  & $\cdots$ &       $\cdots$  &     0.57 &   0.53$\pm$0.26\\
$[$N~{\sc ii}$]$   & 5755       & 2p$^{2}$\,$^{1}$D$_{2}$--2p$^{2}$\,$^{1}$S$_{0}$                         &     0.43 &   0.40$\pm$:    &     1.12 &   1.04$\pm$0.22 &     1.26 &   1.15$\pm$0.21 & $\cdots$ &       $\cdots$  &     0.34 &   0.31$\pm$:   \\
C~{\sc iv}         & 5805       & 3s\,$^{2}$S--3p\,$^{2}$P$^{\rm o}$                                       & $\cdots$ &       $\cdots$  & $\cdots$ &       $\cdots$  &    16.1  &   14.6$\pm$1.6  & $\cdots$ &       $\cdots$  & $\cdots$ &       $\cdots$ \\
He~{\sc i}         & 5876       & 2p\,$^{3}$P$^{\rm o}$--3d\,$^{3}$D                                       &    15.0  &   13.8$\pm$1.2  &    16.1  &   14.8$\pm$1.4  &    17.0  &   15.3$\pm$1.3  &    15.5  &   15.0$\pm$1.4  &    14.8  &   13.5$\pm$1.2 \\
$[$O~{\sc i}$]$    & 6300       & 2p$^{4}$\,$^{3}$P$_{2}$--2p$^{4}$\,$^{1}$D$_{2}$                         &     3.42 &   3.08$\pm$0.61 &     1.49 &   1.34$\pm$0.30 &     3.75 &   3.28$\pm$0.65 & $\cdots$ &       $\cdots$  &     1.54 &   1.37$\pm$0.27\\
$[$S~{\sc iii}$]$  & 6312$^{\rm j}$ & 3p$^{2}$\,$^{1}$D$_{2}$--3p$^{2}$\,$^{1}$S$_{0}$                     &     1.97 &   1.78$\pm$0.38 &     1.38 &   1.24$\pm$0.34 &     2.53 &   2.22$\pm$0.52 & $\cdots$ &       $\cdots$  &     0.92 &   0.82$\pm$0.26\\
$[$O~{\sc i}$]$    & 6363       & 2p$^{4}$\,$^{3}$P$_{1}$--2p$^{4}$\,$^{1}$D$_{2}$                         &     1.27 &   1.14$\pm$0.54 &     0.39 &   0.35$\pm$0.18 &     1.25 &   1.09$\pm$0.53 & $\cdots$ &       $\cdots$  &     0.48 &   0.43$\pm$:   \\
$[$N~{\sc ii}$]$   & 6548       & 2p$^{2}$\,$^{3}$P$_{1}$--2p$^{2}$\,$^{1}$D$_{2}$                         &    10.8  &   9.62$\pm$1.05 &     8.77 &   7.77$\pm$0.95 &    22.0  &   18.9$\pm$2.1  &    72.3  &   68.1$\pm$7.4  &     4.25 &   3.72$\pm$0.40\\
H~{\sc i}          & 6563$^{\rm e}$ & 2p\,$^{2}$P$^{\rm o}$--3d\,$^{2}$D                                   &   290    &    286$\pm$13   &   256    &    284$\pm$12   &   281    &    287$\pm$12   &   290    &    285$\pm$14   &   299    &    282$\pm$13  \\
$[$N~{\sc ii}$]$   & 6583       & 2p$^{2}$\,$^{3}$P$_{2}$--2p$^{2}$\,$^{1}$D$_{2}$                         &    30.8  &   27.3$\pm$2.0  &    24.6  &   21.8$\pm$1.8  &    63.8  &   54.7$\pm$4.0  &   255    &    240$\pm$18   &    11.4  &   9.93$\pm$0.72\\
He~{\sc i}         & 6678$^{\rm k}$ & 2p\,$^{1}$P$^{\rm o}_{1}$--3d\,$^{1}$D$_{2}$                         &     4.08 &   3.60$\pm$0.49 &     4.26 &   3.75$\pm$0.58 &     4.24 &   3.61$\pm$0.49 & $\cdots$ &       $\cdots$  &     4.15 &   3.60$\pm$0.49\\
$[$S~{\sc ii}$]$   & 6716       & 3p$^{3}$\,$^{4}$S$^{\rm o}_{3/2}$--3p$^{3}$\,$^{2}$D$^{\rm o}_{5/2}$     &     2.12 &   1.87$\pm$0.34 &     1.00 &   0.87$\pm$0.18 &     3.23 &   2.74$\pm$0.50 &    12.1  &   11.3$\pm$2.0  &     0.90 &   0.78$\pm$0.14\\
$[$S~{\sc ii}$]$   & 6731       & 3p$^{3}$\,$^{4}$S$^{\rm o}_{3/2}$--3p$^{3}$\,$^{2}$D$^{\rm o}_{3/2}$     &     4.40 &   3.87$\pm$0.52 &     1.45 &   1.27$\pm$0.19 &     5.29 &   4.49$\pm$0.60 &    16.8  &   15.8$\pm$2.1  &     1.53 &   1.32$\pm$0.17\\
$[$Ar~{\sc v}$]$   & 7005       & 3p$^{2}$\,$^{3}$P$_{2}$--3p$^{2}$\,$^{1}$D$_{2}$                         & $\cdots$ &       $\cdots$  & $\cdots$ &       $\cdots$  & $\cdots$ &       $\cdots$  & $\cdots$ &       $\cdots$  &     0.44 &   0.38$\pm$:   \\
He~{\sc i}         & 7065       & 2p\,$^{3}$P$^{\rm o}$--3s\,$^{3}$S                                       &     9.04 &   7.82$\pm$1.17 &    10.2  &   8.81$\pm$1.40 &     8.83 &   7.34$\pm$1.10 & $\cdots$ &       $\cdots$  &     6.70 &   5.68$\pm$0.85\\
$[$Ar~{\sc iii}$]$ & 7136       & 3p$^{4}$\,$^{3}$P$_{2}$--3p$^{4}$\,$^{1}$D$_{2}$                         &    12.8  &   11.1$\pm$1.10 &     9.00 &   7.72$\pm$0.90 &    19.8  &   16.4$\pm$1.7  &    13.7  &   12.7$\pm$1.6  &     8.10 &   6.84$\pm$0.68\\
He~{\sc i}         & 7281       & 2p\,$^{1}$P$^{\rm o}_{1}$--3s\,$^{1}$S$_{0}$                             & $\cdots$ &       $\cdots$  & $\cdots$ &       $\cdots$  & $\cdots$ &       $\cdots$  & $\cdots$ &       $\cdots$  &     0.57 &   0.47$\pm$:   \\
$[$O~{\sc ii}$]$   & 7320       & 2p$^{3}$\,$^{2}$D$^{\rm o}_{5/2}$--2p$^{3}$\,$^{2}$P$^{\rm o}_{3/2}$     &     6.76 &   5.78$\pm$0.74 &     8.39 &   7.13$\pm$1.05 &     3.65 &   2.98$\pm$0.38 &     6.84 &   6.31$\pm$0.82 &     1.92 &   1.60$\pm$0.20\\
$[$O~{\sc ii}$]$   & 7330       & 2p$^{3}$\,$^{2}$D$^{\rm o}_{3/2}$--2p$^{3}$\,$^{2}$P$^{\rm o}_{3/2}$     &     5.75 &   4.90$\pm$0.73 &     8.16 &   6.94$\pm$1.21 &     3.58 &   2.92$\pm$0.44 &    11.2  &   10.3$\pm$1.5  &     1.44 &   1.20$\pm$0.17\\
$[$Ar~{\sc iii}$]$ & 7751       & 3p$^{4}$\,$^{3}$P$_{1}$--3p$^{4}$\,$^{1}$D$_{2}$                         &     2.08 &   1.73$\pm$0.43 &     1.18 &   0.98$\pm$0.30 &     3.63 &   2.88$\pm$0.72 &     6.07 &   5.53$\pm$1.37 &     1.17 &   0.96$\pm$0.23\\
H~{\sc i}          & 8750       & 3d\,$^{2}$D--12f\,$^{2}$F$^{\rm o}$                                      &     1.57 &   1.25$\pm$0.37 & $\cdots$ &       $\cdots$  & $\cdots$ &       $\cdots$  & $\cdots$ &       $\cdots$  &     1.63 &   1.25$\pm$0.37\\
H~{\sc i}          & 9015       & 3d\,$^{2}$D--10f\,$^{2}$F$^{\rm o}$                                      &     4.54 &   3.59$\pm$1.03 & $\cdots$ &       $\cdots$  & $\cdots$ &       $\cdots$  & $\cdots$ &       $\cdots$  &     1.48 &   1.13$\pm$0.32\\
$[$S~{\sc iii}$]$  & 9069       & 3p$^{2}$\,$^{3}$P$_{1}$--3p$^{2}$\,$^{1}$D$_{2}$                         &    37.8  &   30.0$\pm$2.4  & $\cdots$ &       $\cdots$  & $\cdots$ &       $\cdots$  & $\cdots$ &       $\cdots$  &    13.0  &   9.86$\pm$0.78\\
H~{\sc i}          & 9229       & 3d\,$^{2}$D--9f\,$^{2}$F$^{\rm o}$                                       &     2.84 &   2.23$\pm$0.78 & $\cdots$ &       $\cdots$  & $\cdots$ &       $\cdots$  & $\cdots$ &       $\cdots$  &     2.15 &   1.64$\pm$0.57\\
He~{\sc ii}        & 9345       & 5g\,$^{2}$G--8h\,$^{2}$H$^{\rm o}$                                       &     6.81 &   5.34$\pm$1.72 & $\cdots$ &       $\cdots$  & $\cdots$ &       $\cdots$  & $\cdots$ &       $\cdots$  &     4.98 &   3.77$\pm$1.21\\
$[$S~{\sc iii}$]$  & 9531$^{\rm l}$ & 3p$^{2}$\,$^{3}$P$_{1}$--3p$^{2}$\,$^{1}$D$_{2}$                     &    55.4  &   43.3$\pm$6.1  & $\cdots$ &       $\cdots$  & $\cdots$ &       $\cdots$  & $\cdots$ &       $\cdots$  &    12.2  &   9.22$\pm$1.20\\
\\
$c$(H$\beta$) & &                         & \multicolumn{2}{c}{0.172} & \multicolumn{2}{c}{0.177} & \multicolumn{2}{c}{0.220} & \multicolumn{2}{c}{0.087} & \multicolumn{2}{c}{0.196}\\
\multicolumn{2}{l}{$\log$\,$F$(H$\beta$)$^{\rm m}$}  &  & \multicolumn{2}{c}{$-$15.34} & \multicolumn{2}{c}{$-$15.07} & \multicolumn{2}{c}{$-$15.45} & \multicolumn{2}{c}{$-$15.29} & \multicolumn{2}{c}{$-$15.21}\\
\hline
\multicolumn{13}{l}{NOTE. -- Fluxes and intensities are normalized 
such that H$\beta$=100.  Colon ``:'' indicates that the uncertainty 
in line intensity is large ($>$100\%).}
\end{tabular}
\begin{description}
\item[$^{\rm a}$] A blend of the O~{\sc ii} $\lambda$3726 
(2p$^{3}$\,$^{4}$S$^{\rm o}_{3/2}$--2p$^{3}$\,$^{2}$D$^{\rm 
o}_{3/2}$) and $\lambda$3729 (2p$^{3}$\,$^{4}$S$^{\rm 
o}_{3/2}$--2p$^{3}$\,$^{2}$D$^{\rm o}_{5/2}$) doublet.
\item[$^{\rm b}$] Blended with the He~{\sc i} $\lambda$3888 
(2s\,$^{3}$S--3p\,${3}$P$^{\rm o}$) line.
\item[$^{\rm c}$] Blended with H~{\sc i} $\lambda$3970 
(2p\,$^{2}$P$^{\rm o}$--7d\,$^{2}$D) and He~{\sc i} $\lambda$3965 
(2s\,$^{1}$S--4p\,$^{1}$P$^{\rm o}$).
\item[$^{\rm d}$] Blended with [S~{\sc ii}] $\lambda$4076; probably 
also blended with the weak O~{\sc ii} M10 3p\,$^{4}$D$^{\rm 
o}$--3d\,$^{4}$F and C~{\sc iii} M16 4f\,$^{3}$F$^{\rm 
o}$--5g\,$^{3}$G lines. 
\item[$^{\rm e}$] Corrected for the flux from the blended He~{\sc 
ii} line.
\item[$^{\rm f}$] Blended with the N~{\sc iii} 
$\lambda\lambda$4634,4642 lines; probably also blended with O~{\sc 
ii} M1 $\lambda\lambda$4639,4642. 
\item[$^{\rm g}$] Blended with the O~{\sc ii} M2 
3s\,$^{4}$P--3p\,$^{4}$D$^{\rm o}$ lines. 
\item[$^{\rm h}$] Corrected for the flux from the blended He~{\sc i} 
$\lambda$4713 (2p\,$^{3}$P$^{\rm o}$--4s\,$^{3}$S) line.
\item[$^{\rm i}$] Blended with [N~{\sc i}] $\lambda$5200 
(2p$^{3}$\,$^{4}$S$^{\rm o}_{3/2}$--2p$^{3}$\,$^{2}$D$^{\rm 
o}_{5/2}$).
\item[$^{\rm j}$] Corrected for the flux from the blended He~{\sc 
ii} $\lambda$6311 (5g\,$^{2}$G--16h\,$^{2}$H$^{\rm o}$) line. 
\item[$^{\rm k}$] Corrected for the flux from the blended He~{\sc 
ii} $\lambda$6683 (5g\,$^{2}$G--13h\,$^{2}$H$^{\rm o}$) line.
\item[$^{\rm l}$] Flux underestimated due to the second-order 
contamination beyond 9200\,{\AA}. 
\item[$^{\rm m}$] In units of erg\,cm$^{-2}$\,s$^{-1}$, as measured 
in the extracted spectrum.
\end{description}
\end{center}
\end{table*}

We extracted a 1D spectrum on the fully calibrated 2D frame of 
each PN for spectral analysis.  As an example, Figure~\ref{fig4} 
shows the 1D spectra for PN12 and PN17 in our sample.  In the 
common wavelength region (5080--7850\,{\AA}) covered by the R1000B 
and R1000R grisms, differences in the fluxes of emission lines 
(He~{\sc i} $\lambda\lambda$5876,6678,7065, [N~{\sc ii}] 
$\lambda\lambda$6548,6583, H$\alpha$, [S~{\sc ii}] 
$\lambda$6716,6731, [Ar~{\sc iii}] $\lambda$7136, [O~{\sc ii}] 
$\lambda\lambda$7320,7330) detected in both spectra of PN9, PN11, 
PN13 and PN17 are mostly less than 5\%.  We corrected for the 
effect of second-order contamination in the red part of the R1000B 
spectrum (Figure\ref{fig4}, top) following the method of 
\citet{fang15}.  For the R1000R grism, the second-order 
contamination exists beyond 9200\,{\AA} (Figure~\ref{fig4}, 
bottom).  Fortunately, this contamination only affects the [S~{\sc 
iii}] $\lambda$9531 emission line.  The [S~{\sc iii}] $\lambda$9069 
nebular line was unaffected. 

Despite careful data reduction, detection of emission lines in one 
(target PN18) of the two PNe in M32 failed due to its close 
proximity (7\farcs4) to the bright nucleus of M32, although this 
target is the brightest in our sample.  The other M32 PN (PN16) 
is 15\farcs4 from M32's centre and has good data quality.  We thus
analyzed ten PNe (PN8--PN17; Table~\ref{targets}) in this paper.

\section{Results and Analysis} 
\label{sec3}

\subsection{Emission Line Fluxes} 
\label{sec3:part1}

The emission line fluxes were measured from the extracted 1D spectra
by integrating over line profiles.  The observed line fluxes of all 
targets, normalized to $F$(H$\beta$)=100, are presented in 
Table~\ref{lines}, where the observed H$\beta$ fluxes (in 
erg\,cm$^{-2}$\,s$^{-1}$) are also presented.  The R1000R spectrum 
was scaled according to the H$\alpha$ line flux in the R1000B 
spectrum.  We derived the logarithmic extinction parameter, 
$c$(H$\beta$), by comparing the observed and theoretical ratios of 
hydrogen Balmer lines, H$\alpha$/H$\beta$ and H$\gamma$/H$\beta$. 
The theoretical Case~B H~{\sc i} line ratios were adopted from 
\citet{sh95} at an electron temperature of 10\,000~K and a density 
of 10$^{4}$~cm$^{-3}$.  The $c$(H$\beta$) values of our PNe are small 
(0.08-0.25) and are presented in Table~\ref{lines}.  The observed 
line fluxes were then dereddened using the formula 
\begin{equation}
\label{eq1}
I(\lambda) = 10^{c(\rm{H}\beta) [1 + {\it f}(\lambda)]} F(\lambda),
\end{equation}
where $f$($\lambda$) is the extinction curve of \citet{ccm89} 
with a total-to-selective extinction ratio $R_{V}$ = 3.1.  The 
extinction corrected line intensities, all normalized to 
$I$(H$\beta$) = 100, along with the measurement errors, are 
presented in Table~\ref{lines}.  Given the excellent observing 
conditions (seeing$<$1\farcs0) and the slit width (1\farcs0), 
light loss in strong emission lines is expected to be negligible.

\subsection{Plasma Diagnostics} 
\label{sec3:part2}

We carried out plasma diagnostics of PNe using the ratios of the 
extinction-corrected fluxes of the collisionally excited lines 
(CELs; also often called forbidden lines) of heavy elements in 
Table~\ref{lines}.  The [S~{\sc ii}] $\lambda$6716/$\lambda$6731 
ratio is a common density diagnostic.  Where available, intensity 
ratio of the fainter [Ar~{\sc iv}] $\lambda\lambda$4711,4740 lines 
was also used to derive the electron density; here the flux of the 
blended He~{\sc i} $\lambda$4713 line was corrected for using the 
theoretical He~{\sc i} line ratios calculated by \citet{porter12}. 
The electron temperature was derived from the [O~{\sc iii}] 
($\lambda$4959+$\lambda$5007)/$\lambda$4363 nebular-to-auroral line 
ratio.  The [N~{\sc ii}] temperature was also determined whenever 
the [N~{\sc ii}] $\lambda$5755 line was detected.  References for 
the atomic data utilized in plasma diagnostics as well as the 
ionic-abundance determinations (in Section~\ref{sec3:part3}) are 
summarized in Table~\ref{atomic_data}, where sources of the 
effective recombination coefficients for the optical recombination 
lines (ORLs) analyzed in the paper are also given.  Results of 
plasma diagnostics are presented in Table~\ref{temden}.

\begin{table}
\begin{center}
\caption{References for Atomic Data}
\label{atomic_data}
\begin{tabular}{lll}
\hline\hline
Ion & \multicolumn{2}{c}{CELs}\\
\cline{2-3}
    & Transition Probabilities & Collision Strengths\\
\hline
N$^{+}$   & \citet{bhs95}     & \citet{stafford94}\\
O$^{+}$   & \citet{zeippen87} & \citet{pradhan06}\\
O$^{2+}$  & \citet{sz00}      & \citet{lb94}\\
Ne$^{2+}$ & \citet{lb05}      & \citet{mb00}\\
S$^{+}$   & \citet{keenan93}  & \citet{ram96}\\
S$^{2+}$  & \citet{mz82a}     & \citet{tg99}\\
Ar$^{2+}$ & \citet{bh86}      & \citet{gmz95}\\
Ar$^{3+}$ & \citet{mz82b}     & \citet{ram97}\\
Ar$^{4+}$ & \citet{mz82a}     & \citet{mendoza83}\\
\hline
Ion & \multicolumn{2}{c}{ORLs}\\
\cline{2-3}
    & Effective Recombination Coeff. & Comments\\
\hline
H~{\sc i}   & \citet{sh95}     & Case~B\\
He~{\sc i}  & \citet{porter12} & Case~B\\
He~{\sc ii} & \citet{sh95}     & Case~B\\
C~{\sc ii}  & \citet{davey00}  & Case~B\\
\hline
\end{tabular}
\end{center}
\end{table}

Only in two PNe (PN12 and PN13) did we find that $T_{\rm 
e}$([N~{\sc ii}]) is reasonably lower than $T_{\rm e}$([O~{\sc 
iii}]).  In the other targets (except PN16), $T_{\rm e}$([N~{\sc 
ii}])$>T_{\rm e}$([O~{\sc iii}]).  This might be due to high-density 
clumps in PNe \citep{morisset16}: at high densities ($N_{\rm 
e}\gtrsim$10$^{5}$\,cm$^{-3}$), emission of the [N~{\sc ii}] 
$\lambda\lambda$6548,6583 nebular lines can be suppressed due to 
collisional deexcitation (while emission of the $\lambda$5755 
auroral line is unaffected), and consequently the [N~{\sc ii}] 
temperature is overestimated. 
Results of plasma diagnostics based on the CELs are visually 
demonstrated in Figure~\ref{fig5}, where the diagnostic curves of 
different forbidden-line ratios are plotted for each PN.  The {\sc 
fortran} code {\sc equib}, which was originally developed by 
\citet{ha81} to solve the statistical equilibrium equations of 
multi-level atoms to derive level populations and line emissivities 
under given nebular physical conditions, was used for the plasma 
diagnostics. 

Other temperature-sensitive ratios are [O~{\sc ii}] 
$\lambda$3727/($\lambda$7320+$\lambda$7330) and [S~{\sc ii}] 
($\lambda$6716+$\lambda$6731)/$\lambda$4072, where $\lambda$3727 
is a blend of the [O~{\sc ii}] $\lambda\lambda$3726,3729 doublet, 
and $\lambda$4072 is a blend of [S~{\sc ii}] 
$\lambda\lambda$4068,4076.  However, not in all PNe did we detect 
these faint auroral lines to a desired S/N.  For the four PNe for 
which the OSIRIS R1000R spectroscopy was obtained, we also derived 
the temperature using the [S~{\sc iii}] 
($\lambda$9069+$\lambda$9531)/$\lambda$6312 line ratio.  Since 
the [S~{\sc iii}] $\lambda$9531 nebular line was affected by the 
second-order contamination (see Section~\ref{sec2:part3}), we 
assumed a theoretical ratio $\lambda$9531/$\lambda$9069 = 2.48 
\citep{mz82a,mendoza83} to derive the intrinsic flux of this 
[S~{\sc iii}] line.  Besides the traditional CEL diagnostics, we 
also determined the electron temperatures using the He~{\sc i} 
ORLs.  These He~{\sc i} temperatures were generally lower than 
those derived from the CELs (Table~\ref{temden}), consistent with 
Paper~II.  The principles of PN plasma diagnostics based on the 
He~{\sc i} ORLs are described in \citet{zhang05}.

\begin{table*}
\begin{center}
\begin{minipage}{145mm}
\caption{Plasma Diagnostics}
\label{temden}
\begin{tabular}{lccccc}
\hline\hline
Diagnostic Ratio & PN8  & PN9  & PN10 & PN11 & PN12\\
\hline
                 & \multicolumn{5}{c}{$T_\mathrm{e}$ (K)} \\
$[$O~{\sc iii}$]$ ($\lambda$4959+$\lambda$5007)/$\lambda$4363  &    10\,680$\pm$400  & 10\,200$\pm$300  &    10\,960$\pm$370  & 11\,300$\pm$400  & 10\,500$\pm$340 \\
$[$N~{\sc ii}$]$  ($\lambda$6548+$\lambda$6583)/$\lambda$5755  &    20\,000$\pm$5000 & 12\,200$\pm$1300 &    16\,300$\pm$3900 & 12\,200$\pm$2300 &    9200$\pm$2200\\
$[$O~{\sc ii}$]$  $\lambda$3727/($\lambda$7320+$\lambda$7330)  & $>$20\,000          & 17\,200$\pm$6000 & $>$20\,000          &    7500$\pm$2500 &    7300$\pm$2700\\
$[$S~{\sc iii}$]$ ($\lambda$9069+$\lambda$9531)/$\lambda$6312  &           $\cdots$  & 10\,700$\pm$1100 &           $\cdots$  & 12\,700$\pm$1500 &        $\cdots$ \\
$[$S~{\sc ii}$]$  ($\lambda$6716+$\lambda$6731)/$\lambda$4072\,$^{\rm a}$  & $\cdots$  &        $\cdots$  &           $\cdots$  &        $\cdots$  &    9200$\pm$2800\\
He~{\sc i} $\lambda$5876/$\lambda$4471                         &    10\,000$\pm$4000 &        $\cdots$  &       5600$\pm$3000 &    8000$\pm$3500 &    9400$\pm$3000\\
He~{\sc i} $\lambda$6678/$\lambda$4471                         &       9300$\pm$4000 &    3400$\pm$3000 &    12\,600$\pm$7500 & 12\,900$\pm$8000 & 12\,500$\pm$8000\\
\\
                  & \multicolumn{5}{c}{$N_\mathrm{e}$ (cm$^{-3}$)}\\
$[$S~{\sc ii}$]$  $\lambda$6716/$\lambda$6731                  &       1400$\pm$1000 &    5200$\pm$2000 &       3900$\pm$1200 &    8700$\pm$3800 & 21\,000$\pm$5000\\
$[$Ar~{\sc iv}$]$ $\lambda$4711/$\lambda$4740                  &    16\,200$\pm$5400 &    7600$\pm$3800 &       2500$\pm$:    &    8000$\pm$2700 &        $\cdots$ \\
$[$O~{\sc ii}$]$  $\lambda$3727/($\lambda$7320+$\lambda$7330)  & $>$20\,000          & 14\,000$\pm$8000 & $\sim$20\,000       &    4000$\pm$2000 & 12\,000$\pm$8000\\
\rule[0mm]{0mm}{1.0mm}\\
\cline{2-6}
 & PN13 & PN14 & PN15 & PN16 & PN17\\
\hline
                 & \multicolumn{5}{c}{$T_\mathrm{e}$ (K)} \\
$[$O~{\sc iii}$]$ ($\lambda$4959+$\lambda$5007)/$\lambda$4363  &       9100$\pm$300  &  11\,000$\pm$370  &     9100$\pm$250  &  10\,200$\pm$700  & 12\,100$\pm$450 \\
$[$N~{\sc ii}$]$  ($\lambda$6548+$\lambda$6583)/$\lambda$5755  &       8000$\pm$2000 &  19\,200$\pm$4400 &  11\,400$\pm$2300 &         $\cdots$  & 13\,600$\pm$2000\\
$[$O~{\sc ii}$]$  $\lambda$3727/($\lambda$7320+$\lambda$7330)  &       8300$\pm$3000 & $>$20\,000        &  10\,200$\pm$2900 & $>$20\,000        & 10\,900$\pm$2600\\
$[$S~{\sc iii}$]$ ($\lambda$9069+$\lambda$9531)/$\lambda$6312  & 10\,020$\pm$930 &      $\cdots$  &         $\cdots$  &         $\cdots$  & 11\,400$\pm$2500\\
$[$S~{\sc ii}$]$  ($\lambda$6716+$\lambda$6731)/$\lambda$4072\,$^{\rm a}$ & 6000$\pm$2600 &      $\cdots$  &         $\cdots$  & $>$20\,000        &        $\cdots$ \\
He~{\sc i} $\lambda$5876/$\lambda$4471                         &    $>$20\,000       &     2700$\pm$:    &     2800$\pm$:    &     5100$\pm$2500 &    3200$\pm$:   \\
He~{\sc i} $\lambda$6678/$\lambda$4471                         &    14\,700$\pm$8000 &     6300$\pm$3000 &  10\,200$\pm$5000 &         $\cdots$  &    5200$\pm$2000\\
\\
                  & \multicolumn{5}{c}{$N_\mathrm{e}$ (cm$^{-3}$)}\\
$[$S~{\sc ii}$]$  $\lambda$6716/$\lambda$6731                  & 19\,600$\pm$8000 &       2600$\pm$1400 & 3700$\pm$1900 &       2040$\pm$1000 & 4700$\pm$1500\\
$[$Ar~{\sc iv}$]$ $\lambda$4711/$\lambda$4740                  &        $\cdots$  &           $\cdots$  & 8800$\pm$3400 &           $\cdots$  & 6100$\pm$1800\\
$[$O~{\sc ii}$]$  $\lambda$3727/($\lambda$7320+$\lambda$7330)  & $\sim$20\,000    & $>$20\,000          & 6300$\pm$3000 & $>$20\,000          & 6000$\pm$3000\\
$[$Cl~{\sc iii}$]$ $\lambda$5517/$\lambda$5537                 &        $\cdots$  &           $\cdots$  &     $\cdots$  &           $\cdots$  & 5500$\pm$2400\\
\hline
\multicolumn{6}{l}{NOTE. -- The colon ``:'' indicates very large 
uncertainty.}
\end{tabular}
\begin{description}
\item[$^{\rm a}$] A blend of [S~{\sc ii}] 
$\lambda\lambda$4068,\,4076; also blended with O~{\sc ii} M10 
3p\,$^{4}$D$^{\rm o}$--3d\,$^{4}$F and C~{\sc iii} M16 
4f\,$^{3}$F$^{\rm o}$--5g\,$^{3}$G lines.
\end{description}
\end{minipage}
\end{center}
\end{table*}

Uncertainties in the electron temperatures and densities presented 
in Table~\ref{temden} were estimated based on the measurement errors 
of emission line fluxes through propagation.  Weaker lines generally 
have larger measurement errors, as a result introducing larger 
uncertainties in temperatures/densities. A typical example is 
[N~{\sc ii}] $\lambda$5755, whose intensity is 10\%--30\% that of 
[O~{\sc iii}] $\lambda$4363 for most targets in our sample.  Errors 
in the [N~{\sc ii}] temperatures are systematically higher that 
those in the [O~{\sc iii}] temperatures, which are thus best 
measured.

\subsection{Ionic Abundances} 
\label{sec3:part3}

Using the relative intensities of emission lines in 
Table~\ref{lines} and the electron temperatures and densities in 
Table~\ref{temden}, we calculated the ionic abundances of our PNe. 
The {\sc equib} program was used to calculate the ionic abundances 
of He, C, N, O, Ne, S, Cl and Ar relative to hydrogen, which are 
presented in Table~\ref{ionic}.  The deep GTC spectroscopy enabled 
detection of several faint diagnostic lines, including the [O~{\sc 
iii}] $\lambda$4363, [N~{\sc ii}] $\lambda$5755, and [S~{\sc iii}] 
$\lambda$6312 auroral lines.  It is thus possible to consider 
multiple ionization zones within a nebula, i.e., to assign different 
temperatures/densities when calculating abundances of ionic species 
with different ionization stages (i.e., the ionization potentials). 
This is a more realistic paradigm of nebular analysis, and reduces 
the uncertainties in resultant abundances that may arise from the 
temperatures assumed. 

The [O~{\sc iii}] temperature was used to derive O$^{2+}$/H$^{+}$, 
Ne$^{2+}$/H$^{+}$, Ar$^{2+}$/H$^{+}$, Ar$^{3+}$/H$^{+}$ and 
Cl$^{2+}$/H$^{+}$.  We adopted the [N~{\sc ii}] temperature to 
calculate N$^{+}$/H$^{+}$ and O$^{+}$/H$^{+}$ for PN12 and PN13, 
where we found the [N~{\sc ii}] temperature is lower than that 
derived from the [O~{\sc iii}] line ratio.  For the other PNe, we 
considered the recipe of \citet[][also \citealt{kh01}]{dufour15} 
for the electron temperature in the low-ionization region: if 
He~{\sc ii} $\lambda$4686 was detected, we adopted an [N~{\sc ii}] 
temperature of 10\,300~K derived by \citet{kaler86}; otherwise, we 
assumed a temperature of 10\,000~K.

\begin{figure*}
\begin{center}
\includegraphics[width=17.75cm,angle=0]{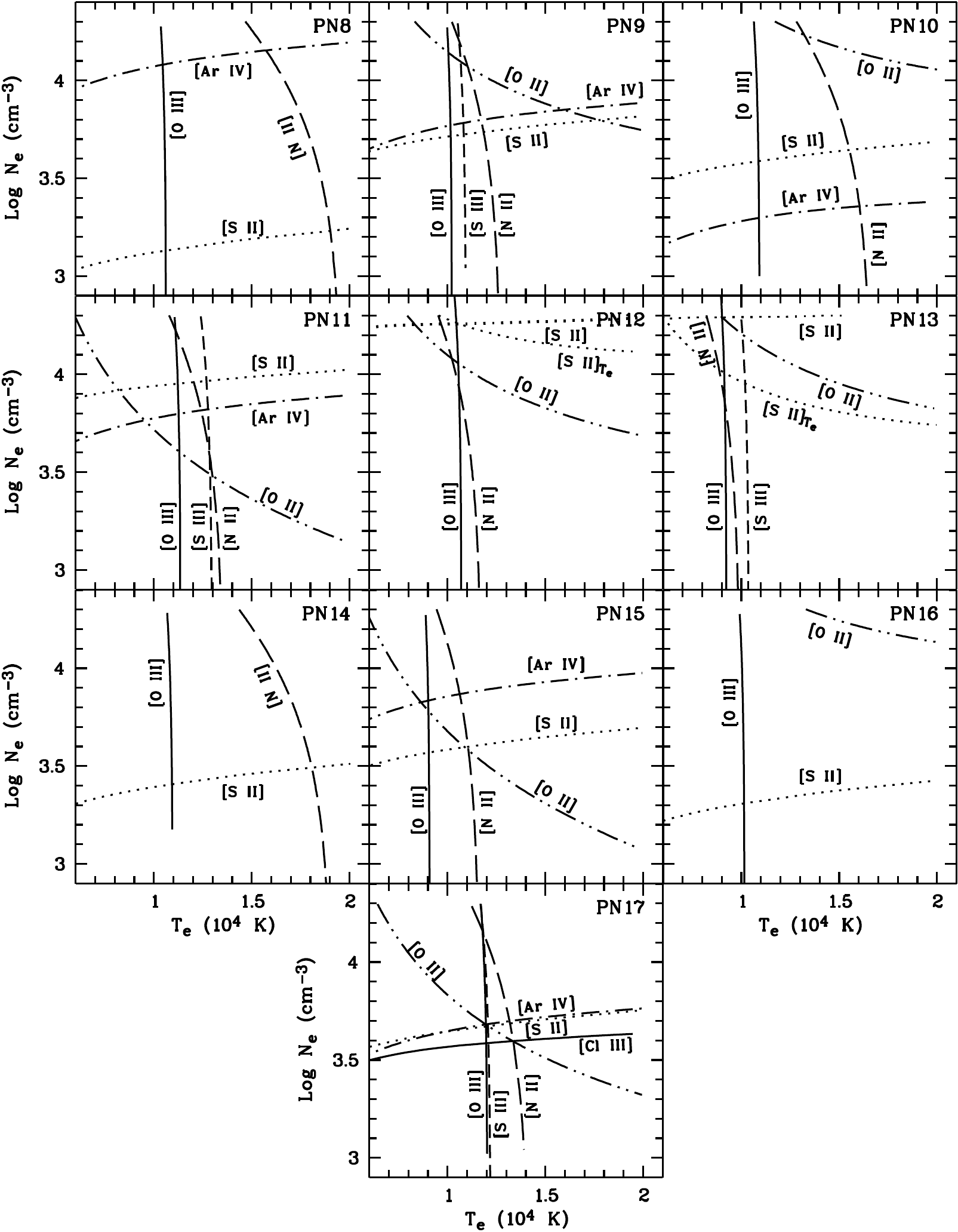}
\caption{Plasma diagnostic diagrams. Different line types represent 
the temperature or density diagnostics using different line ratios 
(see Table~\ref{temden} for line ratios).  For PN12 and PN13, the 
dotted line labeled with [S~{\sc ii}]$_{T_{\rm e}}$ is the 
temperature-diagnostic curve using the [S~{\sc ii}] 
($\lambda$6716+$\lambda$6731)/$\lambda$4072 intensity ratio.  The 
dotted line labeled with [S~{\sc ii}] is density-diagnostic curve 
for all PNe.} 
\label{fig5}
\end{center}
\end{figure*}

The [S~{\sc iii}] temperature derived for the four PNe (PN9, PN11, 
PN13 and PN17) is in general meaningfully different from the 
[O~{\sc iii}] temperature, and thus were used to calculate the 
S$^{2+}$/H$^{+}$ ratio.  For the other PNe, we adopted the electron 
temperature that was used to calculate the N$^{+}$/H$^{+}$ ratio to 
derive S$^{2+}$/H$^{+}$.  In the calculations of S$^{+}$/H$^{+}$, 
we adopted the [S~{\sc ii}] temperature (9200~K) for PN12; for the 
other targets, we adopted the temperature that was used to calculate 
the N$^{+}$/H$^{+}$ ratio.  Although temperatures (or the lower 
limits) were also derived from the [O~{\sc ii}] 
$\lambda$3727/($\lambda$7320+$\lambda$7330) line ratio, they were 
much too different from the [S~{\sc iii}] temperatures 
(Table~\ref{temden}), given that the ionization potential of 
O$^{+}$ (35.12~eV) is close to that of S$^{2+}$ (34.83~eV); the 
differences between the [O~{\sc ii}] temperatures and those derived 
from the [O~{\sc iii}] lines are also questionable.  The only 
exception is PN17, whose [O~{\sc ii}] temperature seems reasonable 
compared to those derived from the [O~{\sc iii}] and [S~{\sc iii}] 
line ratios.  We adopted the electron density derived from the 
[S~{\sc ii}] $\lambda$6716/$\lambda$6731 ratio for the 
ionic-abundance calculations of the low-ionization species.  Where 
available, the density yielded by [Ar~{\sc iv}] 
$\lambda$4711/$\lambda$4740 was assumed for the high-ionization 
species, otherwise the [S~{\sc ii}] density was used.  For PN17, 
the density derived from [Cl~{\sc iii}] $\lambda$5517/$\lambda$5537 
was used to derive Cl$^{2+}$/H$^{+}$. 

Care must be taken when deriving the O$^{+}$/H$^{+}$ ratio, 
although O$^{+}$ is not the dominant ionization stage of oxygen in 
PNe.  We noticed that the [O~{\sc ii}] $\lambda$3727 (a blend of 
$\lambda\lambda$3726,3729) nebular line yielded a very different O$^{+}$/H$^{+}$ ratio from that derived from the [O~{\sc ii}] 
$\lambda\lambda$7320,7330 auroral lines, if the same electron 
density was assumed.  Such difference in O$^{+}$/H$^{+}$ even 
reached one order of magnitude for some PNe in our sample.  This 
might be because $\lambda$3727 and $\lambda\lambda$7320,7330 
actually come from regions with very different densities.  Besides 
its dependence on temperature, the [O~{\sc ii}] 
$\lambda$3727/($\lambda$7320+$\lambda$7330) line ratio also has 
non-negligible dependence on the density, as can be seen in 
Figure~\ref{fig5}.  If these two [O~{\sc ii}] lines come from 
different nebular regions, the diagnosed temperature can be 
unrealistically high. 

If we adopt the [S~{\sc ii}] density for [O~{\sc ii}] $\lambda$3727 
and assumed a higher density (e.g., 20\,000~cm$^{-3}$; 
Table~\ref{temden}) for $\lambda$7325 
(=$\lambda$7320+$\lambda$7330), the two O$^{+}$/H$^{+}$ ratios 
derived can be brought to the same level (Table~\ref{ionic}). 
Thus the electron densities (or the lower limit, see 
Table~\ref{temden}) yielded by the [O~{\sc ii}] diagnostic ratio 
was assumed for the $\lambda$7325 line.  We then derived a 
weighted average from the two O$^{+}$/H$^{+}$ ratios, with the 
weights proportional to the intensities of the $\lambda$3727 and 
$\lambda$7325 lines.  The averaged O$^{+}$/H$^{+}$ ratio was 
adopted and then used for the determination of elemental abundances.

Ne$^{2+}$/H$^{+}$ derived from [Ne~{\sc iii}] $\lambda$3868 was 
adopted; the other [Ne~{\sc iii}] nebular line $\lambda$3967 is 
blended with H~{\sc i} $\lambda$3970.  For the four PNe (PN9, PN11, 
PN13 and PN17) where both $\lambda$6312 and $\lambda$9069 of 
[S~{\sc iii}] were detected, a line intensity-weighted average of 
the S$^{2+}$/H$^{+}$ ratios derived from the two [S~{\sc iii}] lines 
was adopted.  For the other PNe in our sample, S$^{2+}$/H$^{+}$ 
derived from $\lambda$6312 was adopted.  The total intensity of 
the [Ar~{\sc iv}] $\lambda\lambda$4711,4740 doublet was used to 
derive Ar$^{3+}$/H$^{+}$.  The flux of [Ar~{\sc iv}] $\lambda$4711 
was corrected for the blended He~{\sc i} $\lambda$4713 line.  The 
effective recombination coefficients of the He~{\sc i} lines 
calculated by \citet{porter12} was used to derive the 
He$^{+}$/H$^{+}$ ratios.  The He$^{2+}$/H$^{+}$ ratio was derived 
from He~{\sc ii} $\lambda$4686 using the hydrogenic effective 
recombination coefficients from \citet{sh95}.  We also detected 
C~{\sc ii} $\lambda$4267 (M6 3d\,$^{2}$D--4f\,$^{2}$F$^{\rm o}$) 
in the spectra of PN11 and PN12, and the C$^{2+}$/H$^{+}$ ratio was 
derived using this line (Table~\ref{ionic}).  The Case~B effective 
recombination coefficients of the C~{\sc ii} lines were adopted 
from \citet{davey00}, and an electron temperature of 10\,000~K and 
a density of 10$^{4}$ cm$^{-3}$ were assumed. 

The uncertainties in ionic abundances (in the brackets in 
Table~\ref{ionic}), were estimated from the measurement errors in 
line fluxes.  Extra errors in abundances could be introduced by 
the electron temperatures and densities adopted in abundance 
determinations, although we have considered multiple ionization 
zones by deriving the abundances of low- and high-ionization 
species using different temperatures/densities.  However, these 
errors in general have minor contribution to the total uncertainty 
budget, and were not included in the final uncertainties of the 
ionic abundances.  In Paper~II we have estimated that errors in the 
[O~{\sc iii}] temperature typically introduce $\sim$10\% 
uncertainties in the resultant ionic abundances, while in this work 
such errors are probably even lower.  It is evident that the ORLs 
of heavy elements (C~{\sc ii}, O~{\sc ii}, N~{\sc ii} and Ne~{\sc 
ii}) observed in PNe could be emitted by nebular regions as cold 
as $\lesssim$1000~K \citep[e.g.,][]{liu12,fl13,mcnabb13}.  For an 
ORL (like C~{\sc ii} $\lambda$4267) excited by radiative 
recombination, its emissivity (i.e., effective recombination 
coefficient) generally decreases with the electron temperature 
\citep[e.g.,][]{of06,fl13}.  Thus C$^{2+}$/H$^{+}$ derived here 
could be overestimated due to the temperature (10\,000~K) assumed. 
According to the calculations of \citet{davey00}, the effective 
recombination coefficient of the C~{\sc ii} $\lambda$4267 line 
decreases by a factor of 9.4 as the temperature increases from 
1000~K to 10\,000~K.

\subsection{Elemental Abundances} 
\label{sec3:part4}

The total elemental abundances (relative to hydrogen) were derived 
based on the ionic abundances presented in Table~\ref{ionic}.  The 
helium abundance is a sum of the ionic ratios, He/H = 
He$^{+}$/H$^{+}$ + He$^{2+}$/H$^{+}$.  For heavy elements, the 
total abundances were derived mostly using the ionization 
correction factors (ICFs) of \citet{kb94}.  Elemental abundances 
are presented in Table~\ref{elemental}, and the ICFs used to 
correct for the unseen ions are presented in Table~\ref{icfs}. 

In the cases where both S$^{+}$/H$^{+}$ and S$^{2+}$/H$^{+}$ were 
derived, S/H = ICF(S)$\times$(S$^{+}$/H$^{+}$ + S$^{2+}$/H$^{+}$) 
was used, where ICF(S) was adopted from \citet[][Equation~A36 
therein]{kb94}.  For PN16 where only S$^{+}$/H$^{+}$ was obtainable,
the empirical fitting formula of \citet[][Equation~A38 
therein]{kb94} was used to derive S$^{2+}$/H$^{+}$.  If both 
Ar$^{2+}$ and Ar$^{3+}$ were observed, we used Ar/H = 
ICF(Ar)$\times$(Ar$^{2+}$/H$^{+}$ + Ar$^{3+}$/H$^{+}$), where 
ICF(Ar) is from \citet[][Equation~A30 therein]{kb94}.  In typical 
physical conditions of PNe, concentration of argon in Ar$^{4+}$ is 
supposed to be negligible compared to Ar$^{2+}$ and Ar$^{3+}$.  If 
only Ar$^{2+}$ was observed (in PN12, PN13, PN14 and PN16 in our 
sample), Equation~A32 in \citet{kb94} was used to derive ICF(Ar). 
Only Cl$^{2+}$ was observed in our spectra (of PN8, PN11, PN12 and 
PN17), and we assumed Cl/Cl$^{2+}$ $\approx$S/S$^{2+}$ as in 
\citet{wl07}, according to similarity in ionization potentials. 
C/H was derived for two PNe in our sample assuming ICF(C) = 
O/O$^{2+}$ (Table~\ref{elemental}).  \citet{delgado14} developed a 
new set of formulae for the ICFs of PNe by computing a large grid 
of photoionization models.  These ICFs have validity application 
ranges defined by the ionic fractions of helium, 
He$^{2+}$/(He$^{+}$ + He$^{2+}$), and oxygen, O$^{2+}$/(O$^{+}$ + 
O$^{2+}$).  We expect that these ICFs are adequate estimates of 
elemental abundances in PNe.  However, not all our targets have 
the helium or oxygen ionic fractions located within these validity 
ranges.  Besides, the new ICFs do not differ much from the 
classical methods of \citet{kb94} for most of the elements 
\citep{garcia16}.  In particular, adopting the new ICFs of 
\citet{delgado14} did not eliminate the ``sulfur anomaly'' in PNe, 
which is discussed in Section~\ref{sec4:part1}.

\begin{table*}
\begin{center}
\caption{Ionic Abundances}
\label{ionic}
\begin{tabular}{llccccc}
\hline\hline
Ion & Line  & \multicolumn{5}{c}{Abundance (X$^{i+}$/H$^{+}$)}\\
\cline{3-7}
    & (\AA) &  PN8  &  PN9  &  PN10 &  PN11 &  PN12\\
\hline
He$^{+}$  & 4471 &  0.105$\pm$0.016                  & 0.087$\pm$0.014                  & 0.090$\pm$0.013                  & 0.099$\pm$0.016                  & 0.109$\pm$0.015                  \\
          & 5876 &  0.104$\pm$0.012                  & 0.111$\pm$0.012                  & 0.090$\pm$0.010                  & 0.095$\pm$0.008                  & 0.102$\pm$0.011                  \\
          & 6678 &  0.113$\pm$0.016                  & 0.102$\pm$0.014                  & 0.082$\pm$0.011                  & 0.089$\pm$0.012                  & 0.100$\pm$0.014                  \\
Adopted$^{\rm a}$ 
          &      &  0.104$\pm$0.012                  & 0.111$\pm$0.012                  & 0.090$\pm$0.010                  & 0.095$\pm$0.008                  & 0.102$\pm$0.011                  \\
He$^{2+}$ & 5411 &       $\cdots$                    &      $\cdots$                    &      $\cdots$                    & 1.47($\pm$0.40)$\times$10$^{-2}$ &      $\cdots$                    \\
          & 4686 &  1.02($\pm$0.28)$\times$10$^{-3}$ & 1.28($\pm$0.41)$\times$10$^{-3}$ &      $\cdots$                    & 1.10($\pm$0.10)$\times$10$^{-2}$ & 0.74($\pm$0.15)$\times$10$^{-3}$ \\
C$^{2+}$  & 4267 &                 $\cdots$          &                $\cdots$          &      $\cdots$                    & 1.47($\pm$0.26)$\times$10$^{-3}$ & 8.17($\pm$2.98)$\times$10$^{-4}$ \\
N$^{+}$   & 5755 &  2.26($\pm$0.50)$\times$10$^{-5}$ & 1.53($\pm$0.34)$\times$10$^{-5}$ & 8.30($\pm$1.83)$\times$10$^{-6}$ & 6.09($\pm$1.30)$\times$10$^{-6}$ & 1.02($\pm$0.20)$\times$10$^{-5}$ \\
          & 6548 &  6.89($\pm$0.74)$\times$10$^{-6}$ & 1.01($\pm$0.12)$\times$10$^{-5}$ & 4.12($\pm$0.44)$\times$10$^{-6}$ & 5.05($\pm$0.54)$\times$10$^{-6}$ & 1.32($\pm$0.15)$\times$10$^{-5}$ \\
          & 6583 &  7.95($\pm$0.70)$\times$10$^{-6}$ & 1.08($\pm$0.10)$\times$10$^{-5}$ & 4.08($\pm$0.36)$\times$10$^{-6}$ & 5.29($\pm$0.46)$\times$10$^{-6}$ & 1.39($\pm$0.12)$\times$10$^{-5}$ \\
Adopted$^{\rm b}$ 
          &      &  7.95($\pm$0.70)$\times$10$^{-6}$ & 1.08($\pm$0.10)$\times$10$^{-5}$ & 4.08($\pm$0.36)$\times$10$^{-6}$ & 5.29($\pm$0.46)$\times$10$^{-6}$ & 1.39($\pm$0.12)$\times$10$^{-5}$ \\
O$^{+}$   & 3727 &  2.15($\pm$0.24)$\times$10$^{-5}$ & 3.58($\pm$0.32)$\times$10$^{-5}$ & 1.47($\pm$0.16)$\times$10$^{-5}$ & 1.46($\pm$0.18)$\times$10$^{-5}$ & 5.82($\pm$0.47)$\times$10$^{-5}$ \\
          & 7320 &  5.54($\pm$0.77)$\times$10$^{-5}$ & 5.01($\pm$0.64)$\times$10$^{-5}$ & 3.40($\pm$0.44)$\times$10$^{-5}$ & 1.30($\pm$0.20)$\times$10$^{-5}$ & 4.23($\pm$0.54)$\times$10$^{-5}$ \\
          & 7330 &  5.69($\pm$0.86)$\times$10$^{-5}$ & 5.14($\pm$0.75)$\times$10$^{-5}$ & 3.24($\pm$0.49)$\times$10$^{-5}$ & 1.24($\pm$0.20)$\times$10$^{-5}$ & 5.67($\pm$0.64)$\times$10$^{-5}$ \\
          & 7325 &  5.61($\pm$0.86)$\times$10$^{-5}$ & 5.07($\pm$0.75)$\times$10$^{-5}$ & 3.33($\pm$0.49)$\times$10$^{-5}$ & 1.27($\pm$0.20)$\times$10$^{-5}$ & 4.88($\pm$0.64)$\times$10$^{-5}$ \\
Adopted$^{\rm c}$ 
          &      &  3.58($\pm$0.42)$\times$10$^{-5}$ & 3.82($\pm$0.45)$\times$10$^{-5}$ & 1.92($\pm$0.25)$\times$10$^{-5}$ & 1.45($\pm$0.20)$\times$10$^{-5}$ & 5.68($\pm$0.68)$\times$10$^{-5}$ \\
O$^{2+}$  & 4363 &  3.60($\pm$0.43)$\times$10$^{-4}$ & 4.09($\pm$0.50)$\times$10$^{-4}$ & 2.68($\pm$0.32)$\times$10$^{-4}$ & 3.30($\pm$0.39)$\times$10$^{-4}$ & 1.69($\pm$0.20)$\times$10$^{-4}$ \\
          & 4959 &  3.53($\pm$0.14)$\times$10$^{-4}$ & 3.96($\pm$0.15)$\times$10$^{-4}$ & 2.61($\pm$0.10)$\times$10$^{-4}$ & 3.17($\pm$0.12)$\times$10$^{-4}$ & 1.65($\pm$0.06)$\times$10$^{-4}$ \\
          & 5007 &  3.63($\pm$0.10)$\times$10$^{-4}$ & 4.13($\pm$0.10)$\times$10$^{-4}$ & 2.71($\pm$0.06)$\times$10$^{-4}$ & 3.34($\pm$0.06)$\times$10$^{-4}$ & 1.71($\pm$0.05)$\times$10$^{-4}$ \\
Adopted$^{\rm d}$ 
          &      &  3.63($\pm$0.10)$\times$10$^{-4}$ & 4.13($\pm$0.10)$\times$10$^{-4}$ & 2.71($\pm$0.06)$\times$10$^{-4}$ & 3.34($\pm$0.06)$\times$10$^{-4}$ & 1.71($\pm$0.05)$\times$10$^{-4}$ \\
Ne$^{2+}$ & 3868 &  9.40($\pm$0.63)$\times$10$^{-5}$ & 8.60($\pm$0.57)$\times$10$^{-5}$ & 6.37($\pm$0.43)$\times$10$^{-5}$ & 7.34($\pm$0.50)$\times$10$^{-5}$ & 2.74($\pm$0.20)$\times$10$^{-5}$ \\
          & 3967 &  9.25($\pm$0.72)$\times$10$^{-5}$ & 8.29($\pm$0.64)$\times$10$^{-5}$ & 6.07($\pm$0.46)$\times$10$^{-5}$ & 8.76($\pm$0.68)$\times$10$^{-5}$ & 4.12($\pm$0.31)$\times$10$^{-5}$ \\
Adopted$^{\rm e}$ 
          &      &  9.40($\pm$0.63)$\times$10$^{-5}$ & 8.60($\pm$0.57)$\times$10$^{-5}$ & 6.37($\pm$0.43)$\times$10$^{-5}$ & 7.34($\pm$0.50)$\times$10$^{-5}$ & 2.74($\pm$0.20)$\times$10$^{-5}$ \\
S$^{+}$   & 6716 &  9.00($\pm$1.86)$\times$10$^{-8}$ & 2.43($\pm$0.50)$\times$10$^{-7}$ & 7.10($\pm$1.65)$\times$10$^{-8}$ & 2.32($\pm$0.47)$\times$10$^{-7}$ & 1.39($\pm$0.28)$\times$10$^{-7}$ \\
          & 6731 &  8.99($\pm$1.69)$\times$10$^{-8}$ & 2.43($\pm$0.32)$\times$10$^{-7}$ & 7.10($\pm$1.11)$\times$10$^{-8}$ & 2.33($\pm$0.28)$\times$10$^{-7}$ & 2.10($\pm$0.21)$\times$10$^{-7}$ \\
Adopted   &      &  8.99($\pm$1.69)$\times$10$^{-8}$ & 2.43($\pm$0.32)$\times$10$^{-7}$ & 7.10($\pm$1.12)$\times$10$^{-8}$ & 2.33($\pm$0.30)$\times$10$^{-7}$ & 2.10($\pm$0.22)$\times$10$^{-7}$ \\
S$^{2+}$  & 6312 &  4.69($\pm$1.17)$\times$10$^{-6}$ & 2.39($\pm$0.44)$\times$10$^{-6}$ & 1.82($\pm$0.50)$\times$10$^{-6}$ & 2.03($\pm$0.46)$\times$10$^{-6}$ & 2.38($\pm$0.38)$\times$10$^{-6}$ \\
          & 9069 &                 $\cdots$          & 3.65($\pm$0.30)$\times$10$^{-6}$ &                $\cdots$          & 2.59($\pm$0.21)$\times$10$^{-6}$ &                $\cdots$          \\
          & 9531 &                 $\cdots$          & 5.28($\pm$0.78)$\times$10$^{-7}$ &                $\cdots$          & 2.94($\pm$0.41)$\times$10$^{-7}$ &                $\cdots$          \\
Adopted$^{\rm f}$ 
          &      &  4.69($\pm$1.17)$\times$10$^{-6}$ & 3.57($\pm$0.31)$\times$10$^{-6}$ & 1.82($\pm$0.50)$\times$10$^{-6}$ & 2.54($\pm$0.22)$\times$10$^{-6}$ & 2.38($\pm$0.38)$\times$10$^{-6}$ \\
Cl$^{2+}$ & 5537 &  1.83($\pm$0.46)$\times$10$^{-7}$ &                $\cdots$          &                $\cdots$          & 4.57($\pm$1.80)$\times$10$^{-8}$ & 6.69($\pm$2.72)$\times$10$^{-8}$ \\
Ar$^{2+}$ & 7136 &  1.30($\pm$0.12)$\times$10$^{-6}$ & 1.11($\pm$0.13)$\times$10$^{-6}$ & 7.46($\pm$0.77)$\times$10$^{-7}$ & 1.01($\pm$0.13)$\times$10$^{-6}$ & 1.20($\pm$0.11)$\times$10$^{-6}$ \\
          & 7751 &  8.32($\pm$2.07)$\times$10$^{-7}$ & 6.79($\pm$2.01)$\times$10$^{-7}$ & 5.40($\pm$1.35)$\times$10$^{-7}$ & 8.70($\pm$2.16)$\times$10$^{-7}$ & 8.35($\pm$2.00)$\times$10$^{-7}$ \\
Adopted$^{\rm g}$ 
          &      &  1.30($\pm$0.12)$\times$10$^{-6}$ & 1.11($\pm$0.13)$\times$10$^{-6}$ & 7.46($\pm$0.77)$\times$10$^{-7}$ & 1.01($\pm$0.13)$\times$10$^{-6}$ & 1.20($\pm$0.11)$\times$10$^{-6}$ \\
Ar$^{3+}$ & 4711 &  2.35($\pm$0.58)$\times$10$^{-7}$ & 2.10($\pm$0.63)$\times$10$^{-7}$ & 1.81($\pm$0.41)$\times$10$^{-7}$ & 5.52($\pm$0.83)$\times$10$^{-7}$ &                $\cdots$          \\
          & 4740 &  4.90($\pm$0.60)$\times$10$^{-7}$ & 2.39($\pm$0.44)$\times$10$^{-7}$ & 1.81($\pm$0.38)$\times$10$^{-7}$ & 5.36($\pm$0.88)$\times$10$^{-7}$ &                $\cdots$          \\
Adopted
          &      &  4.90($\pm$0.60)$\times$10$^{-7}$ & 2.39($\pm$0.44)$\times$10$^{-7}$ & 1.81($\pm$0.38)$\times$10$^{-7}$ & 5.36($\pm$0.88)$\times$10$^{-7}$ &                $\cdots$          \\
\rule[-1mm]{0mm}{5.0mm}\\
\cline{3-7}
    & (\AA) &  PN13 &  PN14 &  PN15 &  PN16 &  PN17\\
\hline
He$^{+}$  & 4471 &  0.104$\pm$0.015                  & 0.086$\pm$0.013                  & 0.090$\pm$0.013                  & 0.095$\pm$0.013                  & 0.081$\pm$0.012                  \\
          & 5876 &  0.089$\pm$0.008                  & 0.096$\pm$0.010                  & 0.099$\pm$0.008                  & 0.096$\pm$0.012                  & 0.088$\pm$0.008                  \\
          & 6678 &  0.088$\pm$0.013                  & 0.092$\pm$0.014                  & 0.089$\pm$0.013                  &      $\cdots$                    & 0.088$\pm$0.011                  \\
Adopted$^{\rm a}$ 
          &      &  0.089$\pm$0.008                  & 0.096$\pm$0.010                  & 0.099$\pm$0.008                  & 0.096$\pm$0.012                  & 0.088$\pm$0.008                  \\
He$^{2+}$ & 5411 &       $\cdots$                    &      $\cdots$                    &      $\cdots$                    &      $\cdots$                    & 2.00($\pm$0.36)$\times$10$^{-2}$ \\
          & 4686 &       $\cdots$                    &      $\cdots$                    & 1.73($\pm$0.20)$\times$10$^{-3}$ & 1.80($\pm$0.22)$\times$10$^{-2}$ & 2.21($\pm$0.20)$\times$10$^{-2}$ \\
N$^{+}$   & 5755 &  6.21($\pm$:)$\times$10$^{-6}$    & 8.82($\pm$1.87)$\times$10$^{-6}$ & 2.47($\pm$0.45)$\times$10$^{-5}$ &                $\cdots$          & 1.65($\pm$:)$\times$10$^{-6}$    \\
          & 6548 &  8.91($\pm$0.97)$\times$10$^{-6}$ & 3.61($\pm$0.44)$\times$10$^{-6}$ & 1.48($\pm$0.16)$\times$10$^{-5}$ & 3.81($\pm$0.41)$\times$10$^{-5}$ & 1.42($\pm$0.15)$\times$10$^{-6}$ \\
          & 6583 &  8.61($\pm$0.63)$\times$10$^{-6}$ & 3.43($\pm$0.28)$\times$10$^{-6}$ & 1.45($\pm$0.11)$\times$10$^{-5}$ & 4.57($\pm$0.40)$\times$10$^{-5}$ & 1.29($\pm$0.10)$\times$10$^{-6}$ \\
Adopted$^{\rm b}$ 
          &      &  8.61($\pm$0.63)$\times$10$^{-6}$ & 3.43($\pm$0.28)$\times$10$^{-6}$ & 1.45($\pm$0.11)$\times$10$^{-5}$ & 4.57($\pm$0.40)$\times$10$^{-5}$ & 1.29($\pm$0.10)$\times$10$^{-6}$ \\
O$^{+}$   & 3727 &  1.13($\pm$0.13)$\times$10$^{-4}$ & 9.96($\pm$1.10)$\times$10$^{-6}$ & 4.08($\pm$0.45)$\times$10$^{-5}$ & 3.57($\pm$0.48)$\times$10$^{-5}$ & 1.25($\pm$0.11)$\times$10$^{-5}$ \\
          & 7320 &  9.55($\pm$1.22)$\times$10$^{-5}$ & 3.74($\pm$0.55)$\times$10$^{-5}$ & 2.94($\pm$0.37)$\times$10$^{-5}$ & 5.10($\pm$0.66)$\times$10$^{-5}$ & 1.20($\pm$0.15)$\times$10$^{-5}$ \\
          & 7330 &  9.89($\pm$1.27)$\times$10$^{-5}$ & 4.43($\pm$0.61)$\times$10$^{-5}$ & 3.52($\pm$0.43)$\times$10$^{-5}$ & 1.02($\pm$0.15)$\times$10$^{-4}$ & 1.10($\pm$0.16)$\times$10$^{-5}$ \\
          & 7325 &  9.70($\pm$1.25)$\times$10$^{-5}$ & 4.05($\pm$0.60)$\times$10$^{-5}$ & 3.20($\pm$0.43)$\times$10$^{-5}$ & 7.39($\pm$1.05)$\times$10$^{-5}$ & 1.16($\pm$0.17)$\times$10$^{-5}$ \\
Adopted$^{\rm c}$ 
          &      &  1.10($\pm$0.14)$\times$10$^{-4}$ & 2.03($\pm$0.33)$\times$10$^{-5}$ & 4.01($\pm$0.45)$\times$10$^{-5}$ & 4.25($\pm$0.76)$\times$10$^{-5}$ & 1.24($\pm$0.12)$\times$10$^{-5}$ \\
O$^{2+}$  & 4363 &  4.31($\pm$0.52)$\times$10$^{-4}$ & 1.76($\pm$0.20)$\times$10$^{-4}$ & 6.07($\pm$0.72)$\times$10$^{-4}$ & 5.47($\pm$0.86)$\times$10$^{-4}$ & 2.84($\pm$0.30)$\times$10$^{-4}$ \\
          & 4959 &  4.23($\pm$0.16)$\times$10$^{-4}$ & 1.70($\pm$0.07)$\times$10$^{-4}$ & 6.00($\pm$0.23)$\times$10$^{-4}$ & 5.32($\pm$0.30)$\times$10$^{-4}$ & 2.73($\pm$0.10)$\times$10$^{-4}$ \\
          & 5007 &  4.40($\pm$0.08)$\times$10$^{-4}$ & 1.79($\pm$0.04)$\times$10$^{-4}$ & 6.15($\pm$0.11)$\times$10$^{-4}$ & 5.52($\pm$0.13)$\times$10$^{-4}$ & 2.88($\pm$0.05)$\times$10$^{-4}$ \\
Adopted$^{\rm d}$ 
          &      &  4.40($\pm$0.10)$\times$10$^{-4}$ & 1.79($\pm$0.05)$\times$10$^{-4}$ & 6.15($\pm$0.12)$\times$10$^{-4}$ & 5.52($\pm$0.15)$\times$10$^{-4}$ & 2.88($\pm$0.05)$\times$10$^{-4}$ \\
Ne$^{2+}$ & 3868 &  7.25($\pm$0.47)$\times$10$^{-5}$ & 3.15($\pm$0.20)$\times$10$^{-5}$ & 1.41($\pm$0.10)$\times$10$^{-4}$ & 1.22($\pm$0.13)$\times$10$^{-4}$ & 5.29($\pm$0.40)$\times$10$^{-5}$ \\
          & 3967 &  7.70($\pm$0.60)$\times$10$^{-5}$ & 3.34($\pm$0.27)$\times$10$^{-5}$ & 1.58($\pm$0.10)$\times$10$^{-4}$ & 1.48($\pm$0.15)$\times$10$^{-4}$ & 6.90($\pm$0.56)$\times$10$^{-5}$ \\
Adopted$^{\rm e}$ 
          &      &  7.25($\pm$0.47)$\times$10$^{-5}$ & 3.15($\pm$0.20)$\times$10$^{-5}$ & 1.41($\pm$0.10)$\times$10$^{-4}$ & 1.22($\pm$0.13)$\times$10$^{-4}$ & 5.29($\pm$0.40)$\times$10$^{-5}$ \\
S$^{+}$   & 6716 &  7.26($\pm$1.32)$\times$10$^{-7}$ & 5.75($\pm$1.19)$\times$10$^{-8}$ & 3.52($\pm$0.64)$\times$10$^{-7}$ & 8.16($\pm$1.95)$\times$10$^{-7}$ & 5.78($\pm$1.03)$\times$10$^{-8}$ \\
          & 6731 &  7.26($\pm$1.10)$\times$10$^{-7}$ & 5.75($\pm$0.86)$\times$10$^{-8}$ & 3.52($\pm$0.47)$\times$10$^{-7}$ & 8.16($\pm$1.62)$\times$10$^{-7}$ & 5.78($\pm$0.85)$\times$10$^{-8}$ \\
Adopted   &      &  7.26($\pm$1.20)$\times$10$^{-7}$ & 5.75($\pm$0.90)$\times$10$^{-8}$ & 3.52($\pm$0.56)$\times$10$^{-7}$ & 8.16($\pm$1.95)$\times$10$^{-7}$ & 5.78($\pm$0.94)$\times$10$^{-8}$ \\
\end{tabular}
\end{center}
\end{table*}

\addtocounter{table}{-1}
\begin{table*}
\begin{center}
\caption{(Continued)}
\label{ionic}
\begin{tabular}{llccccc}
\hline\hline
Ion & Line  & \multicolumn{5}{c}{Abundance (X$^{i+}$/H$^{+}$)}\\
\cline{3-7}
    & (\AA) &  PN13 &  PN14 &  PN15 &  PN16 &  PN17\\
\hline
S$^{2+}$  & 6312 &  6.05($\pm$1.32)$\times$10$^{-6}$ & 1.98($\pm$0.54)$\times$10$^{-6}$ & 7.89($\pm$1.85)$\times$10$^{-6}$ &                $\cdots$          & 1.07($\pm$0.23)$\times$10$^{-6}$ \\
          & 9069 &  7.40($\pm$0.60)$\times$10$^{-6}$ &                $\cdots$          &                $\cdots$          &                $\cdots$          & 1.75($\pm$0.16)$\times$10$^{-6}$ \\
          & 9531 &  1.94($\pm$0.27)$\times$10$^{-6}$ &                $\cdots$          &                $\cdots$          &                $\cdots$          & 2.96($\pm$0.38)$\times$10$^{-7}$ \\
Adopted$^{\rm f}$ 
          &      &  7.32($\pm$1.10)$\times$10$^{-6}$ & 1.98($\pm$0.54)$\times$10$^{-6}$ & 7.89($\pm$1.85)$\times$10$^{-6}$ &                $\cdots$          & 1.70($\pm$1.25)$\times$10$^{-6}$ \\
Cl$^{2+}$ & 5517 &                 $\cdots$          &                $\cdots$          &                $\cdots$          &                $\cdots$          & 4.06($\pm$4.05)$\times$10$^{-8}$ \\
          & 5537 &                 $\cdots$          &                $\cdots$          &                $\cdots$          &                $\cdots$          & 4.06($\pm$1.92)$\times$10$^{-8}$ \\
Ar$^{2+}$ & 7136 &  1.28($\pm$0.15)$\times$10$^{-6}$ & 5.71($\pm$0.67)$\times$10$^{-7}$ & 1.90($\pm$0.20)$\times$10$^{-6}$ & 1.11($\pm$0.16)$\times$10$^{-6}$ & 4.16($\pm$0.42)$\times$10$^{-7}$ \\
          & 7751 &  8.34($\pm$2.07)$\times$10$^{-7}$ & 3.03($\pm$0.93)$\times$10$^{-7}$ & 1.39($\pm$0.35)$\times$10$^{-6}$ & 2.03($\pm$0.64)$\times$10$^{-6}$ & 2.43($\pm$0.58)$\times$10$^{-7}$ \\
Adopted$^{\rm g}$ 
          &      &  1.28($\pm$0.15)$\times$10$^{-6}$ & 5.71($\pm$0.67)$\times$10$^{-7}$ & 1.90($\pm$0.20)$\times$10$^{-6}$ & 1.11($\pm$0.16)$\times$10$^{-6}$ & 4.16($\pm$0.42)$\times$10$^{-7}$ \\
Ar$^{3+}$ & 4711 &                 $\cdots$          &                $\cdots$          & 2.75($\pm$0.43)$\times$10$^{-7}$ &                $\cdots$          & 4.94($\pm$0.88)$\times$10$^{-7}$ \\
          & 4740 &                 $\cdots$          &                $\cdots$          & 3.35($\pm$0.54)$\times$10$^{-7}$ &                $\cdots$          & 5.35($\pm$1.01)$\times$10$^{-7}$ \\
Adopted
          &      &                 $\cdots$          &                $\cdots$          & 3.35($\pm$0.54)$\times$10$^{-7}$ &                $\cdots$          & 5.35($\pm$1.01)$\times$10$^{-7}$ \\
Ar$^{4+}$ & 7005 &                 $\cdots$          &                $\cdots$          &                $\cdots$          &                $\cdots$          & 4.59($\pm$:)$\times$10$^{-8}$    \\
\hline
\end{tabular}
\end{center}
\begin{description}
\item[$^{\rm a}$] The He$^{+}$/H$^{+}$ abundance ratio derived from 
the He~{\sc i} $\lambda$5876 line is adopted.
\item[$^{\rm b}$] The N$^{+}$/H$^{+}$ abundance ratio derived from 
[N~{\sc ii}] $\lambda$6583 is adopted.
\item[$^{\rm c}$] A weighted average value of the O$^{+}$/H$^{+}$ 
ratios derived from the [O~{\sc ii}] $\lambda$3727 nebular and 
$\lambda$7325 (= $\lambda$7320+$\lambda$7330) auroral lines are 
adopted, with the assigned weights proportional to the intensities 
of these two lines.  See text for details.
\item[$^{\rm d}$] The O$^{2+}$/H$^{+}$ ratio derived from [O~{\sc 
iii}] $\lambda$5007 is adopted. 
\item[$^{\rm e}$] The Ne$^{2+}$/H$^{+}$ ratio derived from [Ne~{\sc 
iii}] $\lambda$3868 is adopted.
\item[$^{\rm f}$] For PN9, PN11, PN13 and PN17, where both the 
$\lambda$6312 and $\lambda$9069 lines are observed, a weighted 
average of the S$^{2+}$/H$^{+}$ ratios derived from these two 
[S~{\sc iii}] lines is adopted, with the weights proportional to 
the dereddened line fluxes. 
\item[$^{\rm g}$] The Ar$^{2+}$/H$^{+}$ ratio derived from [Ar~{\sc 
iii}] $\lambda$7136 is adopted, because measurements of this line 
are much better than [Ar~{\sc iii}] $\lambda$7751 lying at the red 
end of R1000B grism. 
\end{description}
\end{table*}

\begin{table*}
\begin{center}
\begin{minipage}{162mm}
\caption{Elemental Abundances}
\label{elemental}
\begin{tabular}{lcccccccc}
\hline\hline
Elem. & \multicolumn{8}{c}{X/H}\\
\cline{2-9}
 & \multicolumn{2}{c}{PN8} & \multicolumn{2}{c}{PN9} & \multicolumn{2}{c}{PN10} & \multicolumn{2}{c}{PN11}\\
\hline
He & 0.105$\pm$0.014                  & 11.02    & 0.112$\pm$0.017                  & 11.05    & 0.091$\pm$0.014                  & 10.96    & 0.106$\pm$0.015                  & 11.02\\ 
C  &                $\cdots$          & $\cdots$ &                $\cdots$          & $\cdots$ &                $\cdots$          & $\cdots$ & 1.65($\pm$0.51)$\times$10$^{-3}$ &  9.22\\ 
N  & 8.90($\pm$1.34)$\times$10$^{-5}$ &  7.95    & 1.28($\pm$0.20)$\times$10$^{-4}$ &  8.11    & 6.16($\pm$1.05)$\times$10$^{-5}$ &  7.79    & 1.37($\pm$0.27)$\times$10$^{-4}$ &  8.14\\ 
O  & 4.01($\pm$0.45)$\times$10$^{-4}$ &  8.60    & 4.55($\pm$0.53)$\times$10$^{-4}$ &  8.66    & 2.90($\pm$0.34)$\times$10$^{-4}$ &  8.46    & 3.74($\pm$0.44)$\times$10$^{-4}$ &  8.57\\ 
Ne & 1.04($\pm$0.21)$\times$10$^{-4}$ &  8.02    & 9.47($\pm$2.28)$\times$10$^{-5}$ &  7.98    & 6.82($\pm$1.64)$\times$10$^{-5}$ &  7.83    & 8.24($\pm$2.02)$\times$10$^{-5}$ &  7.92\\ 
S  & 7.64($\pm$1.92)$\times$10$^{-6}$ &  6.88    & 6.35($\pm$1.60)$\times$10$^{-6}$ &  6.80    & 3.32($\pm$0.84)$\times$10$^{-6}$ &  6.52    & 5.87($\pm$1.48)$\times$10$^{-6}$ &  6.77\\ 
Cl & 3.00($\pm$1.15)$\times$10$^{-7}$ &  5.47    &                $\cdots$          & $\cdots$ &                $\cdots$          & $\cdots$ & 1.03($\pm$0.38)$\times$10$^{-7}$ &  5.01\\ 
Ar & 1.97($\pm$0.58)$\times$10$^{-6}$ &  6.30    & 1.47($\pm$0.44)$\times$10$^{-6}$ &  6.17    & 9.92($\pm$3.03)$\times$10$^{-7}$ &  6.00    & 1.60($\pm$0.50)$\times$10$^{-6}$ &  6.20\\ 
\cline{2-9}
 & \multicolumn{2}{c}{PN12} & \multicolumn{2}{c}{PN13} & \multicolumn{2}{c}{PN14} & \multicolumn{2}{c}{PN15}\\ 
\hline
He & 0.103$\pm$0.014                  & 11.01    & 0.089$\pm$0.013                  & 10.95    & 0.096$\pm$0.014                  & 10.98    & 0.101$\pm$0.014                  & 11.00   \\ 
C  & 1.09($\pm$0.34)$\times$10$^{-3}$ &  9.04    &                $\cdots$          & $\cdots$ &                $\cdots$          & $\cdots$ &                $\cdots$          & $\cdots$\\ 
N  & 5.58($\pm$1.10)$\times$10$^{-5}$ &  7.75    & 4.29($\pm$0.84)$\times$10$^{-5}$ &  7.63    & 3.37($\pm$0.66)$\times$10$^{-5}$ &  7.53    & 2.39($\pm$0.47)$\times$10$^{-4}$ &  8.38   \\ 
O  & 2.28($\pm$0.27)$\times$10$^{-4}$ &  8.36    & 5.51($\pm$0.65)$\times$10$^{-4}$ &  8.74    & 2.00($\pm$0.23)$\times$10$^{-4}$ &  8.30    & 6.63($\pm$0.78)$\times$10$^{-4}$ &  8.82   \\ 
Ne & 3.67($\pm$0.90)$\times$10$^{-5}$ &  7.56    & 9.07($\pm$2.22)$\times$10$^{-5}$ &  7.96    & 3.51($\pm$0.86)$\times$10$^{-5}$ &  7.54    & 1.52($\pm$0.37)$\times$10$^{-4}$ &  8.18   \\ 
S  & 3.11($\pm$0.78)$\times$10$^{-6}$ &  6.49    & 8.60($\pm$2.17)$\times$10$^{-6}$ &  6.93    & 3.13($\pm$0.80)$\times$10$^{-6}$ &  6.50    & 1.48($\pm$0.38)$\times$10$^{-5}$ &  7.17   \\ 
Cl & 8.74($\pm$3.22)$\times$10$^{-8}$ &  4.94    &                $\cdots$          & $\cdots$ &                $\cdots$          & $\cdots$ &                $\cdots$          & $\cdots$\\ 
Ar & 2.24($\pm$0.68)$\times$10$^{-6}$ &  6.35    & 2.39($\pm$0.73)$\times$10$^{-6}$ &  6.38    & 1.07($\pm$0.33)$\times$10$^{-6}$ &  6.03    & 2.38($\pm$0.73)$\times$10$^{-6}$ &  6.38   \\ 
\cline{2-9}
 & \multicolumn{2}{c}{PN16} & \multicolumn{2}{c}{PN17} & \multicolumn{2}{c}{Orion$^{\rm a}$} & \multicolumn{2}{c}{Solar$^{\rm b}$}\\
\hline
He & 0.114$\pm$0.050                  & 11.06    & 0.110$\pm$0.015                  & 11.04    & 0.098                 & 10.99 & 0.085                 & 10.93\\
C  &                $\cdots$          & $\cdots$ &                $\cdots$          & $\cdots$ & 2.63$\times$10$^{-4}$ &  8.42 & 2.69$\times$10$^{-4}$ &  8.43\\
N  & 7.15($\pm$1.42)$\times$10$^{-4}$ &  8.85    & 3.62($\pm$0.71)$\times$10$^{-5}$ &  7.56    & 5.37$\times$10$^{-5}$ &  7.73 & 6.76$\times$10$^{-5}$ &  7.83\\
O  & 6.66($\pm$0.78)$\times$10$^{-4}$ &  8.82    & 3.48($\pm$0.41)$\times$10$^{-4}$ &  8.54    & 5.13$\times$10$^{-4}$ &  8.71 & 4.89$\times$10$^{-4}$ &  8.69\\
Ne & 1.51($\pm$0.40)$\times$10$^{-4}$ &  8.18    & 6.41($\pm$1.57)$\times$10$^{-5}$ &  7.81    & 1.12$\times$10$^{-4}$ &  8.05 & 8.51$\times$10$^{-5}$ &  7.93\\
S  & 1.26($\pm$0.32)$\times$10$^{-5}$ &  7.10    & 3.85($\pm$0.97)$\times$10$^{-6}$ &  6.58    & 1.66$\times$10$^{-5}$ &  7.22 & 1.32$\times$10$^{-5}$ &  7.12\\
Cl &                $\cdots$          & $\cdots$ & 8.95($\pm$3.30)$\times$10$^{-8}$ &  4.95    & 2.88$\times$10$^{-7}$ &  5.46 & 3.16$\times$10$^{-7}$ &  5.50\\
Ar & 2.08($\pm$0.64)$\times$10$^{-6}$ &  6.32    & 1.03($\pm$0.31)$\times$10$^{-6}$ &  6.01    & 4.17$\times$10$^{-6}$ &  6.62 & 2.51$\times$10$^{-6}$ &  6.40\\
\hline
\multicolumn{9}{l}{NOTE. -- For each PN, abundances both in linear 
form and logarithm, $\log$(X/H)\,+\,12, are presented.} 
\end{tabular}
\begin{description}
\item[$^{\rm a}$] \citet{esteban04}.
\item[$^{\rm b}$] \citet{asplund09}.
\end{description}
\end{minipage}
\end{center}
\end{table*}

\begin{table}
\begin{center}
\caption{Ionization Correction Factors}
\label{icfs}
\begin{tabular}{lrrrrrrr}
\hline\hline
Elem. & \multicolumn{5}{c}{ICF}\\
\cline{2-6}
      & PN8  & PN9  & PN10 & PN11 & PN12\\
\hline
C  & $\cdots$ & $\cdots$ & $\cdots$ &  1.123 & 1.339\\
N  & 11.20  & 11.91  & 15.11  & 25.85  & 4.020\\
O  &  1.006 &  1.008 &  1.000 &  1.076 & 1.005\\
Ne &  1.106 &  1.101 &  1.071 &  1.123 & 1.339\\
S  &  1.599 &  1.629 &  1.753 &  2.077 & 1.202\\
Cl &  1.630 & $\cdots$ & $\cdots$ &  2.263 & 1.308\\
Ar &  1.098 &  1.092 &  1.071 &  1.040 & 1.870\\
\cline{2-6}
      & PN13 & PN14 & PN15 & PN16 & PN17\\
\hline
C  & $\cdots$ & $\cdots$ & $\cdots$ & $\cdots$ & $\cdots$\\
N  & 4.986 & 9.804 & 16.50  & 15.65  & 28.13 \\
O  & 1.000 & 1.000 &  1.012 &  1.121 &  1.162\\
Ne & 1.251 & 1.114 &  1.078 &  1.207 &  1.212\\
S  & 1.269 & 1.536 &  1.802 &  1.772 &  2.134\\
Cl & $\cdots$ & $\cdots$ & $\cdots$ & $\cdots$ & 2.204\\
Ar & 1.251 & 1.870 &  1.064 &  1.870 &  1.037\\
\hline
\end{tabular}
\end{center}
\end{table}

The uncertainties in the brackets following the elemental 
abundances in Table~\ref{elemental} were estimated from the errors 
in ionic abundances (Table~\ref{ionic}) through propagation.  The 
possible errors brought about by ionization corrections were not 
considered, although they could be significant for some heavy 
elements.  The dominant ionization stage of helium in PNe is 
He$^{+}$, and thus error in He/H mainly comes from He$^{+}$/H$^{+}$.
The [O~{\sc ii}] and [O~{\sc iii}] nebular lines are among the best 
observed in the optical spectrum of a PN.  Although determination 
of O$^{+}$/H$^{+}$ is usually less accurate than O$^{2+}$/H$^{+}$ 
due to the detection sensitivity of instruments in the optical 
wavelength region, concentration of oxygen in O$^{2+}$ is much 
higher than in O$^{+}$.  Besides, ICF(O) is always close to unity 
(Table~\ref{icfs}).  He/H and O/H thus are the most accurate among 
all elements analyzed in this paper.  N/H and Ne/H were derived 
based on the ionic and elemental abundances of oxygen and are 
expected to be reliable.  Determinations of S/H in this work were 
improved for some PNe by using the strong [S~{\sc iii}] 
$\lambda$9069 nebular line and adopting the [S~{\sc iii}] 
temperatures.  Errors in S/H and Ar/H introduced by the ICFs are 
non-negligible, but difficult to quantify.  Thus they were not 
considered in the error estimation.  Uncertainty in Cl/H is large 
because only Cl$^{2+}$ was observed.  Uncertainty in carbon is also 
considerable since only the weak C~{\sc ii} $\lambda$4267 line was 
observed in our GTC spectrum.

\subsection{Emission Features of [WR] Central Stars}
\label{sec3:part5}

\begin{figure*}
\begin{center}
\includegraphics[width=5.0cm,angle=-90]{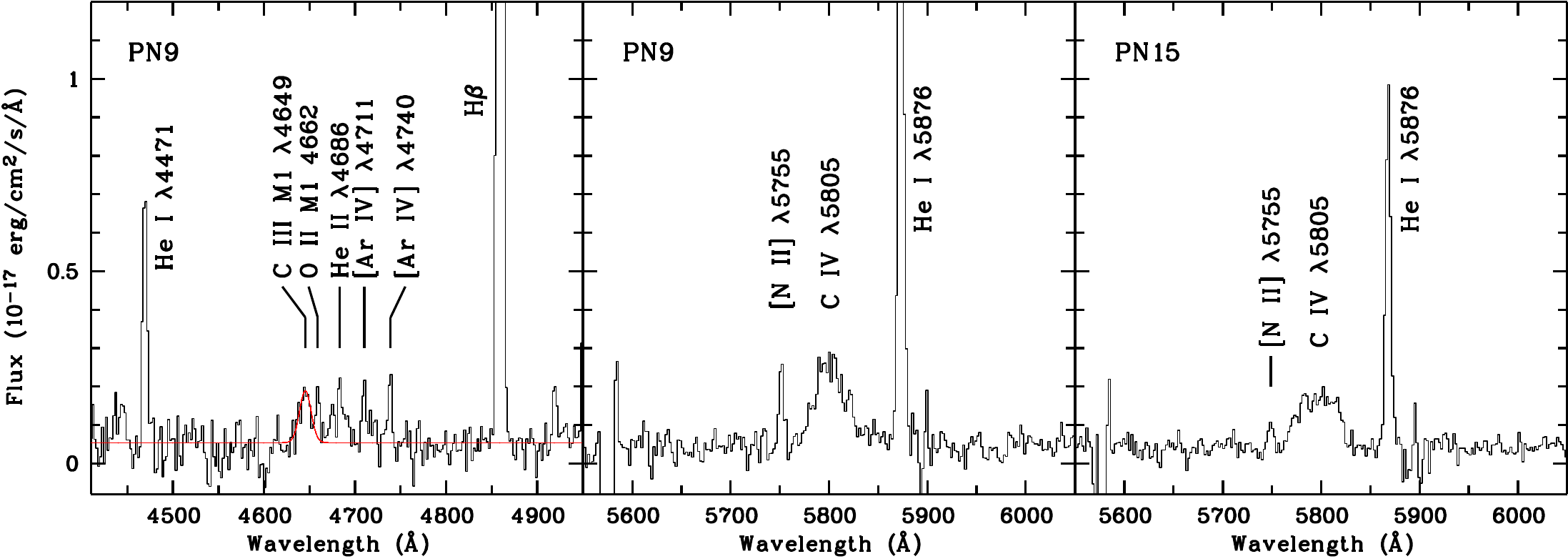}
\caption{Spectrum of PN9 and PN15 showing the broad C~{\sc iv} 
$\lambda$5805 emission line (middle and right panels) of the central 
stars.  The C~{\sc iii} $\lambda$4649 line (left panel) might also 
be due to the central star emission and is fitted by a Gaussian 
profile with FWHM $\sim$15\,{\AA}.} 
\label{fig6}
\end{center}
\end{figure*}

We detected the C~{\sc iv} $\lambda$5805 (a blend of the 
$\lambda\lambda$5801.33,5811.98\footnote{Wavelengths of the 
permitted transitions, including ORLs, are adopted from 
\citet{hh94}.} doublet) broad emission in the spectra of PN9 and 
PN15 (Figure~\ref{fig6}).  Including target PN2 in Paper~II, we 
have now observed this C~{\sc iv} line in three PNe in our GTC 
sample.  In PN9, we also detected C~{\sc iii} $\lambda$4649 (a 
blend of $\lambda\lambda$4647.42,4650.25,4651.47 of M1 
2s3s\,$^{3}$S--2s3p\,$^{3}$P$^{\rm o}$ multiplet), whose observed 
flux is 2.2$\times$10$^{-17}$ erg\,cm$^{-2}$\,s$^{-1}$ and line 
width is $\sim$15\,{\AA}.  This C~{\sc iii} line is probably also 
blended with the faint O~{\sc ii} $\lambda\lambda$4649.13,4650.84 
ORLs of M1 2p$^{2}$\,3s\,$^{4}$P--2p$^{2}$\,3p\,$^{4}$D$^{\rm o}$ 
multiplet, because we found a nearby narrow emission feature which 
could be due to $\lambda$4661.63 of the same multiplet of O~{\sc 
ii}.  Line fluxes and widths of C~{\sc iv} $\lambda$5805 of the 
three PNe are presented in Table~\ref{carbon_lines}. 

The broad C~{\sc iii} and C~{\sc iv} lines observed in PN2, PN9 
and PN15 are from their central stars, which are probably of 
Wolf-Rayet ([WR]) type.  The intensity ratio and FWHM of C~{\sc 
iii} $\lambda$4649 and C~{\sc iv} $\lambda$5805 indicate that PN9 
has a [WC4] central star, according to the classification of 
\citet[][also \citealt{cmb98}]{an03}.  The estimated stellar 
temperature of PN9 seems to be slightly higher than the temperature 
range \citep[$\sim$55\,000--91\,000~K,][]{an03} covered by the 
[WC4]-type stars.  C~{\sc iii} $\lambda$4649 was not observed in 
the other two PNe.  Previous GTC spectroscopy have found [WC4]-type 
central stars in two outer-disk PNe \citep[][PN ID 174 and 2496 in 
\citealt{merrett06}]{balick13}. 

How these [WC] central stars formed is not well understood, although
a significant fraction of Galactic PNe have been observed to harbor
such type of stars.  The five M31 PNe with unambiguous detection of 
broad [WC] features are all bright in [O~{\sc iii}], $<$0.9~mag from
the PNLF bright cut-off, indicating that He-burning cores might 
produce visible PNe.  This may help to constrain the post-AGB 
evolutionary models of He burners.  As suggested by 
\citet{balick13}, an AGB final thermal pulse, or a late thermal 
pulse early in post-AGB, might be the channel to these [WC] central 
stars. 

\subsection{Duplication with Recent GTC Spectroscopy}
\label{sec3:part6}

PN14 in our sample was also observed at the GTC by \citet[][PN ID 
M2507]{corradi15}, who also used the OSIRIS R1000B grism but 
with a slit width of 0\farcs8.  Our observing conditions are better 
(seeing $\sim$0\farcs8 and clear nights). The logarithmic extinction 
parameters $c$(H$\beta$) derived from the two observations are also 
quite similar.  Although the extinction laws adopted are different 
(\citealt{ccm89} in this work, and \citealt{sm79} in 
\citealt{corradi15}), it has been proved that in the wavelength 
range covered by the OSIRIS R1000B grism, line fluxes corrected 
using different extinction laws differ very little.  Our observed 
H$\beta$ flux of PN14 is only 4.6\% lower than that given by 
\citet{corradi15} by 4.6\%.  The dereddened fluxes of the [O~{\sc 
ii}] $\lambda$3727 and [O~{\sc iii}] $\lambda$5007 nebular lines of 
PN14 differ from the observations of \citet{corradi15} by $\sim$1\% 
and 2\%, respectively.  However, the dereddened fluxes of the 
[O~{\sc iii}] $\lambda$4363 differ by $\sim$10\%.  Our O/H ratio 
[2.00($\pm$0.23)$\times$10$^{-4}$] of PN14 agrees with that of 
\citet[][1.73($\pm$0.44)$\times$10$^{-4}$]{corradi15} within the 
errors.

\begin{table}
\begin{center}
\caption{The C~{\sc iv} $\lambda$5805 Emission Line}
\label{carbon_lines}
\begin{tabular}{lcc}
\hline\hline
ID   & FWHM  & Flux$^{\rm a}$ \\
     & (\AA) & (erg\,cm$^{-2}$\,s$^{-1}$)\\
\hline
PN2  & 43 & 8.19$\times$10$^{-17}$ \\
PN9  & 33 & 7.35$\times$10$^{-17}$ \\
PN15 & $\sim$49 & 6.74$\times$10$^{-17}$ \\
\hline
\end{tabular}
\begin{description}
\item[$^{\rm a}$] Observed flux measured from the extracted 1D 
spectrum. 
\end{description}
\end{center}
\end{table}

\section{Discussion} 
\label{sec4}

\subsection{Correlation Studies of Abundances} 
\label{sec4:part1}

The abundances of $\alpha$-elements (O, Ne, S, Ar, etc.) of a PN 
reflect the metallicity ($Z$) of the interstellar medium (ISM) 
from which the progenitor star of the PN formed.  The He/H and 
N/O ratios are also indicative of progenitor masses: Type~I PNe 
tend to have higher abundance ratios, while Type~II PNe generally 
exhibit low N/O and less massive progenitors 
\citep{peimbert78,maciel92,kb94}.  The $\alpha$-elements of PNe 
also help to constrain the theory of stellar models and probe into 
the evolution of low- to intermediate-mass stars and consequently 
chemical evolution of galaxies \citep[e.g.,][]{kh01,milingo02a,
milingo02b,kwitter03,henry04}.  Although the exact mechanism(s) 
still unclear, previous studies of Galactic and Magellanic Cloud 
PNe have established that PN morphology is a useful indicator of 
stellar populations \citep[][]{peimbert83,manchado00,shaw01,
stanghellini02a,stanghellini02b,stanghellini03,stanghellini06}. 
At the distance of M31, all PNe are spatially unresolved, and their 
central stars cannot be directly observed.  The stellar population 
of PNe can be inferred from the elemental abundances of the nebulae.

\begin{figure*}
\begin{center}
\includegraphics[width=17.5cm,angle=0]{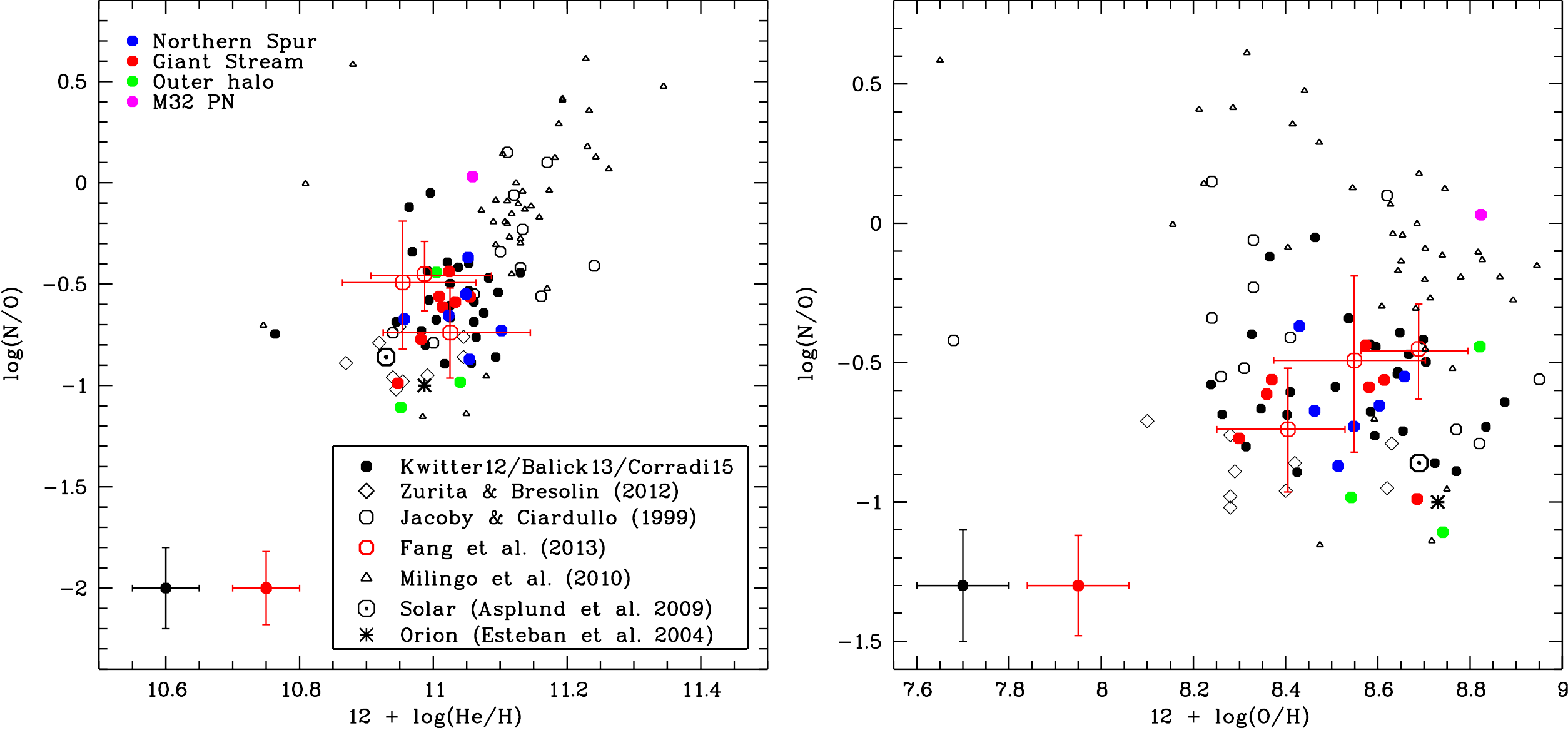}
\caption{N/O versus He/H (left) and O/H (right), both displayed in 
logarithm.  Different symbols represent different samples (see the 
legend).  Explanation of different data sets are given in the text.
The color-filled (blue, red, green and pink) circles are our 
GTC sample, including those observed in Paper~II; color code is 
the same as in Figure~\ref{fig1} (see also the top-left legend). 
Red open circles are the Northern Spur PNe studied in Paper~I. 
The M31 disk PNe observed by \citet[][Kwitter12]{kwitter12}, 
\citet[][Balick13]{balick13} and \citet[][Corradi15]{corradi15} 
are represented by black filled circles. Typical error bars of the 
two samples are indicated at the corner.  Symbols in 
Figures~\ref{fig8}--\ref{fig11} have the same meaning.} 
\label{fig7}
\end{center}
\end{figure*}

\begin{figure*}
\begin{center}
\includegraphics[width=17.5cm,angle=0]{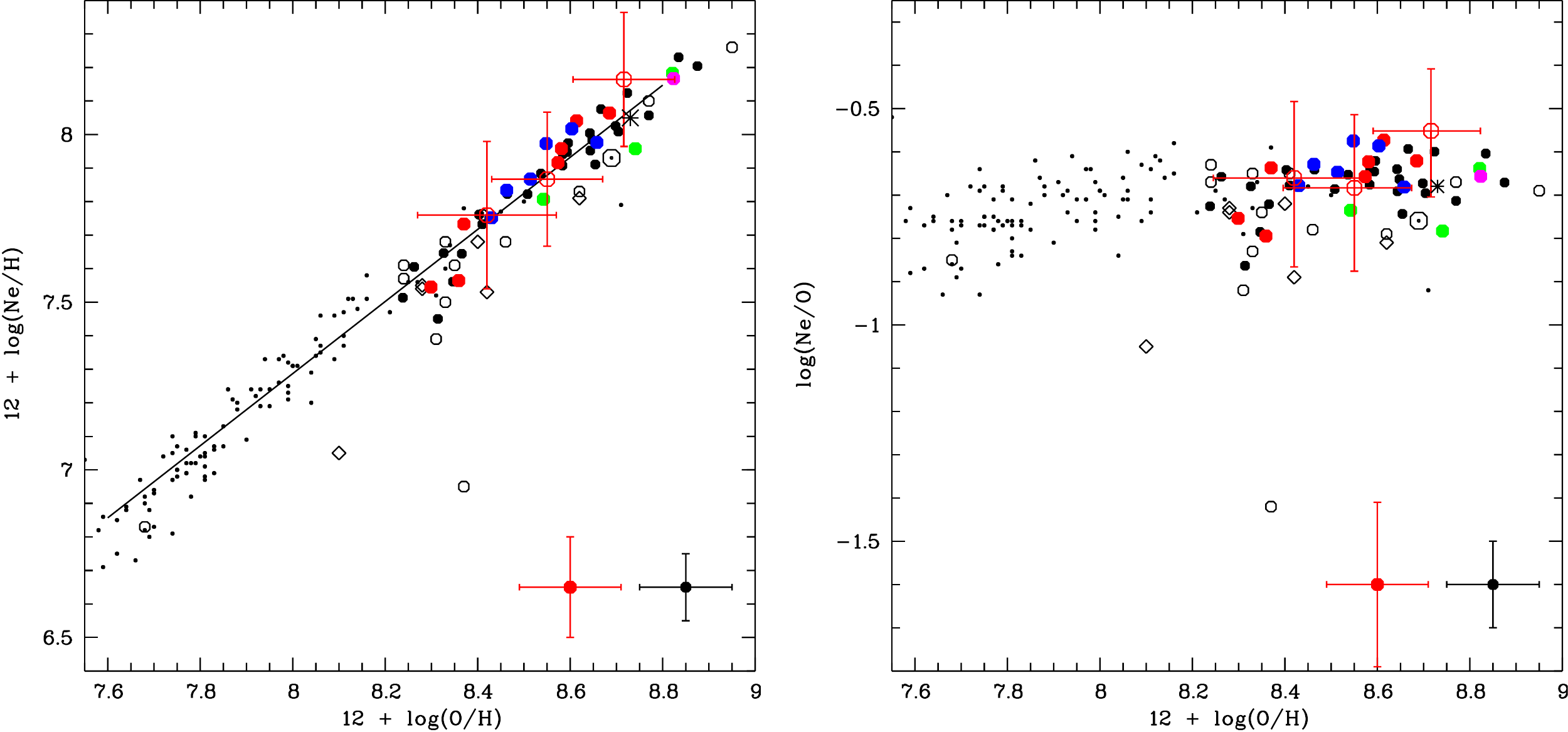}
\caption{Ne/H versus O/H (left) and Ne/O versus O/H (right), both 
displayed in logarithm.  The small black dots are the extragalactic 
H~{\sc ii} regions and metal-poor blue compact galaxies (see 
references in the text), and the solid black line in the left panel 
is a least-squares linear fit to these data.  These symbols in 
Figures~\ref{fig9} and \ref{fig10} have the same meaning.  Color 
codes of the other symbols are the same as in Figure~\ref{fig7}.} 
\label{fig8}
\end{center}
\end{figure*}

\begin{figure*}
\begin{center}
\includegraphics[width=17.5cm,angle=0]{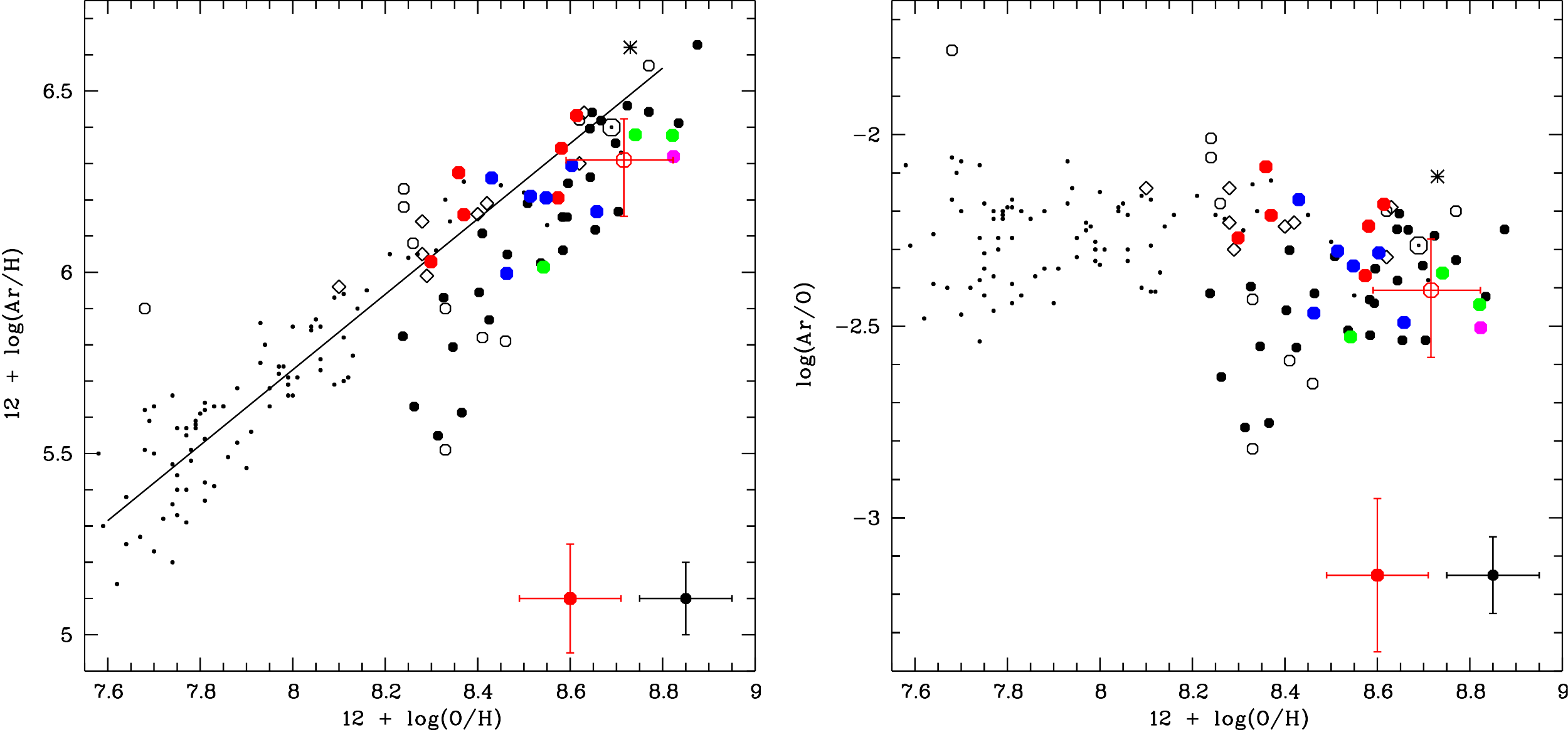}
\caption{Same as Figure~\ref{fig8} but for Ar/H versus O/H (left)
and the Ar/O ratio versus O/H (right).}
\label{fig9}
\end{center}
\end{figure*}

\begin{figure*}
\begin{center}
\includegraphics[width=17.5cm,angle=0]{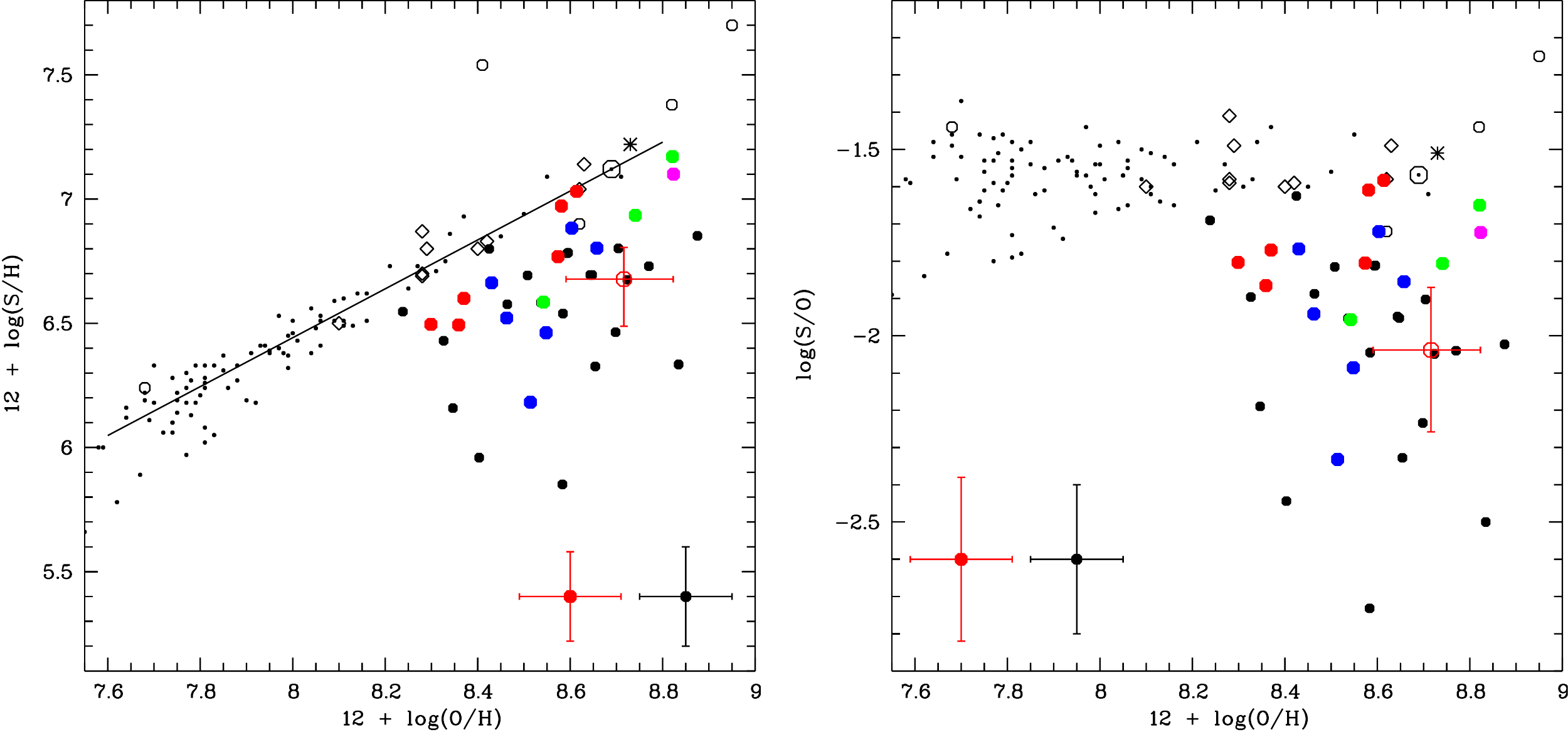}
\caption{Same as Figure~\ref{fig8} but for S/H versus O/H (left)
and the S/O ratio versus O/H (right).}
\label{fig10}
\end{center}
\end{figure*}

\begin{figure}
\begin{center}
\includegraphics[width=1.0\columnwidth,angle=0]{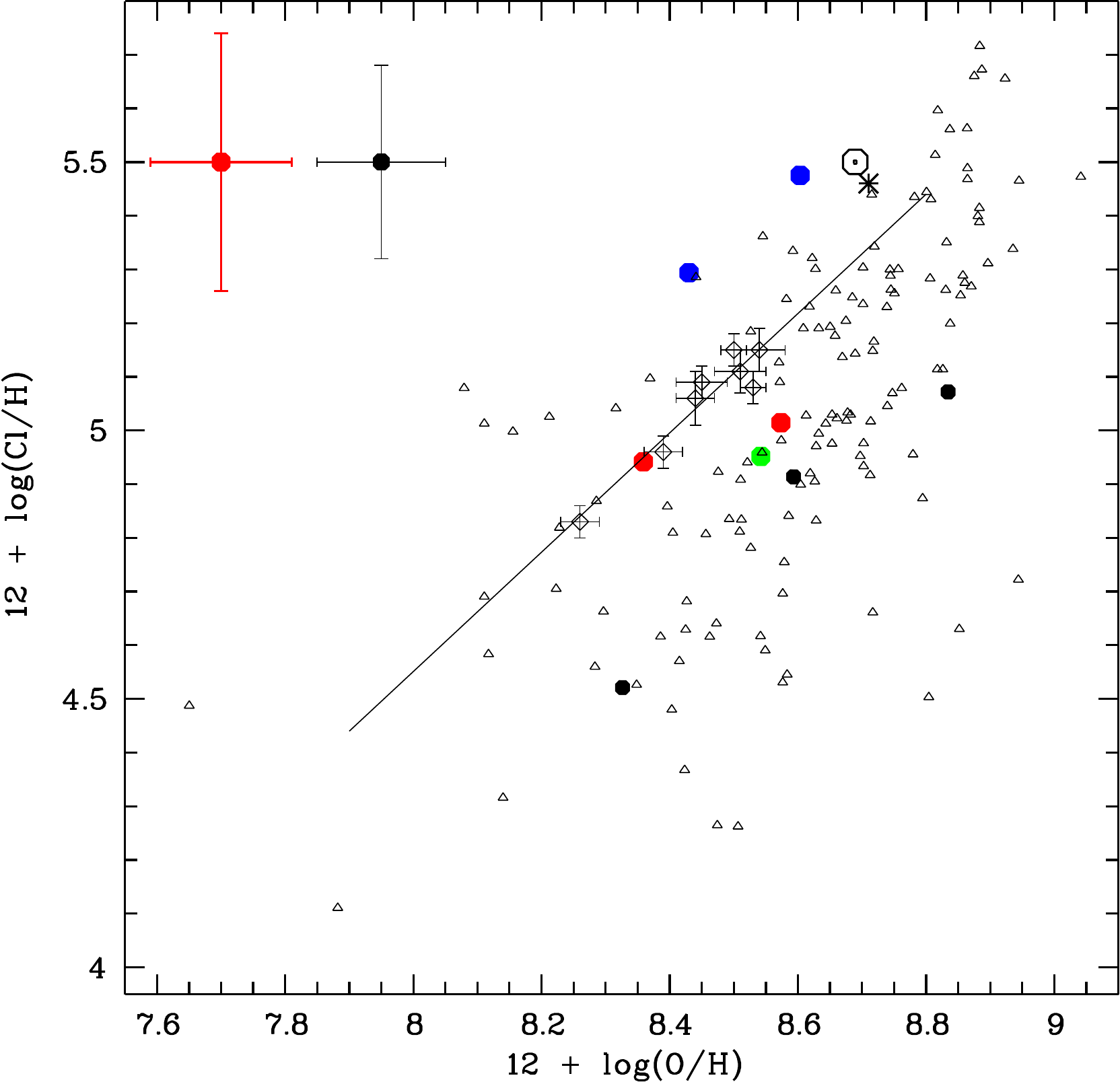}
\caption{The Cl/H versus O/H in logarithm scale.  The small 
open triangles are the Galactic PN samples from \citet{henry04,
henry10} and \citet{milingo10}; the open diamonds are the 
Galactic H~{\sc ii} regions studied by \citet{esteban15}, and 
the black line is a linear fit to this sample, with a slop of 
1.11$\pm$0.15.  The other symbols and color codes are the same 
as in Figure~\ref{fig7}.}
\label{fig11}
\end{center}
\end{figure}

Following the discussion in Paper~II, in this section we made a 
correlation study of the abundances of He, N, O and 
$\alpha$-elements for our extended halo sample (see 
Figures~\ref{fig7}--\ref{fig11}).  The M31 outer-disk PNe observed 
by \citet{kwitter12}, \citet{balick13} and 
\citet{corradi15}\footnote{In the outer-disk sample studied by 
\citet{corradi15}, there are three PNe with large deviation from 
the disk kinematics that might actually be associated with the outer 
halo or some of its substructures, as suggested by the authors.}  
using large telescopes (Gemini-North and the GTC) were included, 
and henceforth mentioned as the disk sample (or the disk PNe) in 
this paper.  Also considered in the study were previous observations 
of PNe and H~{\sc ii} regions in M31, including the bulge and disk 
PNe from \citet{jc99}, and the nine H~{\sc ii} regions from 
\citet{zb12} where the temperature-diagnostic auroral lines were 
observed.  The solar abundances from \citet{asplund09} and the 
abundances of the Orion Nebula from \citet{esteban04} were used as 
benchmarks. 

Figure~\ref{fig7} shows the N/O abundance ratio versus He/H (left) 
and O/H (right) in logarithm.  Our sample (the color-filled circles 
in Figure~\ref{fig7}), including the PNe studied in Papers~I and 
II, all have low N/O ($<$0.5) except PN16, a PN associated with 
M32 and whose N/O ($\sim$1.07$\pm$0.24) is higher than the other 
targets, and hints at the possibility of Type~I nature.  Its He/H 
(=0.114) however, is normal compared to others. 
Our targets and the disk sample show no obvious trend in N/O versus
He/H or O/H, and are clearly separated from the Galactic Type~I 
objects of \citet{milingo10}, whose N/O seems to be correlated 
with He/H and anti-correlated with O/H.  Among the M31 disk sample,
there are two outliers and one of them has N/O close to 1.0.  Most 
of the M31 PNe have higher N/O ratios than H~{\sc ii} regions 
(including the Orion Nebula), but there are three PNe in our sample 
with very low N/O.  Within our sample, the PNe associated with the 
Northern Spur, the Giant Stream, and the outer halo generally cannot 
be distinguished from each other in abundance ratios, although the 
halo target PN13 has the lowest N/O ratio. 

All M31 PNe show positive correlation between neon and oxygen 
(Figure~\ref{fig8}), consistent with the previous observations of 
PNe in the Galaxy, the Large and Small Magellanic Clouds and M31 
\citep{henry89}.  One outlier is from \citet{jc99}.  This 
neon-oxygen positive correlation was defined by the samples of 
H~{\sc ii} regions and metal-poor blue compact galaxies analyzed by 
\citet{it99}, \citet{izotov12} and \citet{kennicutt03}. We noticed 
that 12+$\log$(Ne/H) of the Sun \citep[7.93$\pm$0.10,][]{asplund09}
is slightly lower than, although still agrees within the errors 
with, what is expected from the neon-oxygen correlation, indicating 
that the current solar neon might be underestimated.  This problem 
was investigated through a comparison study of PNe and H~{\sc ii} 
regions by \citet{wl08}.

Argon of our targets are generally correlated with oxygen, although 
the scatter is noticeable (Figure~\ref{fig9}). Our sample also 
seems to have slightly lower argon than what is expected from the 
argon-oxygen correlation defined by the H~{\sc ii} regions and 
metal-poor blue compact galaxies, but this deficiency is more 
obvious in the outer-disk sample, especially in the PNe with lower 
oxygen [12+$\log$(O/H)$<$8.45].  As discussed in Paper~II, sulfur 
abundances of M31 PNe are all lower than what is expected from the 
sulfur-oxygen correlation.  Although the GTC spectra of four PNe 
(PN9, PN11, PN13 and PN17) in our sample have covered the [S~{\sc 
iii}] $\lambda\lambda$9069,9531 lines, their 12+$\log$(S/H) are 
still underabundant by $\sim$0.24--0.39~dex.  This deficiency in 
sulfur, known as the ``sulfur anomaly'', was previously noticed in 
the Galactic PNe \citep{henry04,milingo10}.  So far the most 
plausible explanation seems to be the inadequacy in ICF used to 
correct for the sulfur ions (e.g., S$^{3+}$ and higher ionization 
stages) unobserved in the optical but detectable in the IR, although 
taking into account the IR observations of S$^{3+}$ has alleviated 
but could not eliminate this deficiency \citep{henry12}. Theoretical 
studies shows that sulfur is unlikely destroyed by the 
nucleosynthetic processes in the low- to intermediate-mass stars 
\citep{sk13}.  On the other hand, the M31 H~{\sc ii} regions of 
\citet{zb12} generally agree with the sulfur-oxygen correlation 
(Figure~\ref{fig10}). 

Although strictly speaking chlorine cannot be classified as an 
$\alpha$-element because its two stable isotopes, $^{35}$Cl and 
$^{37}$Cl, are not formed through the $\alpha$-processes but produced
during both hydrostatic and explosive oxygen burning \citep{ww95}, 
it is a secondary product during oxygen burning and created from 
the isotopes of sulfur and argon \citep{clayton03}.  Cl/H was 
derived for five objects in our GTC sample using the [Cl~{\sc iii}] 
$\lambda\lambda$5517,5537 doublet (Table~\ref{elemental}).  It has 
also been derived for three outer-disk PNe.  Together with the 
Galactic samples from \citet{henry04,henry10} and \citet{milingo10}, 
chlorine exhibits a loose correlation with oxygen 
(Figure~\ref{fig11}).  The chlorine-oxygen relation among the M31 
PNe alone has large scatter, probably much affected by large 
uncertainties, given the weakness of the [Cl~{\sc iii}] lines. 
Using the Galactic H~{\sc ii} regions from \citet[][including the 
Orion Nebula]{esteban15} as a baseline for correlation, we found 
that not all M31 PNe are located along this relation within the 
errors.

\subsection{Populations of PNe}
\label{sec4:part2}

In this section, we study the stellar populations of our sample 
by constraining the central star parameters, which again, were 
estimated from the observed nebular spectra. 
In a PN spectrum, the intensities of nebular emission lines, such
as [O~{\sc iii}] $\lambda$5007 and He~{\sc ii} $\lambda$4686,
relative to the H$\beta$, are to some extent representative of
the central star temperature ($T_{\rm eff}$).  For an optically 
thick PN, at a given $T_{\rm eff}$, the central star luminosity 
($L_{\ast}$) can be determined from the nebular H$\beta$ luminosity
\citep[e.g.,][]{of06}.  Based on the photoionization models of a 
large sample of optically thick PNe in the Large and Small 
Magellanic Clouds (hereafter, LMC and SMC), \citet[][also 
\citealt{md91}]{dm91} derived empirical relationships between 
$T_{\rm eff}$ and $L_{\ast}$ and excitation class (EC) in the form 
of polynomials (Equations~3.1 and 3.2 in \citealt{dm91}).  These 
empirical relations work in equivalent to the transformation between 
the observed Hertzsprung-Russell (H-R) diagram, $\log{L({\rm 
H}\beta)}$ versus EC, and the true H-R diagram, 
$\log{(L_{\ast}/L_{\sun})}$ versus $\log{T_{\rm eff}}$.  The EC 
parameter was defined in terms of the $\lambda$5007/H$\beta$ and 
$\lambda$4686/H$\beta$ nebular line ratios.  However, we were aware 
that these relationships were based on the models of the optically 
thick PNe in the Clouds, which are both metal-poor ($Z$=0.008 in 
the LMC, and 0.004 in the SMC), while previous and our current 
spectroscopic observations have demonstrated that the bright PNe 
in the outer-disk and the halo of M31 are all metal-rich 
\citep{kwitter12,balick13,fang13,fang15,corradi15}.  The 
relationships given by \citet{dm91} could be metallicity dependent 
\citep{dopita92}, although the [O~{\sc iii}]/H$\beta$ line ratio is 
less dependent than other lines.  Thus whether they are applicable 
to the PNe in M31 might be questionable.

In order to assess the applicability of the relationship of 
\citet{dm91}, we derived the $T_{\rm eff}$ for the M31 disk PNe 
studied by \citet{kwitter12} and \citet{balick13} using the 
relation, and compared them with the {\sc cloudy} model results 
presented in these two papers.  The differences in $\log{T_{\rm 
eff}}$ are mostly less than 0.06~dex. 
We made the same comparison of the $L_{\ast}$ of the same disk 
sample, and found that the differences in 
$\log{(L_{\ast}/L_{\sun})}$ are mostly less than 0.05~dex. 
The disk sample contains the brightest PNe in M31 with 
$m_{\lambda5007}\sim$20.4--20.9.  This comparison study between 
the two sets of $T_{\rm eff}$ and $L_{\ast}$ thus confirms our 
anticipation that the brightest PNe in M31, those within two 
magnitudes from the bright cut-off of the PNLF, should not be 
quite evolved and probably still optically thick (A.~A.\ Zijlstra, 
private communications).  We noticed  
that the empirically derived $L_{\ast}$ of three less bright PNe 
(with $m_{\lambda5007}$ = 20.72, 20.88 and 20.89) are lower than 
their corresponding {\sc cloudy} models by $\sim$0.2~dex. 
This difference might be due to a possibility that the empirical 
relationship of \citet{dm91} underestimates the stellar 
luminosities of fainter PNe, which might no longer be optically 
thick.  We made a similar comparison study of the stellar 
parameters for the M31 bulge and disk PNe of \citet{jc99}, which 
are systematically fainter ($m_{\lambda5007}\sim$20.73--23.16), 
and found that the empirically derived $\log{(L_{\ast}/L_{\sun})}$ 
are lower than the photoionization models by 0.2~dex in average 
(although scatter in the sample of \citealt{jc99} is large due to 
faintness of the targets).  The largest difference in both 
$\log{T_{\rm eff}}$ and $\log{(L_{\ast}/L_{\sun})}$ is found in 
the brightest PN of \citet[][PN~ID M2496, $m_{\lambda5007}$
=20.42]{balick13}.


\begin{table*}
\begin{center}
\begin{minipage}{140mm}
\caption{Estimated Properties of PN Central Stars and the 
Progenitors$^{\rm a}$}
\label{cspne}
\begin{tabular}{lcclllllllll}
\hline\hline
  &  &  &
\multicolumn{3}{c}{\underline{~~~~~~~~$Z$=0.016\,$^{\rm b}$~~~~~~}} &
\multicolumn{3}{c}{\underline{~~~~~~~~$Z$=0.01\,$^{\rm c}$~~~~~~~}} &
\multicolumn{3}{c}{\underline{~~~~~~~~$Z$=0.02\,$^{\rm c}$~~~~~~~}} \\
ID & $\log{T_{\rm eff}}$ & $\log{(L_{\ast}/L_{\sun})}$ &
$M_{\rm fin}$ & $M_{\rm ini}$ & $t_{\rm ms}$ &
$M_{\rm fin}$ & $M_{\rm ini}$ & $t_{\rm ms}$ & 
$M_{\rm fin}$ & $M_{\rm ini}$ & $t_{\rm ms}$\\
\hline
 PN1 & 4.971 & 3.657 & 0.597    & 1.82 & 1.43 & 0.563    & 1.53 & 1.77 & 0.565    & 1.55 & 2.32\\
     &       &       & $\cdots$ & 1.75 & 1.61 & $\cdots$ & 1.40 & 2.31 & $\cdots$ & 1.42 & 3.05\\
 PN2 & 5.077 & 3.464 & 0.594    & 1.80 & 1.49 & 0.556    & 1.47 & 1.99 & 0.554    & 1.45 & 2.82\\
     &       &       & $\cdots$ & 1.72 & 1.69 & $\cdots$ & 1.32 & 2.71 & $\cdots$ & 1.30 & 3.99\\
 PN3 & 5.126 & 3.696 & 0.618    & 2.00 & 1.10 & 0.583    & 1.70 & 1.31 & 0.579    & 1.67 & 1.85\\
     &       &       & $\cdots$ & 1.97 & 1.15 & $\cdots$ & 1.60 & 1.55 & $\cdots$ & 1.56 & 2.25\\
 PN4 & 4.932 & 3.525 & 0.576    & 1.64 & 1.94 & 0.550    & 1.42 & 2.20 & 0.545    & 1.38 & 3.35\\
     &       &       & $\cdots$ & 1.53 & 2.40 & $\cdots$ & 1.26 & 3.13 & $\cdots$ & 1.21 & 5.08\\
 PN5 & 5.052 & 3.702 & 0.606    & 1.90 & 1.27 & 0.576    & 1.64 & 1.44 & 0.577    & 1.65 & 1.91\\
     &       &       & $\cdots$ & 1.84 & 1.38 & $\cdots$ & 1.53 & 1.76 & $\cdots$ & 1.54 & 2.34\\
 PN6 & 4.989 & 3.448 & 0.597    & 1.82 & 1.43 & 0.563    & 1.53 & 1.77 & 0.565    & 1.55 & 2.32\\
     &       &       & $\cdots$ & 1.75 & 1.61 & $\cdots$ & 1.40 & 2.31 & $\cdots$ & 1.42 & 3.05\\
 PN7 & 4.967 & 3.445 & 0.570    & 1.59 & 2.14 & 0.545    & 1.38 & 2.41 & 0.539    & 1.32 & 3.78\\
     &       &       & $\cdots$ & 1.47 & 2.73 & $\cdots$ & 1.21 & 3.55 & $\cdots$ & 1.15 & 6.04\\
 PN8 & 5.082 & 3.310 & 0.585    & 1.72 & 1.69 & 0.559    & 1.50 & 1.89 & 0.559    & 1.50 & 2.57\\
     &       &       & $\cdots$ & 1.62 & 2.00 & $\cdots$ & 1.35 & 2.53 & $\cdots$ & 1.35 & 3.52\\
 PN9 & 5.073 & 3.501 & 0.589    & 1.76 & 1.60 & 0.558    & 1.49 & 1.92 & 0.558    & 1.49 & 2.62\\
     &       &       & $\cdots$ & 1.67 & 1.85 & $\cdots$ & 1.34 & 2.59 & $\cdots$ & 1.34 & 3.61\\
PN10 & 4.980 & 3.687 & 0.602    & 1.87 & 1.34 & 0.567    & 1.56 & 1.66 & 0.570    & 1.59 & 2.13\\
     &       &       & $\cdots$ & 1.80 & 1.47 & $\cdots$ & 1.44 & 2.12 & $\cdots$ & 1.47 & 2.72\\
PN11 & 5.137 & 3.662 & 0.611    & 1.94 & 1.20 & 0.583    & 1.70 & 1.31 & 0.578    & 1.66 & 1.88\\
     &       &       & $\cdots$ & 1.90 & 1.28 & $\cdots$ & 1.60 & 1.55 & $\cdots$ & 1.55 & 2.29\\
PN12 & 4.761 & 3.639 & 0.584    & 1.71 & 1.72 & 0.554    & 1.45 & 2.06 & 0.553    & 1.44 & 2.87\\
     &       &       & $\cdots$ & 1.61 & 2.04 & $\cdots$ & 1.30 & 2.84 & $\cdots$ & 1.29 & 4.10\\
PN13 & 4.906 & 3.423 & 0.567    & 1.57 & 2.24 & 0.536    & 1.30 & 2.86 & 0.530    & 1.25 & 4.58\\
     &       &       & $\cdots$ & 1.44 & 2.92 & $\cdots$ & 1.11 & 4.52 & $\cdots$ & 1.05 & 7.96\\
PN14 & 4.825 & 3.643 & 0.588    & 1.75 & 1.62 & 0.556    & 1.47 & 1.98 & 0.555    & 1.46 & 2.76\\
     &       &       & $\cdots$ & 1.66 & 1.89 & $\cdots$ & 1.32 & 2.71 & $\cdots$ & 1.31 & 3.89\\
PN15 & 5.067 & 3.380 & 0.580    & 1.68 & 1.83 & 0.551    & 1.43 & 2.16 & 0.546    & 1.38 & 3.28\\
     &       &       & $\cdots$ & 1.57 & 2.21 & $\cdots$ & 1.27 & 3.05 & $\cdots$ & 1.22 & 4.94\\
PN16 & 5.240 & 3.331 & 0.633    & 2.13 & 0.93 & 0.583    & 1.70 & 1.31 & 0.579    & 1.67 & 1.85\\
     &       &       & $\cdots$ & 2.12 & 0.94 & $\cdots$ & 1.60 & 1.54 & $\cdots$ & 1.56 & 2.25\\
PN17 & 5.166 & 3.464 & 0.610    & 1.93 & 1.21 & 0.574    & 1.62 & 1.49 & 0.576    & 1.64 & 1.94\\
     &       &       & $\cdots$ & 1.88 & 1.30 & $\cdots$ & 1.51 & 1.84 & $\cdots$ & 1.53 & 2.39\\
\hline
\multicolumn{12}{l}{NOTE. -- All masses are in unit of solar mass 
($M_{\sun}$) and ages in giga years (Gyr).}
\end{tabular}
\begin{description}
\item[$^{\rm a}$] For each PN, $M_{\rm ini}$ values in the first 
and second lines were derived using Equations~1 and 2 of the 
linear initial-final mass relations of \citet[][]{catalan08}, 
respectively.  $t_{\rm ms}$ was derived based on the model grids 
computed by \citet{schaller92}.
\item[$^{\rm b}$] $M_{\rm fin}$ were interpolated from the post-AGB 
model tracks of \citet{vw94}. 
\item[$^{\rm c}$] $M_{\rm fin}$ were interpolated from the post-AGB 
model tracks of \citet{mb16}. 
\end{description}
\end{minipage}
\end{center}
\end{table*}

The EC of our target PNe are mostly between $\sim$3 and 7.6.  We 
derived their central star temperatures and luminosities using the 
relationships of \citet{dm91}.  Despite the applicability of these 
relationships assessed above, the stellar luminosity thus derived 
might still be underestimated for 1) bright, young (and thus 
probably dusty) PNe due to absorption of a large fraction of 
stellar flux by dust, and 2) the PNe that are optically thin to the 
ionizing radiation of the central stars.  Since the extinction of 
our targets are low (Section~\ref{sec3:part1}), the first 
possibility could be discounted.  The relatively faint PNe in M31 
might not be optically thick, and their stellar luminosities 
derived using the empirical relationship might be underestimated. 
According to the discussion above, we revised up the 
$\log{(L_{\ast}/L_{\sun})}$ of the fainter targets 
($m_{\lambda5007}\geq$21) in our sample by 0.2~dex to compensate 
for the possible underestimation.  The stellar temperatures and 
luminosities of our PNe are presented in Table~\ref{cspne} (the 
second and third columns). 

The model-based $\log{(L_{\ast}/L_{\sun})}$ versus EC relation 
of \citet{dm91} actually is similar to the method of 
\citet{zp89}, who derived the central star luminosities of PNe 
using the H$\beta$ flux only.  With the assumption that PNe are 
optically thick to the ionizing stellar radiation, the methodology 
of \citet{zp89} is analogous to the Zanstra method for determining 
stellar temperatures \citep{zanstra31}.  This assumption is 
justified by the very high optical depth of the H~{\sc i} Lyman 
emission lines: each recombination of hydrogen (H$^{+}$ + e$^{-}$) 
eventually results in emission of Balmer lines and the Ly$\alpha$ 
photon \citep{of06}.  Following the method of \citet{zp89} and the 
tabulation of $T_{\rm eff}$ versus $L_{\ast}$ in 
\citet[][p.169]{pottasch84}, we derived the stellar luminosities 
that are close to those derived using the empirical relationship 
of \citet{dm91}.

\begin{figure}
\begin{center}
\includegraphics[width=1.0\columnwidth,angle=0]{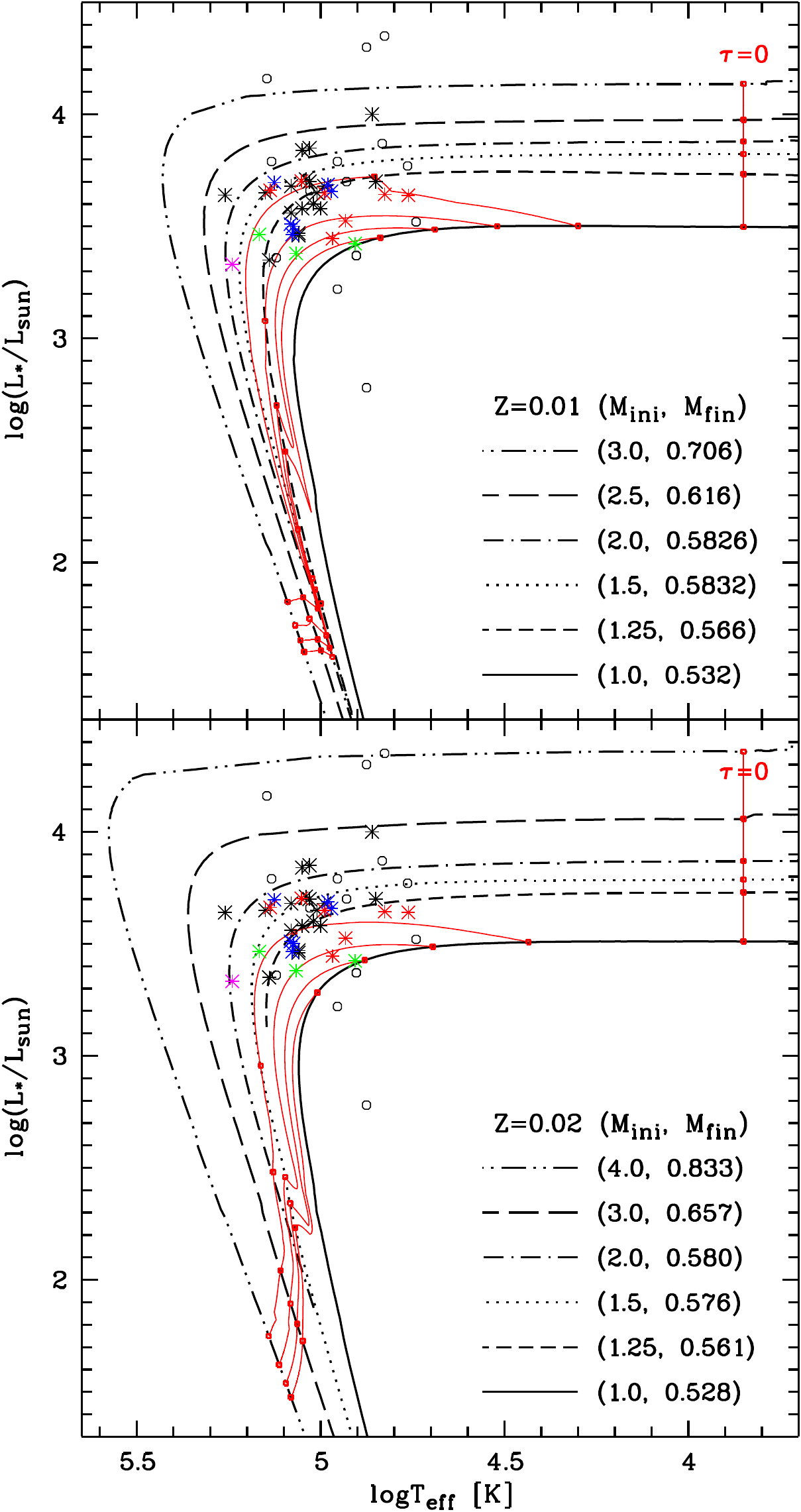}
\caption{Central star positions of M31 PNe in the 
$\log{(L_{\ast}/L_\sun)}$ vs. $\log{T_{\rm eff}}$ diagram.  The 
black asterisks are the disk PNe observed by \citet{kwitter12} 
and \citet{balick13}; our targets are represented by the colored 
asterisks (color codes are the same as in previous figures). The 
black open circles are the M31 bulge and disk PNe studied by 
\citet{jc99}.  Over-plotted in the diagram are the model tracks of 
the H-burning post-AGB sequences calculated by \citet{mb16} at 
$Z$=0.01 (top) and 0.02 (bottom); different line types represent 
different initial and final masses.  Red curves are isochrones for 
the evolutionary ages ($\tau$ = 0, 5000, 10\,000, 15\,000, and 
20\,000~yr) since the zero point of post-AGB defined at 
$\log{T_{\rm eff}}$ =3.85.} 
\label{fig12}
\end{center}
\end{figure}

\begin{figure}
\begin{center}
\includegraphics[width=1.0\columnwidth,angle=0]{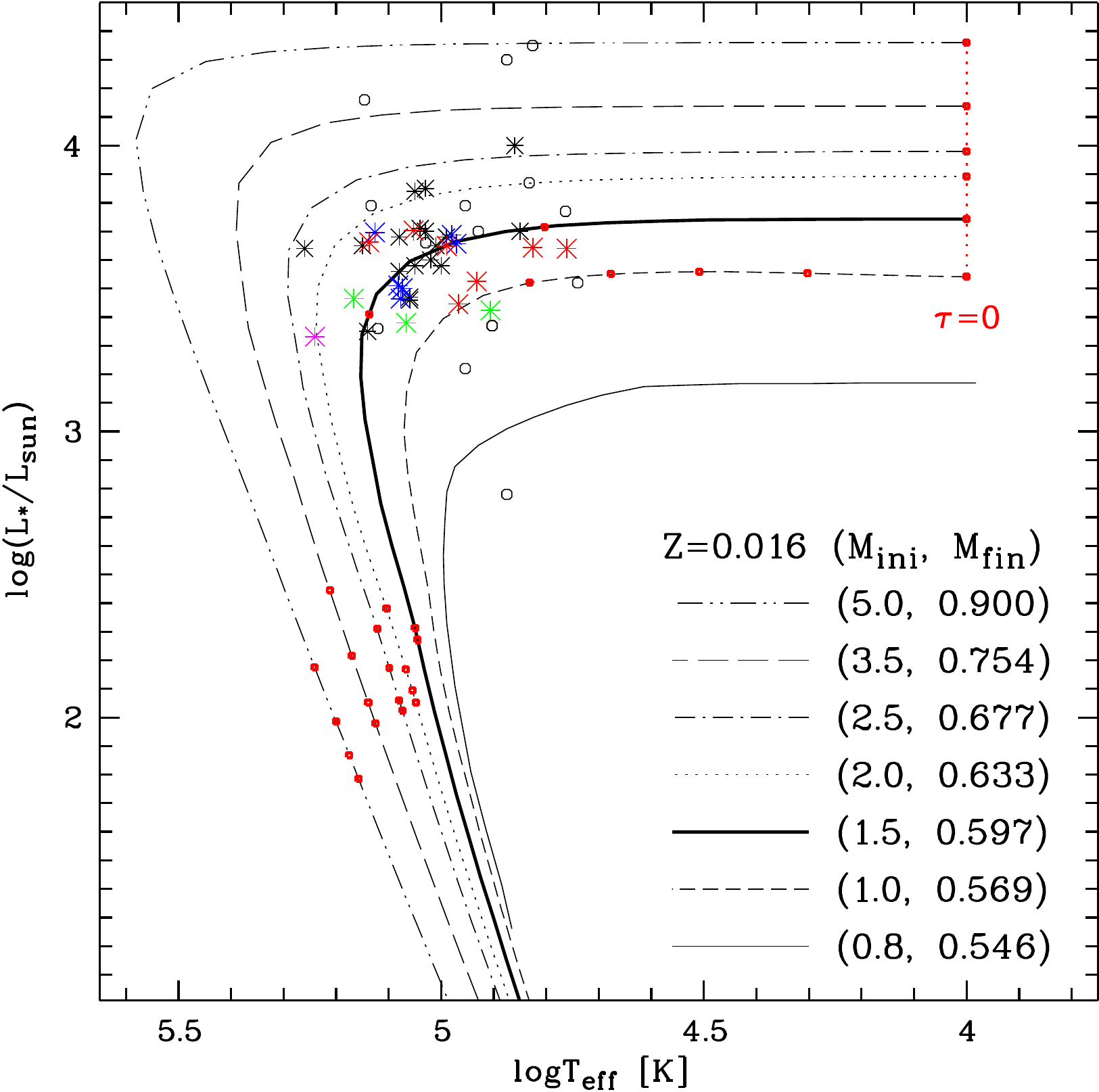}
\caption{Same as Figure~\ref{fig14} but over-plotted with the 
H-burning post-AGB evolutionary models of \citet[][$Z$=0.016]{vw94}.
The model track with 0.546\,$M_{\sun}$ core mass is adopted from 
\citet{schoen83}.  Red dots on the model tracks have the same 
meaning as in Figure~\ref{fig12}.} 
\label{fig13}
\end{center}
\end{figure}

The $\log{T_{\rm eff}}$ versus EC relation of \citet{dm91} is 
dependent on the spectral energy distribution (SED) of the ionizing 
source.  Blackbodies (i.e., Planck function), as assumed in the 
photoionization models of \citet{dm91}, can result in different 
$\log{T_{\rm eff}}$ versus EC relations from the cases where more 
realistic models (like those of \citealt{rauch03}) are used.  At a 
given intensity ratio of two nebular lines, e.g., $I$(He~{\sc ii} 
$\lambda$4686)/$I$(H$\beta$), the stellar atmospheres of 
\citet{rauch03} can predict higher $T_{\rm eff}$ than does a 
blackbody.  In this regard, we tried to constrain the central star 
temperatures and luminosities by fitting the observed line ratios 
(including $\lambda$4363/$\lambda$5007, He~{\sc i}/He~{\sc ii}, 
[O~{\sc ii}]/[O~{\sc iii}], [Ar~{\sc iii}]/[Ar~{\sc iv}], [Ne~{\sc 
iii}]/[O~{\sc iii}], and [S~{\sc ii}]/[S~{\sc iii}]) to those 
predicted by the photoionization models calculated at a huge number 
of grids, the Mexican Million Models dataBase 
\citep[3MdB,][]{morisset15}.  The 3MdB models were computed with 
the {\sc cloudy} program \citep{ferland13} and include both the 
radiation- and the matter-bounded cases, as well as stellar SEDs 
of both blackbody and Rauch's atmospheres.  The tolerance of 
fitting was set to be $\leq$20\% for all line ratios.  Using the 
3MdB, we managed to obtain $T_{\rm eff}$ for nine objects in our 
sample, and they generally agree within the errors with the 
corresponding values derived using the empirical relationship of 
\citet{dm91}.  However, no optimum constraint on stellar luminosity 
was achieved. 

The emergent spectrum of a PN can be affected by many factors, 
including central star parameters ($T_{\rm eff}$, $L_{\ast}$, 
$\log{g}$), density profile of the nebular shell, nebular 
abundances, nebular radius, and filling factor; fixing other 
parameters and varying only $T_{\rm eff}$ and $L_{\ast}$, as was 
usually done with {\sc cloudy} for extragalactic PNe, could be 
unrealistic and the derived parameters problematic.  In particular, 
determining the central star luminosity without knowing the exact 
covering factor ($\leq$1) of the nebula may underestimate 
$L_{\ast}$, not mentioning the case where the nebula is optically 
thin.

\begin{figure*}
\begin{center}
\includegraphics[width=10cm,angle=-90]{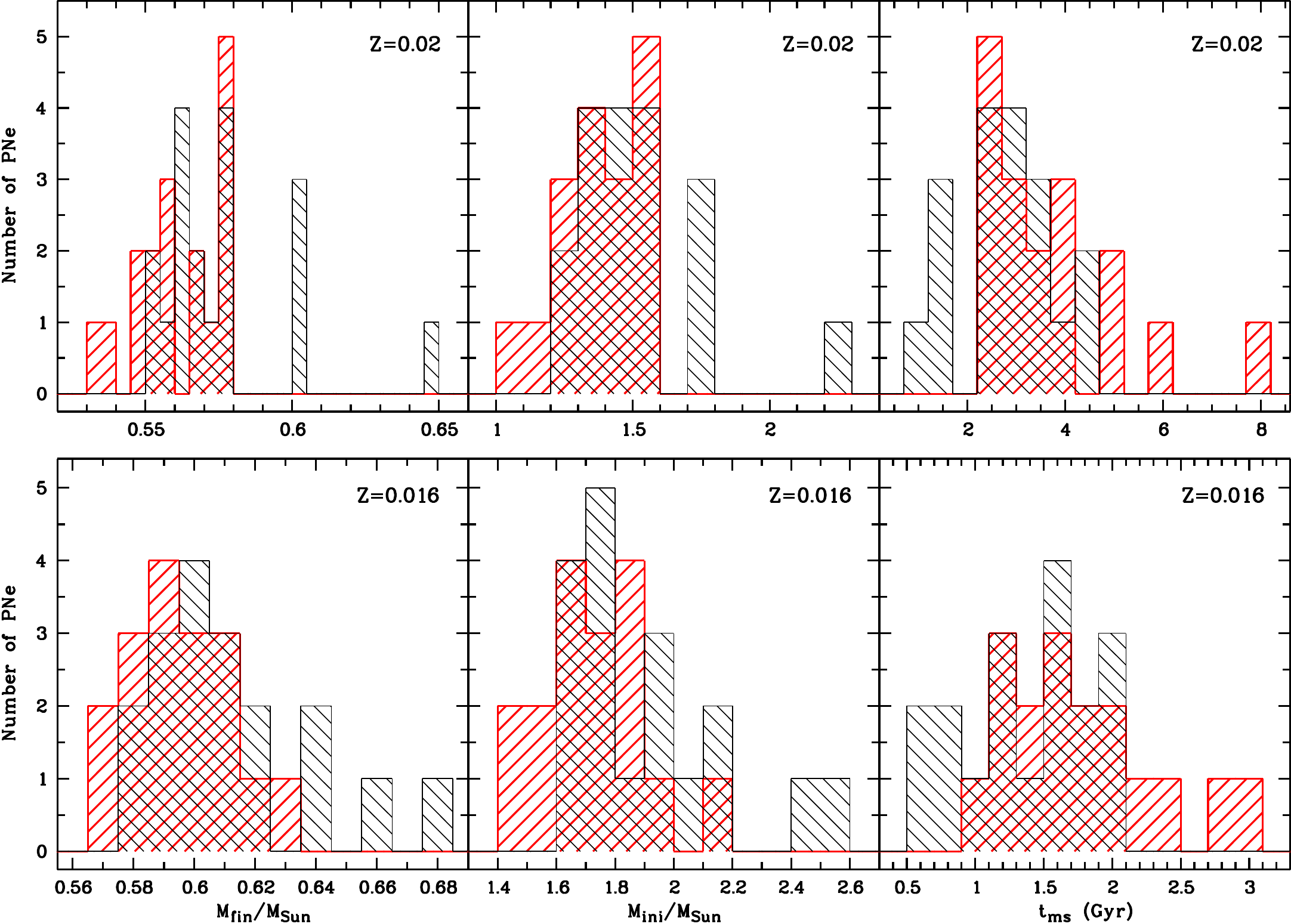}
\caption{Number distributions of the core masses ($M_{\rm fin}$, 
left), initial masses ($M_{\rm ini}$, middle) and main-sequence 
ages ($t_{\rm ms}$, right) of M31 PNe derived using the post-AGB 
evolutionary models of \citet[][$Z$=0.02; top panels]{mb16} 
and \citet[][$Z$=0.016; bottom panels]{vw94}.  $M_{\rm ini}$ 
were derived using the Equation~2 of \citet{catalan08}. Derivation 
of $t_{\rm ms}$ is described in the text.  The red-shaded 
histograms represent our sample, and the black-shaded histograms 
the disk sample.} 
\label{fig14}
\end{center}
\end{figure*}

The central star locations of our PNe in the H-R diagram are shown 
in Figure~\ref{fig12}, along with the new model tracks of the 
H-burning post-asymptotic giant branch (post-AGB) evolutionary 
sequences calculated by \citet[][2016]{mb15} at two metallicities, 
$Z$=0.01 and 0.02, for various initial masses.\footnote{Computations 
of the H-burning post-AGB evolutionary track for the 1.0\,$M_{\sun}$ 
star at $Z$=0.01 and the isochrones presented in Figure~\ref{fig12} 
were recently updated by M.~M.\ Miller~Bertolami (K.\ Gesicki et 
al., in preparation).}  These two metallicities were chosen for 
analysis because they bracket the solar metallicity 
\citep[$Z_{\sun}\sim$0.013--0.018,][]{gn93,gs98,lodders03,
asplund09} and thus are probably suitable for the metallicity 
environment of M31 (as discussed in Section~\ref{sec4:part3}, the 
O/H ratios of the total sample of M31 PNe are mostly located 
between the AGB models with $Z$=0.007 and 0.019). Also over-plotted 
in the H-R diagram are the 18 M31 disk PNe observed by \citet[][16 
PNe]{kwitter12} and \citet[][two PNe]{balick13}.  Our targets are 
located between the model tracks with 0.53--0.58\,$M_{\sun}$ final 
masses, while the disk sample are mostly 0.56--0.657\,$M_{\sun}$, 
with a few objects extending to $<$0.56\,$M_{\sun}$ 
(Figure~\ref{fig12}, bottom).  The 15 disk and bulge PNe from 
\citet{jc99} are also presented in Figure~\ref{fig12}, but with 
large scatter in stellar luminosity.  In order to make a contrast 
study, we also placed all samples in Figure~\ref{fig13} where the 
classical/old post-AGB evolutionary model tracks of \citet{vw94} 
are presented. 

The final core masses ($M_{\rm fin}$) of our targets and the M31 
disk sample were interpolated from the model tracks.  Our targets, 
as constrained by the new post-AGB models, are in the range 
0.536--0.583\,$M_{\sun}$ (Table~\ref{cspne}) with an average mass 
of 0.562($\pm$0.015)\,$M_{\sun}$, while the disk sample are mostly 
0.550--0.602\,$M_{\sun}$ with an average of 
0.571($\pm$0.016)\,$M_{\sun}$.  Based on the old models, the core 
masses of our sample are 0.567--0.633\,$M_{\sun}$, and the disk PNe
are $\sim$0.583--0.660\,$M_{\sun}$.  The initial masses ($M_{\rm 
ini}$) were then derived using the semi-empirical initial-final 
mass relation (IFMR) of white dwarfs given by \citet{catalan08}. 
A comparison between the core masses derived for the two samples 
using the two sets of post-AGB models is presented in 
Figure~\ref{fig14} (left). 

Previously, \citet{kwitter12} derived the $M_{\rm ini}$ of their 
disk sample using Equation~1 in \citet{catalan08}, which is a 
single linear fit to the IFMR in the 1--6\,$M_{\sun}$ range.  In 
order to make a comparison study with the disk sample, we also 
adopted this equation for our sample.  However, in the low-mass 
range (1--2\,$M_{\sun}$), the actual IFMR seems to be flatter than 
what the usual semi-empirical linear fits predict \citep{salaris09,
gesicki14}.  Given that our targets seem to be located between the 
post-AGB evolutionary tracks with the 1 and 2\,$M_{\sun}$ initial 
masses, we also derived the $M_{\rm ini}$ using Equation~2 of 
\citet{catalan08}, which is a linear fit for the $<$2.7\,$M_{\sun}$ 
range and is expected to better describe the IFMR at low masses. 
The $M_{\rm ini}$ thus derived are also presented in 
Table~\ref{cspne} along with those derived using Equation~1 of 
\citet{catalan08}.  The main-sequence lifetimes ($t_{\rm ms}$) were 
then derived from $M_{\rm ini}$ based on the stellar model grids 
computed by \citet[][Tables~45 and 46 therein]{schaller92}. 

Our targets extend to lower core mass regions than the disk sample 
in the H-R diagram, and consequently older main-sequence stellar 
ages.  Compared to the old post-AGB evolutionary models, the new 
models of \citet{mb16} have lowered the core masses and shortened 
the post-AGB evolutionary timescales.  Using the new models, we 
estimated the initial masses of the brightest PNe in our sample 
(PN9, PN10, PN11 and PN16), whose [O~{\sc iii}] luminosities are 
within 1\,magnitude from the PNLF bright cut-off \citep{merrett06}, 
to be $\sim$1.3--1.6\,$M_{\sun}$.  This hints at a tantalizing 
possibility that the brightest PNe may also evolve from very 
low-mass stars.  Figure~\ref{fig14} compares the initial masses and 
main-sequence stellar ages of the two samples of M31 PNe that were 
derived based on the two sets of post-AGB evolutionary models.  It 
is noteworthy that according to the new post-AGB models, our sample 
1) extends to core mass as low as 0.53\,$M_{\sun}$ 
(Figure~\ref{fig14}, top-left), below the lower mass limit 
(0.55\,$M_{\sun}$) for the formation of PNe predicted by the old 
models, and 2) may evolved from young ($\sim$2~Gyr) to intermediate 
age ($\sim$6--8~Gyr) stars (while the old models place our targets 
in the young population, $\lesssim$3~Gyr). 

Within our sample, we did not find obvious systematic difference 
in stellar mass or main-sequence age between the groups of PNe 
associated with different regions, although the three halo PNe 
(PN13, PN15 and PN17) seem to have lower stellar luminosities. 
According to the old stellar models, a post-AGB system with core 
mass lower than 0.55\,$M_{\sun}$ is unable to develop a visible PN 
because its transition time (from the beginning of post-AGB to the 
PN phase) is too long for the star to become hot enough 
($\gtrsim$30\,000~K) to ionize the ever-expanding nebular shell 
\citep[e.g.,][]{schoen83}.  Updated micro- and macro-physics have 
been included in the new post-AGB models of \citet{mb16}, which 
predict higher central star luminosities than the earlier models 
of \citet[][also \citealt{bloecker95} and \citealt{schoen83}]{vw94} 
by $\sim$0.1--0.3~dex at given core masses, and have accelerated 
the post-AGB evolution by a factor of 3 to 8, enabling formation 
of PNe with core masses as low as $\sim$0.53\,$M_{\sun}$.  This 
acceleration in post-AGB evolution, which was previously proposed 
for the existing \citet{bloecker95} model tracks by 
\citet{gesicki14} and has been well confirmed by a recent study 
of the central stars of 32 Galactic bulge PNe \citep{gesicki17,
zijlstra17}, is more significant at lower core masses, which is 
reflected in that the post-AGB timescales predicted by the new 
models are extremely sensitive to core mass, making the low-mass PN 
central stars much more abundant than the more massive ones.

\begin{figure*}
\begin{center}
\includegraphics[width=17.5cm,angle=0]{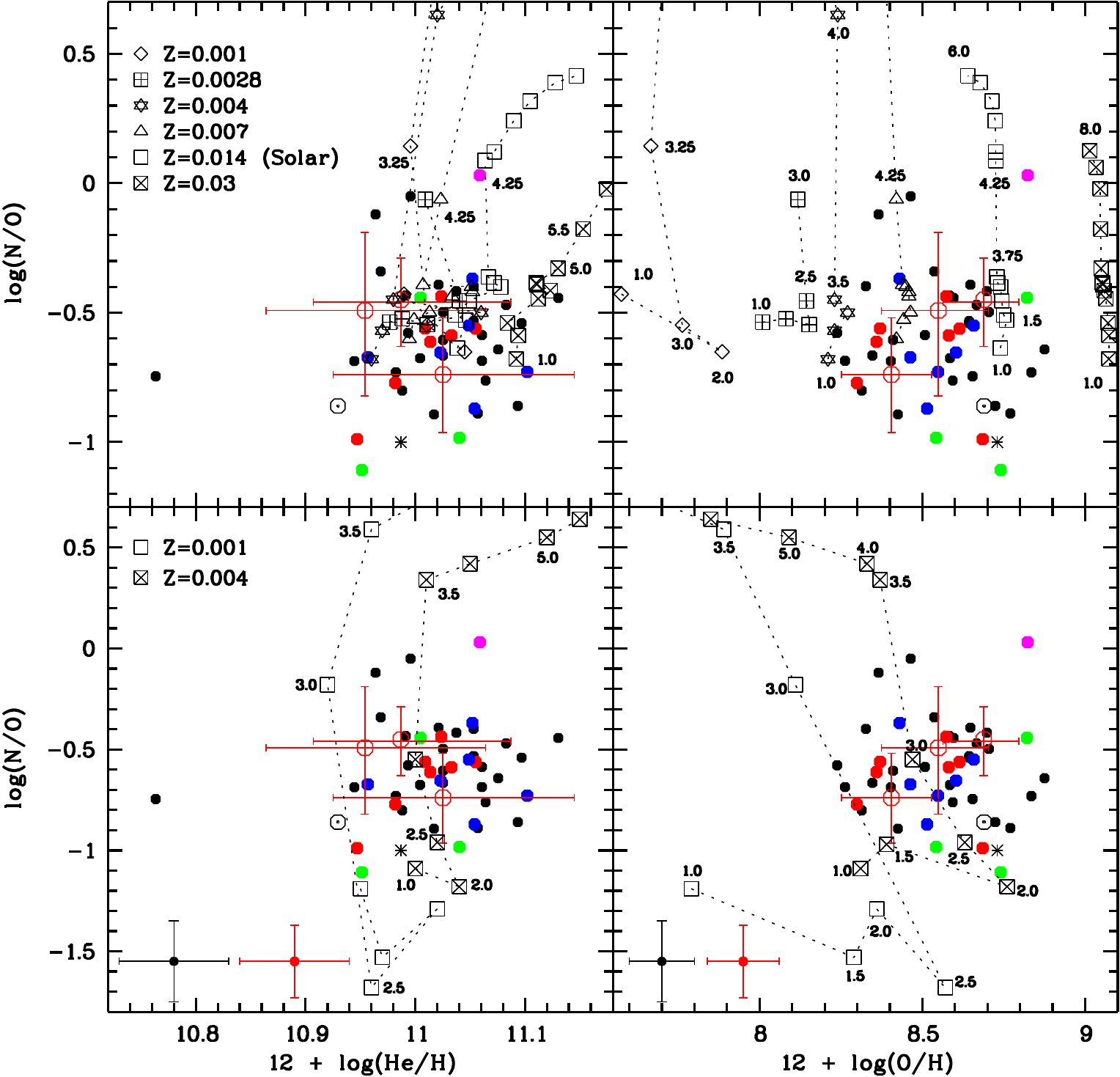}
\caption{The N/O versus the He/H (left) and O/H (right) abundance 
ratios (in logarithm); only our targets and the M31 disk sample 
are presented, along with the abundance ratios of the sun 
\citep{asplund09} and the Orion Nebula \citep{esteban04}. Symbols 
and color codes are the same as in Figure~\ref{fig7}. 
In the top panels: 
AGB model predictions from \citet[][$Z$=0.004]{karakas10}, 
\citet[][$Z$=0.001]{fishlock14} and \citet[][$Z$=0.007, 0.014 and 
0.03]{kl16} for the surface abundances at different metallicities 
are over-plotted for purpose of comparison. 
In the bottom panels: 
AGB model predictions from \citet[][$Z$=0.001]{ventura13} and 
\citet[][$Z$=0.004]{ventura14} are over-plotted. 
Different symbols represent different metallicities; symbols of
the same metallicity are linked by dotted lines to aid
visualization.  Initial mass (in $M_{\sun}$) of the progenitor star 
is labeled for each model.  The vertical scales of the top and 
bottom panels are different to accommodate the AGB model grids.}
\label{fig15}
\end{center}
\end{figure*}

Most of the PNe in our sample are located before the 10\,000~yr 
isochrone (i.e., the post-AGB evolution ages) (Figure~\ref{fig12}). 
This distribution in the H-R diagram is qualitatively in line with 
the fact that these bright M31 PNe should not be quite evolved so 
that they can still be observed today.  A considerable fraction of 
the disk PNe are located before the 5000~yr isochrone; this does 
not necessarily mean that they are very young PNe but might be due 
to overestimated stellar luminosities.  In a similar sense, for 
those PNe located after the 10\,000~yr isochrones, their stellar 
luminosities might be underestimated.  In overall, the locations 
of our sample of M31 PNe in the H-R diagram are reasonable. 


\subsection{Comparison with AGB Model Predictions} 
\label{sec4:part3}

The chemical yields of an AGB star are on dependent on its initial 
mass and metallicity ($Z$).  For an AGB star with $M_{\rm 
ini}\sim$0.8--8\,$M_{\sun}$, its surface abundances can be altered 
due to recurrent mixing events that bring the synthesized material 
to the surface (c.f., reviews \citealt{herwig05}; \citealt{kl14}). 
For stars with $M_{\rm ini}\gtrsim$1.25\,$M_{\sun}$, in the 
thermally pulsing AGB phase, instabilities in the thin He-burning 
shell will drive the third dredge-up, which brings the material 
from He-intershell and enriches the surface with carbon and 
$s$-process elements \citep[e.g.,][]{herwig05}.  Depending on $Z$, 
AGB stars with $M_{\rm ini}\geq$3--4$\,M_{\sun}$ experience the 
second dredge-up and hot bottom burning (HBB), which results in a 
significant increase in the surface nitrogen at the expense of 
carbon and oxygen \citep[e.g.,][]{karakas10}.  The nebular relative 
elemental abundances are thus indicative of $M_{\rm ini}$ and $Z$. 
Analysis of observations against theoretical predictions provides 
insights to stellar astrophysics as well as assessment of the 
stellar evolution models.  A number of theoretical efforts have 
been made to investigate stellar yields from AGB nucleosynthesis 
for various cases of $M_{\rm ini}$ and $Z$ \citep[e.g.,][]{herwig05,
karakas09,karakas14,karakas10,cristallo11,cristallo15,lugaro12,
ventura13,ventura14,fishlock14,shingles15,kl16,pignatari16}.  In 
this section, we make a comparison study between the elemental 
abundances of M31 PNe, as derived from deep spectroscopy, and AGB 
model predictions, aiming at constraining the initial stellar 
masses and investigating the dependence of stellar yields on the 
mass. 

\begin{figure}
\begin{center}
\includegraphics[width=1.0\columnwidth,angle=0]{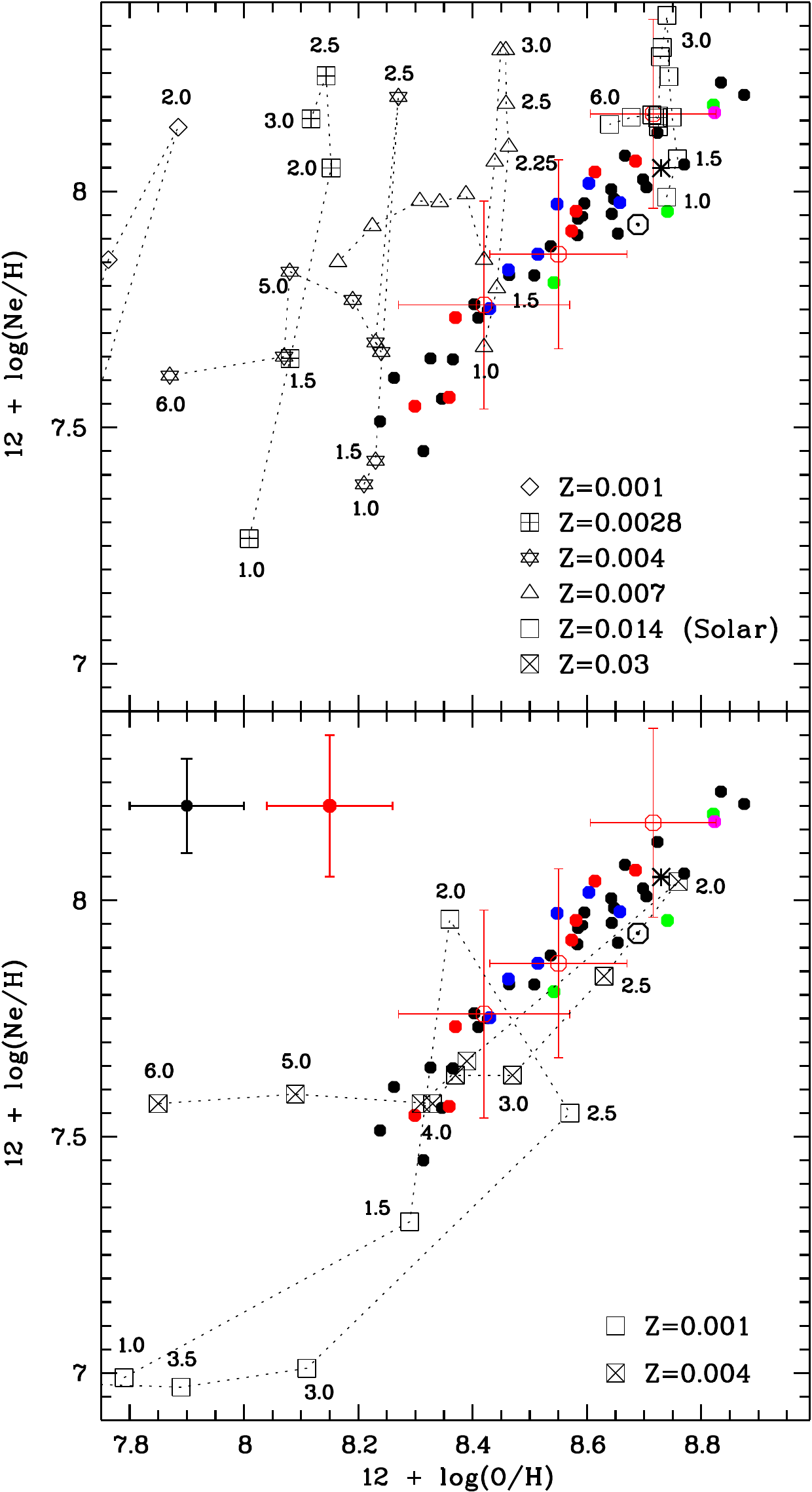}
\caption{Same as Figure~\ref{fig15} but for Ne/H versus O/H.} 
\label{fig16}
\end{center}
\end{figure}

In Figures~\ref{fig15} and \ref{fig16} we compare the He/H, O/H, 
N/O and Ne/H abundance ratios of M31 PNe to those predicted by the 
AGB nucleosynthesis models at different metallicities and initial 
stellar masses.  The PN samples include our PNe (this work and 
Papers~I and II), most of which are associated with substructures, 
and the disk sample observed by \citet[][also \citealt{balick13} 
and \citealt{corradi15}]{kwitter12}.  Sources of the AGB models are 
given in the caption of Figure~\ref{fig15}, and cover a broad range 
of metallicity ($Z$=0.001--0.03).  The He/H and O/H ratios of most 
PNe are located between the AGB models with $Z$=0.004 (SMC) and 
0.014 ($\sim$solar), although the model yields of Karakas et al.\ 
are insensitive to stellar mass at $\leq$3\,$M_{\sun}$; there seems 
to be a few objects with slightly over-solar oxygen, but still 
consistent with the Sun within the errors.  In the AGB models of 
Karakas et al., N/O was assumed to be solar.  Depending on the 
initial mass, all mixing events during the evolution of a low- to 
intermediate-mass star increase nitrogen by a certain amount. 
However, the model predicted N/O of Karakas et al.\ does not extend 
to such low levels as observed in our PN targets (Figure~\ref{fig15},
top).  It is unlikely that these AGB models over-predict N/O, given 
that 1) the N/O ratio only depends on the first dredge-up (because 
all PNe in our sample probably correspond to progenitors with 
$M_{\rm ini}\lesssim$2\,$M_{\sun}$) and 2) for the low-mass AGB 
models, predictions for the first dredge-up more or less all agree.
Thus we conclude that several objects in our sample (as well as the 
disk sample) were probably born in the ISM with very low N/O ratios.

By contrast, N/O predicted by the AGB models of \citet{ventura13,
ventura14} is very sensitive to the initial stellar mass 
(Figure~\ref{fig15}, bottom).  Besides, their model-predicted N/O 
ratios also extend to as low as 0.02 (i.e., $\log$(N/O)=$-$1.68). 
These differences are mainly due to the different initial 
compositions adopted in the two set of models, although 
prescriptions for convection are also different -- the third 
dredge-up and HBB were considered at a lower stellar mass in 
Ventura's models ($\sim$3\,$M_{\sun}$) than in the models of 
Karakas et al.\ ($\gtrsim$4--5\,$M_{\sun}$).  The initial stellar 
masses of the M31 PNe can mostly be constrained by Ventura's models 
to be $\lesssim$3\,$M_{\sun}$, consistent with the estimate of 
\citet[][also \citealt{balick13}]{kwitter12}.  The two sets of AGB 
models thus generally constrain the M31 PNe in metallicity and 
place an upper-limit on their initial masses.  Ventura's AGB models 
also predict higher O/H at given metallicities and stellar masses, 
mainly due to the inclusion of convective boundary mixing (or 
overshooting) at different stellar evolutionary stages, especially 
at the thermally pulsing AGB (TP-AGB) phase, of their models that 
leads to the dredge up of oxygen to the surface.  The broad range 
of $Z$ (0.004--0.014) in M31 PNe can also be seen in the 
$\log$(Ne/H) versus $\log$(O/H) diagram (Figure~\ref{fig16}).

\begin{figure*}
\begin{center}
\includegraphics[width=17.5cm,angle=0]{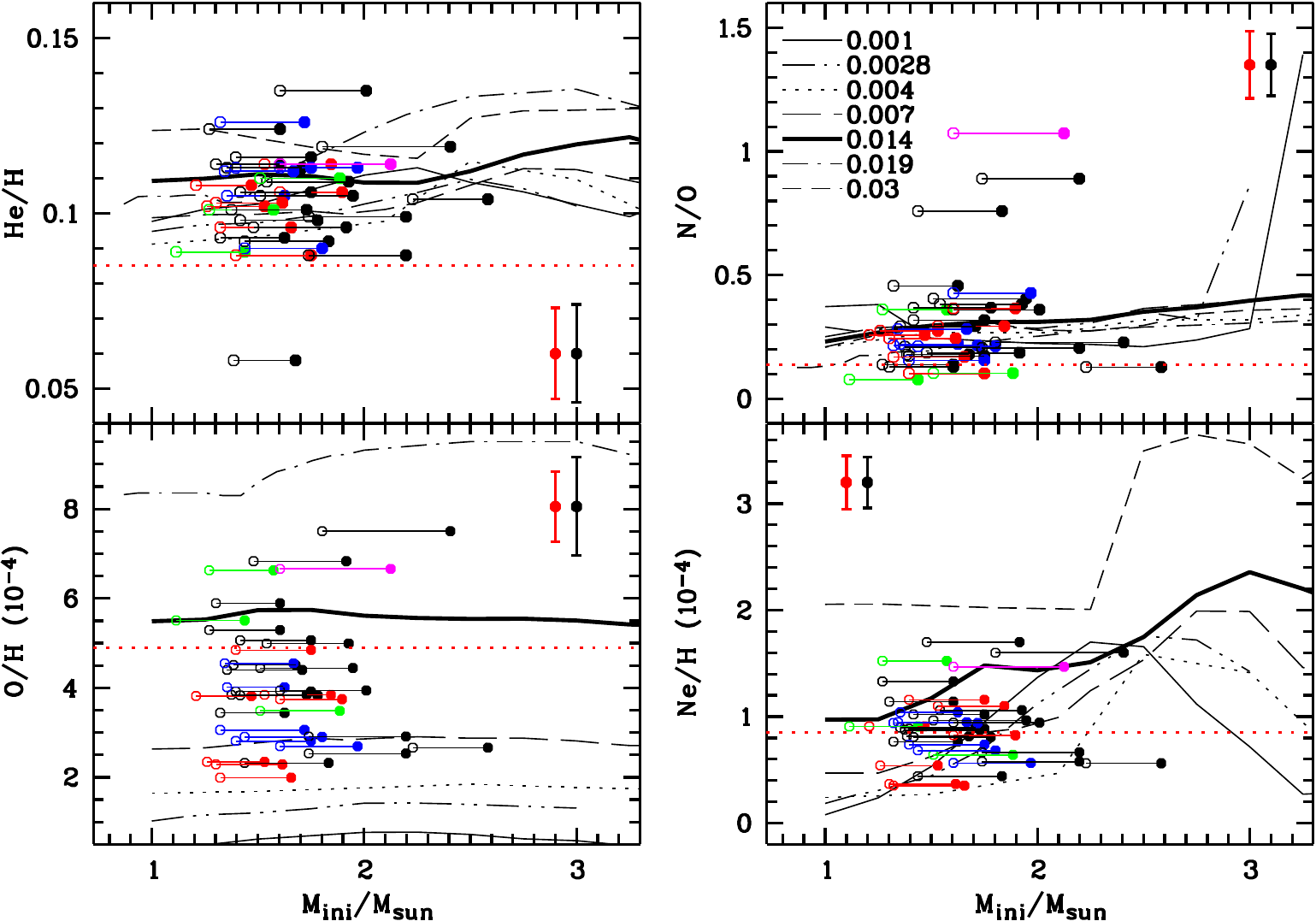}
\caption{Abundance ratios of M31 PNe versus initial mass.  Sample 
includes our targets and the disk PNe from \citet{kwitter12} and 
\citet{balick13}.  For each PN, the initial masses derived from 
the post-AGB evolutionary models of \citet[][filled circle]{vw94} 
and \citet[][open circle]{mb16} are both presented and connected 
by a solid line.  Color codes of data points are the same as in 
Figures~\ref{fig15} and \ref{fig16}.  Representative abundance 
errors of the two samples are indicated.  The AGB yields predicted 
at different metallicities, as represented by different line types 
(see the legend), are over-plotted.  The AGB models are from 
\citet[][$Z$=0.004]{karakas10}, \citet[][$Z$=0.001]{fishlock14}, 
\citet[][$Z$=0.007, 0.014 and 0.03]{kl16}, and 
\citet[][$Z$=0.019]{marigo01}.  The horizontal red-dotted lines 
mark the solar values \citep{asplund09}.} 
\label{fig17}
\end{center}
\end{figure*}

Using the initial masses of the total sample of M31 PNe estimated 
in Section~\ref{sec4:part2}, we demonstrate the dependence of 
abundance ratios on $M_{\rm ini}$ in Figure~\ref{fig17}.  The AGB 
model predictions presented in the top panels of Figures~\ref{fig15}
and \ref{fig16} are also used in Figure~\ref{fig17}.  The 
theoretical AGB yields from \citet{marigo01} at $Z$=0.019 are 
also included.  The $M_{\rm ini}$ were those derived using the 
linear fit to the IFMR at low-masses given by \citet[][Equation~2 
therein]{catalan08}.  For each PN, the initial masses derived based 
on the old post-AGB evolutionary models of \citet{vw94} and the 
new models of \citet{mb16} were both presented.  Generally, the 
abundance ratios of our targets and majority of the disk sample
agree within the uncertainties with the theoretical AGB yields at 
given stellar masses.  In the disk sample, there is an ``outlier'' 
with very low He/H, probably due to large uncertainty in the 
observations \citep{kwitter12}. 

Interestingly, there is no obvious difference in the O/H ratios 
between our halo sample and the outer-disk sample, although the 
initial stellar masses of the disk PNe seem to be higher than 
those of our targets at given O/H.  The O/H ratios of both samples 
of M31 PNe are located between the AGB models with $Z$=0.004 and 
0.019 (Figure~\ref{fig17}, bottom-left); 
these two $Z$ values correspond to [Fe/H]$\sim-$0.5 and 
$\sim$0.1, respectively, generally metal-rich compared to the halo 
metallicities of M31 \citep[][]{ibata07,ibata14}.  Given that 
oxygen is the best observed element in PNe and has been used as a 
proxy for the metallicity of progenitor stars, from which oxygen 
is assumed to be inherited, this distribution of the observed 
O/H with respect to the AGB model predictions confirms the 
metal-rich nature of these M31 PNe.  The insensitivity of the 
model-predicted N/O on stellar mass at low masses, as predicted by 
the AGB models of Karakas et al.\ (Figure~\ref{fig15}, top), is 
better seen in Figure~\ref{fig17} (top-right).  This is because 
the first dredge-up that occurs in AGB stars with $M_{\rm 
ini}\leq$3--4\,$M_{\sun}$ does not increase nitrogen much, and 
oxygen is essentially unchanged.  Figure~\ref{fig17} (top-right) 
also shows that significant increase in N/O occurs at $M_{\rm 
ini}>$3\,$M_{\sun}$ for the metal-poor case $Z$=0.001; in the 
metal-rich AGB stars (e.g., $Z$=0.014), N/O increases 
significantly for $M_{\rm ini}\geq$4\,$M_{\sun}$ \citep{kl16}. 
The M31 PNe discussed in this paper mostly have initial masses 
$\lesssim$2\,$M_{\sun}$.  The dispersion in Ne/H of the total 
sample is much smaller than that in O/H, with an average ratio of 
8.64($\pm$3.35)$\times$10$^{-5}$ for our sample and 
9.47($\pm$3.34)$\times$10$^{-5}$ for the disk sample.  Abundances 
of the PNe are mostly consistent with the AGB model predictions 
within the errors.

\begin{figure*}
\begin{center}
\includegraphics[width=5.2cm,angle=-90]{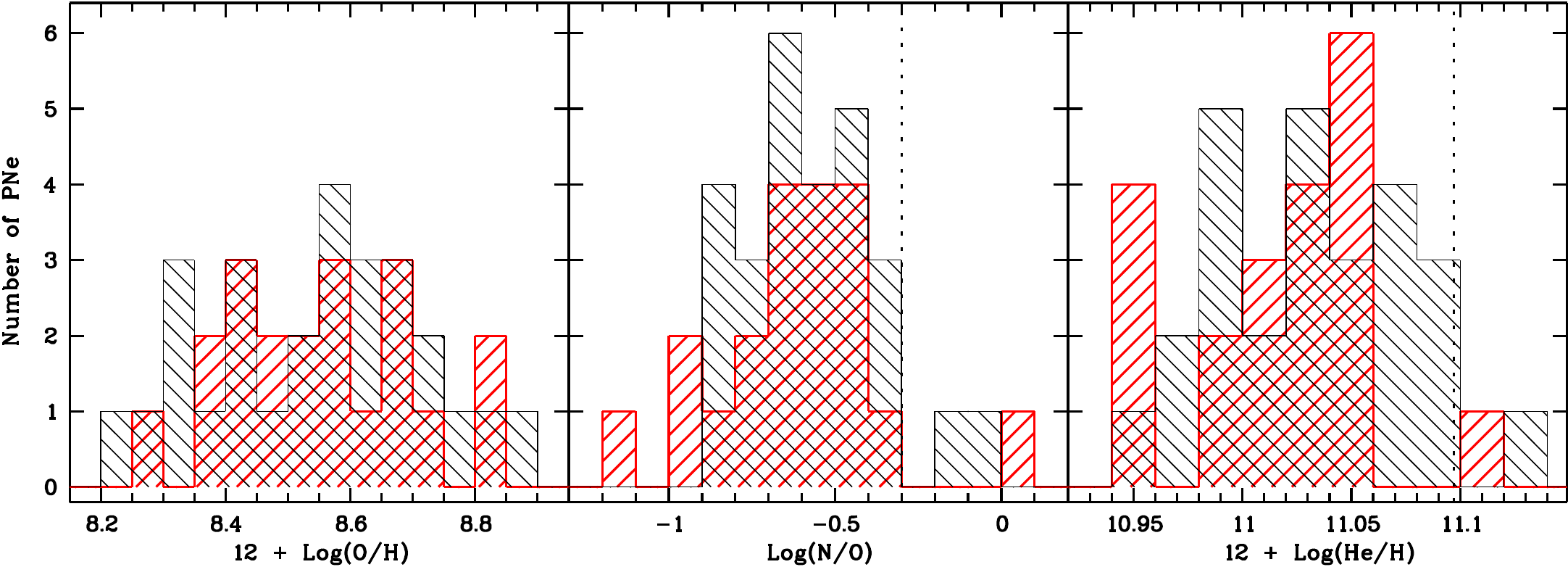}
\caption{Histograms of M31 PNe in the O/H (left), N/O (middle) and 
He/H (right) ratios in logarithm.  The red-shaded histograms 
represent our sample, and the black-shaded histograms the disk PNe. 
The vertical dotted lines mark the original definition of Galactic 
Type~I PNe by \citet{peimbert83}: N/O$>$0.5 or He/H$>$0.125.} 
\label{fig18}
\end{center}
\end{figure*}

Looking at the initial masses derived from the old post-AGB 
evolutionary models, the N/O ratio seems to start to increase at 
$M_{\rm ini}$ close to 2\,$M_{\sun}$ (Figure~\ref{fig17}, 
top-right).  This trend is shifted to lower masses when the new 
post-AGB models are used.  A similar behavior in N/O has also 
been found in a recent study of the Galactic PNe \citep{henry18}. 
This is at variance with the current AGB model prediction that N/O 
increases as a consequence of HBB at $M_{\rm 
ini}\gtrsim$3--5\,$M_{\sun}$ \citep{cristallo11,cristallo15,
karakas14,kl16,ventura15,dicris16}, and tentatively implies HBB 
might actually occur at $M_{\rm ini}<$3\,$M_{\sun}$, or even at 
$\lesssim$2\,$M_{\sun}$.  More detailed investigation is needed to 
assess this possibility. 

There are three PNe with N/O$>$0.5 (a criterion to define the 
Galactic Type~I PNe): PN5 and PN16 from the disk sample of 
\cite{kwitter12}, and target PN16 in our sample.  However, the He/H 
ratios of the three PNe are $<$0.125, at odds with the definition 
of a Type~I PN.  The initial stellar masses of the three PNe are 
all $\lesssim$2.2\,$M_{\sun}$ (Figure~\ref{fig17}, top-right). 
Apart from the convective mixing processes (mostly the first 
dredge-up for low-mass stars), N/O in stars with initial masses of 
1--4\,$M_{\sun}$ may also be affected by the non-standard mixing 
processes such as thermohaline mixing and stellar rotation 
\citep{karakas09}.  This means that AGB stars with high N/O ratios 
may not necessarily have evolved from the intermediate-mass 
(3--8\,$M_{\sun}$) stars, where HBB is needed to enhance the 
surface nitrogen, but from low-mass (1--3\,$M_{\sun}$) stars that 
rotate reasonably rapidly on the main sequence and/or experience 
non-convective mixing processes during the ascent of the first 
giant branch.  However, to what extent N/O can be affected by the 
thermohaline mixing, especially for the low-mass stars, is still 
unclear.  So far very few stellar evolution models have included 
thermohaline mixing and rotation \citep[e.g.,][]{cl10,cantiello10}.
More detailed and quantitative investigation of the effects of 
these extra-mixing processes on the chemical yields of low-mass 
stars is still needed. 

It is worth to mention that the solar abundances quoted in this 
paper are photospheric values \citep[][Table~1 therein]{asplund09}, 
which differ slightly from the initial abundances (or the bulk 
abundances; \citealt{asplund09}, Table~5 therein) due to the 
combined effects of gravitational settling and diffusion.  When 
the Sun becomes a red giant, its photospheric abundances will go 
back to the bulk values because of convective mixing.  During the 
post-main sequence evolution, the first and third dredge-ups change 
the abundances of the Sun.  Stellar evolution models of low-mass 
(1--2\,$M_{\sun}$) stars predict that helium of the Sun can be 
increased by $\sim$0.05~dex, and N/O increased by nearly 0.3~dex 
\citep{mb16}.  Thus the actual locations of the Sun in 
Figures~\ref{fig15}, \ref{fig16} and \ref{fig17} (also 
Figures~\ref{fig7}) should be more consistent with the M31 PNe 
than what they appear to be now. 

Finally, we emphasize that although the central star parameters of 
our sample were estimated using the empirical relationships, which 
again, were based on the photoionization modeling of the Magellanic 
Cloud PNe, their locations in the H-R diagram are generally 
reasonable, and their abundance ratio versus $M_{\rm ini}$ 
relations are consistent with the AGB model predictions. 
Compared to the outer-disk sample, our halo/substructure PNe 
occupy the regions that correspond to relatively lower core masses 
in the H-R diagram, and consequently have older stellar 
ages.  However, detailed photoionization modeling of our targets 
is needed to constrain the central star properties, so that a 
comparison study with the disk PNe can be made in a more consistent 
manner.  This will be presented in a subsequent paper for a more 
extended GTC sample.\footnote{In the semester 2017B (2017 September 
1 -- 2018 February 28) runs, we have obtained GTC OSIRIS long-slit 
spectra for another seven PNe in the outer halo of M31 (GTC program 
\#GTC98-17B, PI: M.~A.\ Guerrero).}

\subsection{Radial Oxygen Abundances}
\label{sec4:part4}

A general comparison of our PN sample and the M31 disk PNe in He/H,
O/H and N/O (in logarithm) is presented in Figure~\ref{fig18}. 
Including the three Northern Spur PNe studied in Paper~I, our 
sample of 20 PNe has a range of 8.30--8.82 in 12+$\log$(O/H) with 
an average of 8.55$\pm$0.15, close to that of the disk sample (27 
PNe).  Our sample has slightly larger scatter in N/O 
(Figure~\ref{fig18}, middle), with an average of 0.28$\pm$0.21.

\begin{figure*}
\begin{center}
\includegraphics[width=14.0cm,angle=0]{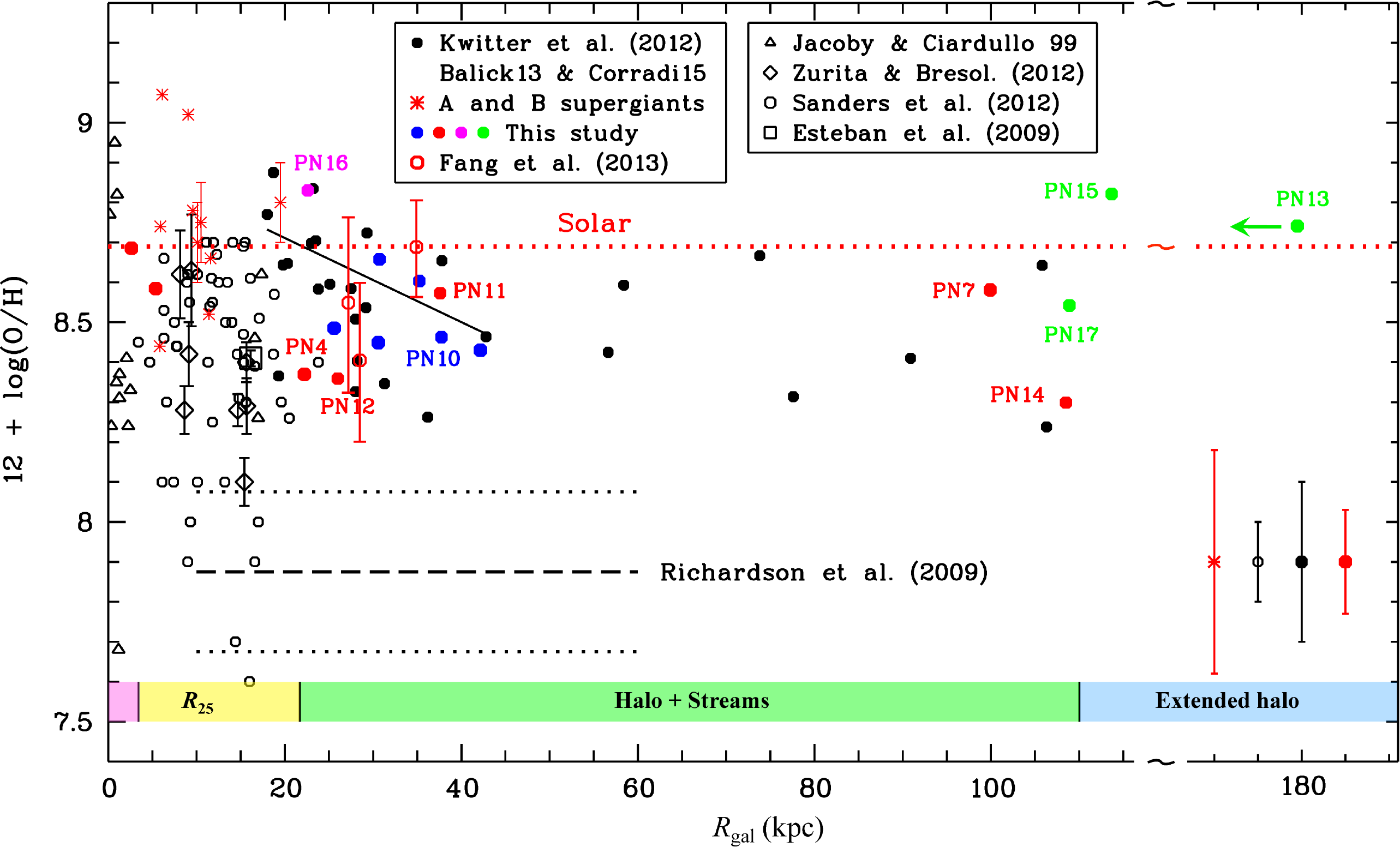}
\caption{Radial distribution of oxygen in M31.  Galactocentric 
distances ($R_{\rm gal}$) have been corrected for the inclination 
of M31 disk.  Our GTC sample are the color-filled circles, 
color-coded in the same manner as in previous figures.  Note that 
the radial distance of PN13 ($\sim$180~kpc) is estimated assuming 
that it is on the extended disk, and is only an upper limit.  Other 
data sets (see the legend) are explained in the text.  The solid 
black line is a linear fit to the outer-disk PNe of 
\citet[][18$\lesssim$$R_{\rm gal}\lesssim$43~kpc]{kwitter12}. 
Typical error bars of different samples are given in the 
bottom-right corner.  The horizontal red-dotted line marks the 
solar value \citep[8.69,][]{asplund09}.  In order for convenience 
in abundance comparison, the distance between 115 and 170~kpc is 
not properly scaled. 
The horizontal black-dashed and dotted lines represent the 
mean metallicity and dispersion of halo stars between 10 and 
60~kpc (estimated from the measurements of \citealt{richardson09}).
The colored bars in the figure bottom mark the radii of M31's bulge 
\citep[pink; $\sim$3.4~kpc,][]{irwin05}, the optical disk (yellow; 
$R_{25}\approx$22~kpc), and the outer halo as well as the farthest 
outreach of the Giant Stream (green; $\sim$8$\degr$, corresponding
to 110~kpc in projection); further out is more extended halo 
(light blue).} 
\label{fig19}
\end{center}
\end{figure*}

The radial distribution of oxygen in M31 represented by our PNe is 
shown in Figure~\ref{fig19}, where literature samples are also 
presented, including the M31 disk and bulge PNe from \citet{jc99} 
and \citet{sanders12}, the outer-disk PNe from \citet{kwitter12}, 
\citet{balick13} and \citet{corradi15}, M31 H~{\sc ii} regions from 
\citet[][object ID K932]{esteban09} and \citet[][nine objects with 
$T_{\rm e}$ determined]{zb12}, and three A-F supergiants from 
\citet{venn00} and seven B-type supergiants from \citet{trundle02}. 
Galactocentric distances ($R_{\rm gal}$) have been rectified for 
the inclination of M31 disk (see caption of Figure~\ref{fig1}). 

In Figure~\ref{fig19} we mark the radii of M31's bulge, the 
inner/optical disk ($R_{25}$, see caption), the halo (as well as 
the boundary of streams), and the extended halo. The bulge radius 
is adopted from \citet{irwin05} and well accommodates the bulge 
PNe studied by \citet{jc99}.  Almost all the disk PNe observed by 
\citet[][also \citealt{balick13} and \citealt{corradi15}]{kwitter12}
are beyond $R_{25}$. 
The oxygen abundances of M31 disk PNe seem to show a marginally 
negative gradient \citep[$-$0.011$\pm$0.004 
dex\,kpc$^{-1}$,][]{kwitter12} within 40~kpc from the centre of 
M31, and tend to flatten out to $\gtrsim$100~kpc.  In our halo 
sample, members of which are mostly associated with the 
substructures of M31, we also found a similar trend of flattening 
in oxygen.  In Paper~II, based on a limited sample, we drew a 
conclusion that the PNe in the Northern Spur and those associated 
with the Giant Stream have homogeneous oxygen abundances (see 
Figure~13 in Paper~II).  Including the ten new targets (PN8--PN17), 
our extended sample displays generally consistent oxygen 
abundances. 

A very prominent feature in Figure~\ref{fig19} is the solar oxygen 
of the outermost nebula PN13.  Its apparent galactocentric radius 
is 42~kpc (Table~\ref{targets}); if we assume that PN13 is in the 
extended disk of M31 \citep{ibata05}, its rectified radius will be 
$\sim$180~kpc, which could be set as an upper-limit distance of 
this PN.  The 12+$\log$(O/H) value of PN15 is 0.13~dex above the 
Sun \citep[8.69,][]{asplund09}, a difference that is close to the 
typical uncertainty in oxygen of our sample.  However, the exact 
galactocentric radii of PN13 and PN15 are unknown because of the 
dubious disk membership according to their kinematics 
(Figure~\ref{fig2}). 

Another halo object PN17 has an oxygen abundance $\sim$0.15~dex 
below the Sun.  A galactocentric radius of 109~kpc is estimated 
for PN17, if we assume that it is in the outer disk.  However, 
considering that its radial velocity deviates significantly from 
the extended disk (Figure~\ref{fig2}, bottom), PN17 is probably a 
halo PN, or even associated with its substructure. 
PN17 is located near the NE Shelf, an overdensity of 
metal-rich RGB, whose stellar populations were found to be similar 
to those of the Giant Stream \citep{ferguson05}.  Numerical 
simulations have suggested that the NE Shelf might be debris from 
the continuation of the Giant Stream \citep{ibata04,fardal08,mr08}.
Given that the oxygen abundance of PN17 is very close to that of 
PN7, which is well associated with the Giant Stream \citep{fang15}, 
we thus suggest that PN17 might be associated with the NE Shelf. 

PN14 is located in the southeast outer halo of M31, and has a 
rectified galactocentric distance of 108~kpc if we assume that it 
is in the extended disk.  Its oxygen abundance is $\sim$0.4~dex 
below the Sun.  According to its radial velocity and location, we 
suggest that PN14 might be associated with the Giant Stream (while 
in \citealt{merrett06} this PN was not identified to be related to 
any substructure).  Although compared to PN7, PN14 is not located 
right on the stellar orbit proposed by \citet[][see also 
Figure~\ref{fig2}]{merrett03}, considering the large spatial 
extension (along the direction orthogonal to the stream) of the 
Giant Stream \citep{mccon03}, this association is possible.  As 
reported in Section~\ref{sec3:part6}, PN14 has been observed by 
\citet[][PN~ID M2507]{corradi15}, who used the same instrument. 
The difference in 12+$\log$(O/H) between the two observations is 
well within the measurement uncertainty; the galactocentric distance
of PN14 given by \citet[][106~kpc]{corradi15} is slightly smaller 
than ours, which is due to a slightly different distance to M31 
adopted.  If we assume PN14 belongs to the Giant Stream, by applying
its three-dimensional structure \citep{mccon03}, we estimate a 
distance of roughly 50~kpc for this PN.  The galactocentric 
distances of PN14 and PN17 are both uncertain.

\begin{figure}
\begin{center}
\includegraphics[width=1.0\columnwidth,angle=0]{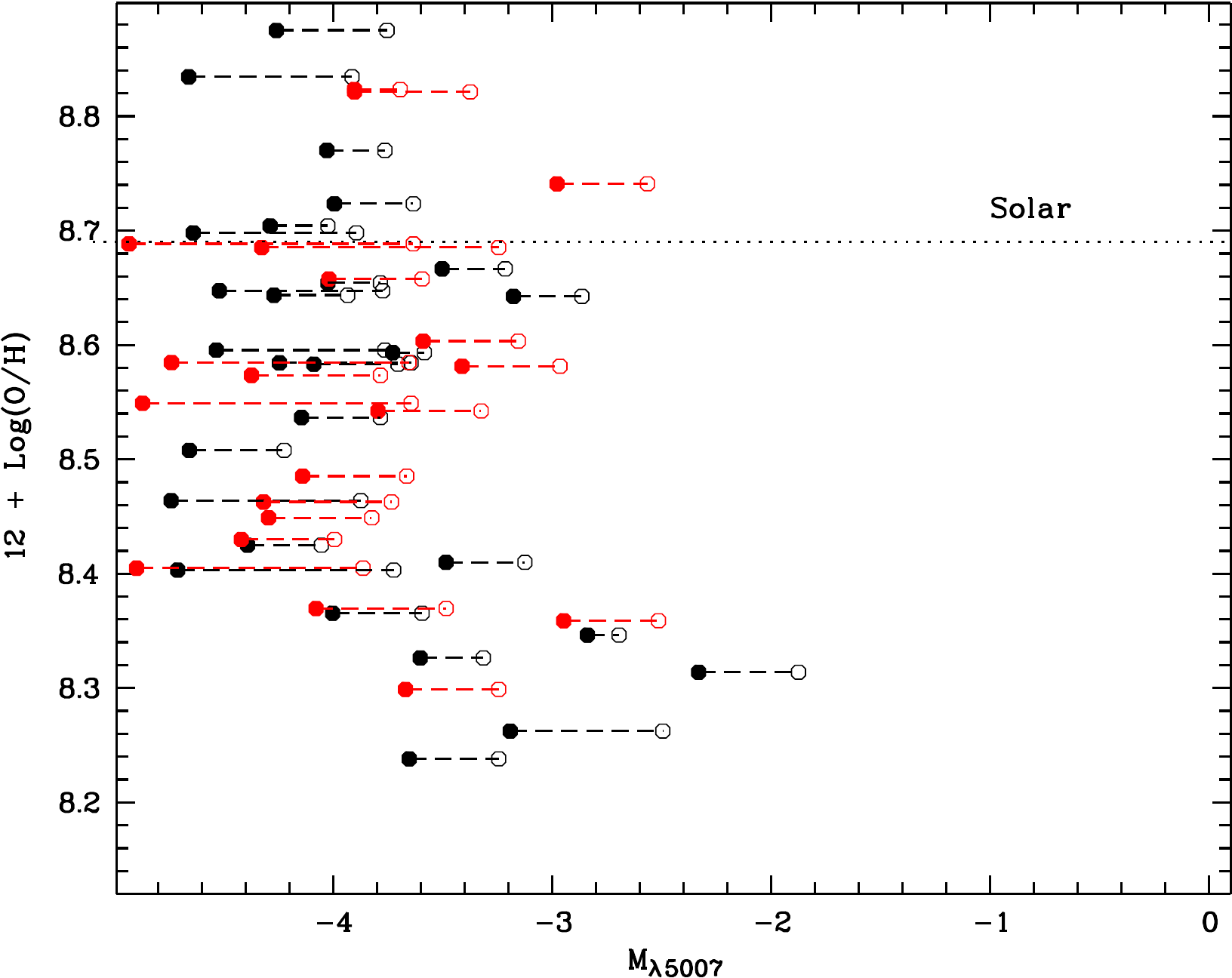}
\caption{Oxygen abundances of M31 PNe versus the absolute 
magnitudes at [O~{\sc iii}] $\lambda$5007.  Our halo/substructure 
PNe are red symbols, and the outer-disk PNe are black.  For each 
PN, the [O~{\sc iii}] magnitudes with and without extinction 
correction (filled and open circles, respectively) are both 
presented and connected by a dash line.  Solar oxygen is marked by 
a horizontal dotted line.} 
\label{fig20}
\end{center}
\end{figure}

Although PN8, PN9 and PN10 have been identified by \citet{merrett06}
as in the Northern Spur (Table~\ref{targets}), their oxygen 
abundances are different: the former two PNe are both close to the 
Sun, while the latter is 0.23~dex below the solar value 
(Figure~\ref{fig19}).  These three PNe, together with the other six
in the Norther Spur, form a sample in this substructure with 
12+$\log$(O/H)$\sim-$0.26--0.  Being 0.13~dex above the Sun, the 
12+$\log$(O/H) value of PN16 is among the highest in our sample. 
The galactocentric distance of PN16 (22.6~kpc) was estimated using 
the distances to M31 (785~kpc) and M32 \citep[763~kpc,][]{karach04} 
and its angular distance ($\sim$0\fdg4) to the centre of M31.  It 
is interesting to note that the oxygen abundance of PN16 seems to 
well fit the trend of the weakly negative gradient represented by 
the disk PNe of \citet{kwitter12}.  

Together with the M31 outer-disk PNe, our halo targets are 
metal-rich ([O/H]$\sim-$0.4 to 0) compared to the mean metallicities
of the dominant stellar populations at similar galactocentric radii 
in the outer halo ([Fe/H]$\lesssim-$1.5 to $\sim-$0.6; e.g., 
\citealt{brown06a,brown08}; \citealt{chapman06,chapman08}; 
\citealt{koch08}; \citealt{richardson09}).  In contrast with the 
high degree of inhomogeneity in the metallicity of M31's extended 
halo (from the most metal-poor population in the outer regions 
with $-2.5<$[Fe/H]$<-1.7$, to the relatively metal-rich population 
in the inner halo where $-0.6<$[Fe/H]$\lesssim$0; 
\citealt{ibata14}), the averaged values of [O/H] of our halo PNe 
in different kinematic groups (i.e., different substructures) are 
quite similar, $\sim-0.2$ to 0, with modest dispersion, and are 
consistently higher than the average [Fe/H] of stars in the 
respective substructures.  The uniformly high oxygen abundances 
of our PNe indicate that they formed from metal-rich ISM, and 
probably belong to the population that is distinct from the 
underlying halo populations of M31. 

All PNe selected for our spectroscopic observations are among 
the brightest in M31, within 2~magnitudes from the bright cut-off 
of the [O~{\sc iii}] PNLF, which allows a reliable spectral 
analysis and accurate abundance determinations (Section~\ref{sec3}).
Whether the metallicity of bright PNe depends on the [O~{\sc iii}] 
luminosity still needs careful studies, although this possible 
relation has previously been investigated for the PNe in LMC 
\citep{richer93}, M31 \citep{jc99}, M32 \citep{rm08}, M33 and the 
Milky Way \citep[e.g.,][]{magrini04}.  In these galaxies, oxygen 
abundances of the bright PNe are found to be independent of the 
[O~{\sc iii}] absolute magnitude 
($M_{\lambda5007}$), and the brightest PNe are representative of 
the whole PN population.  For the bright PNe 
($M_{\lambda5007}\leq-$3.7) in M31's outer disk, oxygen abundance 
has no dependence on $M_{\lambda5007}$, although a slight tendency 
of decreasing oxygen seems to exist in a few objects on the fainter
tail \citep[][Figure~4 therein]{corradi15}.  A very similar trend 
was found in our halo/substructure sample (Figure~\ref{fig20}). 
Here $M_{\lambda5007}$ were derived from the [O~{\sc iii}] apparent
magnitudes ($m_{\lambda5007}$, adopted from \citealt{merrett06}), 
which were corrected for the foreground (and also internal) 
extinction using $c$(H$\beta$). 

Moreover, as predicted by the new post-AGB evolutionary 
models, low-mass central stars of PNe evolve much faster than 
previously thought \citep{mb16,gesicki17}, thus increasing the 
possibility that lower mass (and thus relatively older) stars may 
also form bright PNe that are still visible.  Our population 
analysis (Section~\ref{sec4:part2}; Table~\ref{cspne}) also 
indicates that bright PNe might evolve from low-mass stars, which 
is different from the traditional view that bright PNe evolve from 
rare, high-mass progenitors \citep[e.g.,][]{schoen07}.  Hence, we 
conclude that our sample well represents the population of PNe in 
the halo of M31, regardless of their spatial locations and 
kinematics.  As one of the most efficient coolants in PNe,  the 
[O~{\sc iii}] $\lambda$5007 luminosity is highly dependent on the 
central star effective temperature \citep{dopita92,schoen07}, but 
is also regulated by the nebular metallicity (in particular, the 
fraction of oxygen in O$^{2+}$); thus slight dependency of the 
[O~{\sc iii}] $\lambda$5007 luminosity on oxygen abundance may 
exist. 

The discussion above is based on the assumption that oxygen is 
neither created nor destroyed during the evolution of PN 
progenitors (in our case, low-mass stars; Section~\ref{sec4:part2}),
although third dredge-up and HBB might enhance the oxygen of some 
AGB stars \citep[e.g.,][]{kl14,delgado15}.  In order to compare 
with M31's halo metallicities indicated by [Fe/H], we also assumed 
that [Fe/O]$\approx$0 (i.e., the solar case) for the bright PNe in 
M31.

\subsection{Possible origin of Luminous PNe\\ 
in the Halo of M31} 
\label{sec4:part5} 

Our previous studies of the M31 halo PNe associated with 
substructures were based on two {\it a priori} assumptions: 
1) these PNe represent the stellar populations of the substructures 
where they are located (i.e., they formed {\it in situ}); and 2) 
the PNe associated with substructures have different origins -- and 
thus probably belong to different populations -- from those in the 
disk of M31.  However, so far we did not find any discernible 
difference between our halo sample and the outer-disk PNe in the 
radial distribution of oxygen at $R_{\rm gal}\lesssim$110~kpc 
(Figure~\ref{fig19}). 

Based on the old post-AGB evolutionary models, as were adopted
by \citet{kwitter12}, the stellar ages of our halo sample and the 
outer-disk sample are both constrained to be $\lesssim$2--3~Gyr. 
These M31 PNe thus are the young population, with their stellar 
ages in line with the episode of star formation that occurred 
across the whole disk system of M31 2--4~Gyr ago 
\citep[e.g.,][]{richardson08,bernard12,bernard15a,bernard15b,
williams15}.  The onset of this recent global starburst may 
correspond to an encounter with M33, whose stellar disk also 
experienced enhanced star formation $\sim$2~Gyr ago 
\citep{williams09,bernard12}.  This M31-M33 interaction also 
explains the stellar streams seen in the M31 halo \citep{mccon09}, 
and has been invoked to account for the luminous, oxygen-rich PNe 
in M31's outer disk \citep{balick13,corradi15}. 

\emph{However}, 
using the well assessed new post-AGB evolutionary models, we 
confined the main-sequence ages of our halo sample to be mostly 
$\sim$2--5~Gyr, with the oldest being $\sim$6--8~Gyr; while the 
outer-disk sample are mostly $\lesssim$1--4~Gyr.  We thus 
conjecture that our targets probably formed prior to the encounter 
with M33.  Obviously, our sample represents the population that is 
different from the underlying, smooth, extended (and mostly 
metal-poor) halo component of M31 \citep{ibata07,ibata14}, which 
was formed through repeated accretion of smaller galaxies in the 
distant past.  These bright PNe seem to resemble the younger, 
metal-rich population in the outer stream of M31, as revealed by 
\emph{HST} pencil-beam pointings on the Giant Stream 
\citep{brown06a,bernard15a}.  The metallicity of the stream fields 
enriched continuously from [Fe/H]$\sim-$1.5, to at least solar 
level about 5~Gyr ago \citep{bernard15a}.  This timeline of metal 
enrichment is generally consistent with the stellar ages of our 
metal-rich sample.  $N$-body simulations suggested that the Giant 
Stream and other stream-like features in the halo are debris of a 
massive ($\gtrsim$10$^{9}$--10$^{10}\,M_{\sun}$) progenitor that 
was recently disrupted during the course of merger 
\citep[e.g.,][]{ibata04,geehan06,font06,fardal06,fardal07,fardal08,
fardal13,mr08,sadoun14}.  The extended star-formation history and 
a broad range of metallicity ($-$1.5$\lesssim$[Fe/H]$\lesssim$0.2) 
discovered in the stream fields can be explained by a disk galaxy 
progenitor \citep{brown06a,brown06b,bernard15a}.  If the stellar 
streams in M31's halo indeed have a common origin, our sample of 
halo PNe then probably formed through extended star formation in 
this possibly massive, disk-like progenitor.  Moreover, some 
simulations predict that the remnant of the disrupted satellite 
resides in the NE Shelf \citep[e.g.,][]{fardal08,fardal13,sadoun14};
PN17 in our sample is located in this region and might be 
associated with this substructure (see Section~\ref{sec4:part4}). 

On the other hand, despite the systematic discordance in the 
estimated stellar ages (and also in kinematics) between the two 
samples of M31 PNe, their consistently high oxygen abundances (see 
Figure~\ref{fig19}) signifies something maybe in common. Recently, 
it was suggested that the PNe associated with the outer stellar 
streams might have formed from the same metal-rich ISM as did the 
outer-disk sample, but acquired the kinematics of the streams 
during a subsequent encounter with M33 \citep{balick17}.  This 
postulation hints at a possibility that the halo PNe on the 
streams might have their origins in the M31 disk, where the 
metal-rich ISM mostly resides.  The close interaction between M31 
and M33, combined with the recent impact of the Giant Stream's 
progenitor, could heat up the main disk and redistribute the disk 
material into the substructures we see today \citep{bernard15a}. 

The interpretations presented above are highly speculative, given 
that the M31 halo is extremely complex and the samples of PNe with 
high-quality spectroscopic observations are still very limited. 
In addition, \citet{magrini16} suggested that radial migration of 
the inner-disk PNe in M31 could be important and explain the 
flattening of oxygen gradient (compared to the H~{\sc ii} regions).
This migration mechanism unlikely occurred to our halo PNe because 
to fling a star inside the disk into halo regions far beyond 
$R_{25}$ requires many orbital scatterings due to gravitational 
anomalies, such as dense spiral arms \citep[e.g,][]{bt08}. 

A recent proper motion analysis and cosmological simulations of 
the M31-M33 system suggests it is unlikely that M33 made a recent 
($<$3~Gyr), close ($<$100~kpc) passage about M31 \citep{patel17}; 
this is inconsistent with the scenario proposed by \citet[][also 
\citealt{putman09}]{mccon09}.
If true, the interactions between M31 and the Giant Stream's 
progenitor then seem more plausible to explain the kinematics of 
the metal-rich PNe in the M31 halo.

M32 was speculated to be responsible for the Giant Stream 
\citep{ibata01a,ferguson02,choi02,merrett03}, but kinematical 
studies ruled out this possibility \citep{ibata04}.  The stellar 
orbit proposed by (\citealt{merrett03}; see Figure~\ref{fig2}) that
links the Giant Stream to the Northern Spur is rather generic yet 
highly schematic.  The light-of-sight distance to M32, although 
previously derived \citep[e.g.,][]{jensen03,karach04}, is still 
uncertain.  The N/O ratio of PN16 is higher than our halo PNe, 
also casting doubts on the previous hypothesis that M32 was the 
progenitor of substructures. 

\subsection{Comments on Individual Objects} 
\label{sec4:part8}

Several PNe in our GTC sample are interesting in terms of 
abundances, spatial locations, and kinematics, and worth extra 
attention.  Although most of these PNe have already been discussed
in previous sections, their main characteristics are briefly 
emphasized here. 

{\it PN7} ---
The mostly distant PN so far studied in the extended halo of M31. 
Well located in the southeast extension of the Giant Stream 
and with an oxygen abundance close to the Sun, PN7 is an archetypal 
object that represents the group of metal-rich nebulae associated 
with the outer-halo streams at large galactocentric radii. 

\emph{PN13} ---
An outer-halo PN spatially located on the Giant Stream.  If we 
assume that PN13 is in the extended disk, its rectified 
galactocentric distance will be $\sim$180~kpc, making it the most 
distant solar-metallicity PN in M31 so far observed.  However, its 
disk membership is highly questionable due to kinematics. 

\emph{PN14} ---
It might be associated with the Giant Stream.  Its oxygen is 
$\sim$0.4~dex below the Sun, but is close to those of PN4 and PN12 
on the stream. 

\emph{PN15} ---
Its oxygen is $\sim$0.13~dex above the Sun, making it the most 
metal-rich PN so far observed in the outer halo of M31.  Same as 
PN13, its actual distance to the centre of M31 is unclear. 

\emph{PN16} ---
The only M32 PN analyzed in this work.  Its O/H and N/O ratios are 
both higher than the other PNe in our sample.  PN16 also has the 
highest oxygen abundance among all PNe so far spectroscopically 
studied in M32 \citep{richer99}.  The estimated stellar age 
($t_{\rm ms}\sim$2~Gyr) of PN16 is consistent with the younger 
stellar population discovered in M32 
\citep[2--5~Gyr,][]{monachesi12}.  This bright PN 
($m_{\lambda5007}$=20.78) probably well represents the young 
population of M32.  Given that currently reliable abundance 
measurements of the M32 PNe are extremely sparse, our observations 
provide valuable nebular abundances of this dwarf elliptical galaxy.

\emph{PN17} ---
This halo PN might be associated with the NE Shelf, as judged from 
its O/H, spatial position, and kinematics (see the discussion in 
Section~\ref{sec4:part4}).  If it is true, PN17 will be the first 
PN discovered and studied in this halo substructure.

\section{Summary, Conclusion\\
and Future Work}
\label{sec5}

With the aim of studying the properties and possible origins of 
stellar substructures in the halo of M31, we carry out deep 
spectroscopy of PNe using the 10.4\,m GTC.  Following our previous 
effort \citep{fang15}, where seven PNe associated with the Northern 
Spur and the Giant Stream were targeted, we obtained high-quality 
GTC long-slit spectra of ten PNe that reside at different regions 
and cover a vast area of the M31 system.  These new targets are 
associated with the two known substructures, the eastern and 
southern regions of M31's halo, and dwarf satellite M32.  The 
OSIRIS spectrograph secures a wavelength coverage of 
$\sim$3630--7850\,{\AA} and detection of a number of 
plasma-diagnostic emission lines, including the temperature-sensitive
[O~{\sc iii}] $\lambda$4363 and [S~{\sc iii}] $\lambda$6312 lines 
and the density-diagnostic [S~{\sc ii}] $\lambda\lambda$6716,6731 
doublet.  We also observed the [S~{\sc iii}] 
$\lambda\lambda$9069,9531 nebular lines in four PNe, whose GTC 
spectrum extends beyond 1\,$\mu$m.  Ionic and elemental abundances 
(relative to hydrogen) of helium, oxygen, nitrogen, neon, sulfur, 
argon and chlorine were derived using different temperatures and 
densities according to ionization stages. 

The N/O ratios of our halo sample are mostly $<$0.4, and 
He/H$<$0.126, indicating that they might be Type~II, i.e., of 
relatively low-mass progenitors.  These abundance are generally 
consistent with the outer-disk sample recently observed with the 
8--10\,m telescopes \citep{kwitter12,balick13,corradi15}.  In both 
samples, Ne/H is well correlated with O/H; in some objects, argon 
is under-abundant with respect to the Ar/H versus O/H correlation. 
The ``sulfur anomaly'', originally found in Galactic PNe, also 
exists in the M31 PNe, even for the objects with detection of the 
[S~{\sc iii}] $\lambda\lambda$9069,9531 lines. Although with large 
scatter, Cl/H of our targets are generally in line with the Cl/H 
versus O/H correlation as defined by the Galactic H~{\sc ii} regions 
with best measurements of chlorine.  In an M32 PN, we found a 
relatively high N/O ratio ($\sim$1.07$\pm$0.24). 

The central star temperatures and luminosities of our targets were  
derived using the empirical method.  In the H-R diagram, our sample 
occupies the regions that correspond to relatively lower core 
masses than does the outer-disk sample. 
As constrained using the new post-AGB evolutionary models, 
which have recently been assessed through studies of Galactic PNe, 
the core masses of our sample of halo PNe are 
0.53--0.58\,$M_{\sun}$, and those of the outer-disk sample are 
0.55--0.64\,$M_{\sun}$; the main-sequence ages of our targets are 
then mostly $\sim$2--5~Gyr, with the oldest being $\sim$6--8~Gyr; 
while the outer-disk PNe are mostly $\sim$1--4~Gyr. 
{\it If the new post-AGB evolutionary models are adopted in our 
population analysis, the PNe so far observed in M31 probably all 
evolved from low-mass ($\lesssim$2.2\,$M_{\sun}$) progenitors, 
which formed from the metal-rich ([O/H]$\gtrsim-$0.4) ISM.} 
In the H-R diagram, our GTC targets are mostly located 
before the theoretical isochrone of 10\,000~yr since the beginning 
of post-AGB evolution; given that they are the brightest PNe in 
M31, their locations in the diagram are generally consistent with 
the high probability that these nebulae are not quite evolved in 
the PN phase and still stay around their peak [O~{\sc iii}] 
luminosities.  Our estimated central star parameters thus are, 
although still preliminary, generally reasonable and suggest that 
the brightest PNe in M31 are probably optically thick. 

The He/H, O/H, N/O and Ne/H ratios of both samples of M31 PNe were 
compared with the AGB nucleosynthesis models, and general 
consistency between the observations and theoretical predictions of 
AGB yields was found. From the model-predicted abundance ratios, we 
constrained the upper-limits of the initial stellar masses of both 
samples of M31 PNe to be $<$3\,$M_{\sun}$, in line with our mass 
estimate based on the new post-AGB evolutionary models.  We also 
studied the dependence of abundance ratios on initial mass, and 
found in overall excellent agreement.  In particular, as a 
stellar-mass indicator, the observed N/O ratios mostly agree, 
within the errors, with the AGB models of Karakas et al.\ at low 
masses (1--3\,$M_{\sun}$), except for several outliers.  Although 
still limited by the sample size, N/O of M31 PNe seems to start to 
increase at $\lesssim$2\,$M_{\sun}$, a trend similar to what has 
been recently found in Galactic PNe, indicating that HHB might 
actually occur in very low-mass stars. According to the AGB models, 
O/H of the combined sample of M31 PNe span a broad range that 
encompasses the metallicities of the SMC ($Z$=0.004) and the Sun 
($Z\lesssim$0.02), irrespective of the initial masses.  The O/H 
ratios of our halo sample are similar to those of the outer-disk 
PNe.  From the oxygen abundance, spatial location and kinematics, 
we suggest that PN17 in the eastern halo might belong to the NE 
Shelf. 

Our extended sample of the halo/substructure PNe exhibit uniformly 
high oxygen abundances with modest scatter.  We found nearly-solar 
oxygen in several PNe that are located in the outer halo. 
The most interesting target with solar oxygen is on the southeast 
extension of the Giant Stream, with a sky-projected galactocentric 
distance of $\sim$50~kpc.  In one of the outermost PNe, we even 
found slightly over-solar oxygen ([O/H]$\approx$0.13); there is 
also a halo target in our GTC sample with slightly sub-solar 
oxygen, [O/H]$\sim-$0.4.  Our targets probably belong to the 
population that is different from the underlying, smooth halo 
component of M31, but more like the metal-rich populations in the 
streams.  The estimated stellar ages of our sample (mostly 
$\sim$2--5~Gyr) are consistent with the metal enrichment history 
recently unveiled in the stellar fields of the Giant Stream. 
\emph{If} the substructures where our targets are associated have 
a common origin (i.e., they are tidal debris from a possibly 
massive, disk-like satellite of M31), our deep spectroscopy of 
nebulae confirms the extended star-formation history in this 
satellite that was previously unveiled through \emph{HST} 
photometric studies. 
Alternatively, our observations cannot exclude the possibility 
that our targets might originally formed in the M31 disk, but 
were scattered to the halo regions and gained their current stream 
kinematics as a result of interactions with M31's satellite, which 
can heat up and perturb the stellar disk and redistribute the disk 
material. 
Either of the above two interpretations supports the current 
astrophysical picture that M31's halo evolved through complex 
galactic mergers and interactions. 

Through unprecedentedly deep spectroscopy of a sample of PNe mostly 
associated with the substructures in the halo of M31, we have 
obtained intriguing results which, together with the recent 
spectroscopic observations of the outer-disk sample by other 
research groups at the 8--10\,m class telescopes, may have profound 
effects in our understanding of the M31 system, especially its 
highly structured extended halo, which so far has been rarely 
investigated through quantitative spectroscopy of the ISM.  Our 
findings are qualitatively consistent with the complexity of M31's 
giant halo.  However, one still needs to realize that at this stage 
no definite conclusions can be drawn and all interpretations are 
highly speculative, given the limited PN samples and the complex 
evolutionary history of M31. 

The number of PNe discovered and identified in the outskirts of M31 
has been increased in recently years.  To date our observations 
mainly focus on the PNe in the Northern Spur and the Giant Stream 
substructures that cover the southeast, the eastern and the 
northern regions of the halo.  In the follow-up observations at the 
10\,m GTC, we target the western halo objects and extend further 
north and south, $>$2$\degr$ from the centre of M31.  Through these 
combined efforts, we aim to construct a statistically significant 
and spatially unbiased sample for detailed analysis (including 
photoionization modeling), to make a census study of M31's extended 
halo, and eventually to obtain a grand picture of its evolution.

\acknowledgments

We are grateful to the anonymous referee whose excellent 
comments and suggestions greatly improved this article.  This work 
is supported by grants from the HKRGC (HKU7062/13P) and the Seed 
Fund for Basic Research at the HKU (201611159037).  This work is 
partly supported by grants DGAPA/PAPIIT-107215 and 
CONACyT-CB2015-254132.  R.G.B.\ acknowledges support from the 
Spanish Ministerio de Econom\'{i}a y Competitividad, through 
projects AYA2016-77846-P and AYA2014-57490-P.  X.W.L.\ acknowledges
financial support from the National Key Basic Research Program of 
China 2014CB845700.  M3B is partially supported by ANPCyT 
through grant PICT-2014-2708 and by MinCyT-DAAD bilateral 
cooperation through grant DA/16/07. 
X.F.\ is grateful to Albert A.\ Zijlstra, Bruce Balick, and Romano 
R.~L.\ Corradi for in-depth discussion.  We thank Paolo Ventura for 
providing us the AGB model predictions.  We also thank Michael R.\ 
Merrifield for giving us the permission to reproduce the stellar 
orbit (in Figure~\ref{fig2} of this paper) based on Figure~2 in 
\citet{merrett03}.  This research also made use of NASA's 
Astrophysics Data System (http://adsabs.harvard.edu), the SDSS DR10 
Science Archive Server (SAS; http://data.sdss3.org/), the SDSS DR12 
Finding Chart Tool 
(https://skyserver.sdss.org/dr12/en/tools/chart/), and the SIMBAD 
Astronomical Database (http://simbad.u-strasbg.fr/simbad/). 

\vspace{3mm}
\emph{Software:} {\sc iraf}, {\sc midas}, SAOImage~DS9.

\emph{Facility:} GTC (OSIRIS).



\end{document}